\documentclass[11pt,reqno]{amsart}

\usepackage[margin=1.0in]{geometry}
\usepackage{amsmath}
\usepackage{amsfonts}
\usepackage{mathrsfs}
\usepackage {amssymb,euscript}
\usepackage {amsmath}
\usepackage {amsthm}
\usepackage {amscd}
\usepackage{geometry}
\usepackage{hyperref}
\usepackage{color}
\usepackage{float}
\usepackage{epsfig}
\usepackage{caption}
\usepackage{graphicx}
\usepackage{euscript,bbm,color,slashed,enumerate,bm}
\usepackage{subcaption}
\usepackage{tikz}
\usepackage{tkz-euclide}
\usepackage[T1]{fontenc}
\usepackage{ulem}
\usepackage{verbatim}

\usepackage[foot]{amsaddr}

\usepackage{graphics}
\allowdisplaybreaks

\title[Spacelike Singularities in Cosmological Spacetimes]{Quantitative blow-up estimates for spacelike singularities in gravitational-collapse cosmological spacetimes}
\date{\today}

\author{Xinliang An$^*$$^1$}\author{Haoyang Chen$^{\dag}$$^1$}\author{Taoran He$^{\ddagger}$$^1$}
\address{$^{1}$\small Department of Mathematics, National University of Singapore,
10 Lower Kent Ridge Road, Singapore, 119076 }

\email{$^*$matax@nus.edu.sg}
\email{$^{\dag}$hychen@nus.edu.sg}
\email{$^{\ddagger}$taoran{\textunderscore}he@u.nus.edu}

\newtheorem{lemma}{Lemma}[section]

\newtheorem{proposition}[lemma]{Proposition}
\newtheorem{theorem}[lemma]{Theorem}
\newtheorem{corollary}[lemma]{Corollary}

\newtheorem{remark}{Remark}

\numberwithin{equation}{section}

\begin{document}
\newcommand{\ub}{\underline{u}}
\newcommand{\Cb}{\underline{C}}
\newcommand{\Lb}{\underline{L}}
\newcommand{\Lh}{\hat{L}}
\newcommand{\Lbh}{\hat{\Lb}}
\newcommand{\phib}{\underline{\phi}}
\newcommand{\Phib}{\underline{\Phi}}
\newcommand{\Db}{\underline{D}}
\newcommand{\Dh}{\hat{D}}
\newcommand{\Dbh}{\hat{\Db}}
\newcommand{\omb}{\underline{\omega}}
\newcommand{\omh}{\hat{\omega}}
\newcommand{\ombh}{\hat{\omb}}
\newcommand{\Pb}{\underline{P}}
\newcommand{\chib}{\underline{\chi}}
\newcommand{\chih}{\hat{\chi}}
\newcommand{\chibh}{\hat{\chib}}
\newcommand{\alb}{\underline{\alpha}}
\newcommand{\zeb}{\underline{\zeta}}
\newcommand{\beb}{\underline{\beta}}
\newcommand{\etb}{\underline{\eta}}
\newcommand{\Mb}{\underline{M}}
\newcommand{\oth}{\hat{\otimes}}

\def\a {\alpha}
\def\b {\beta}
\def\ab {\alphab}
\def\bb {\betab}
\def\nab {\nabla}
\def\ub {\underline{u}}
\def\th {\theta}
\def\Lb {\underline{L}}
\def\Hb {\underline{H}}
\def\chib {\underline{\chi}}
\def\chih {\hat{\chi}}
\def\chibh {\hat{\underline{\chi}}}
\def\omegab {\underline{\omega}}
\def\etab {\underline{\eta}}
\def\betab {\underline{\beta}}
\def\alphab {\underline{\alpha}}
\def\Psib {\underline{\Psi}}
\def\hot{\widehat{\otimes}}
\def\Phib {\underline{\Phi}}
\def\thb {\underline{\theta}}
\def\t {\tilde}
\def\st {\tilde{s}}

\def\rhoc{\check{\rho}}
\def\sigmac{\check{\sigma}}
\def\Psic{\check{\Psi}}
\def\kappab{\underline{\kappa}}
\def\betabc {\check{\underline{\beta}}}

\def\d {\delta}
\def\f {\frac}
\def\i {\infty}
\def\l {\bigg(}
\def\r {\bigg)}
\def\S {S_{u,\underline{u}}}
\def\o{\omega}
\def\O{\Omega}\
\def\be{\begin{equation}\begin{split}}
\def\en{\end{split}\end{equation}}
\def\at{a^{\frac{1}{2}}}
\def\af{a^{\frac{1}{4}}}
\def\od{\omega^{\dagger}}
\def\ombd{\underline{\omega}^{\dagger}}
\def\K{K-\frac{1}{|u|^2}}
\def\ut{\frac{1}{|u|^2}}
\def\Kb{K-\frac{1}{(u+\underline{u})^2}}
\def\bf{b^{\frac{1}{4}}}
\def\bt{b^{\frac{1}{2}}}
\def\de{\delta}
\def\ls{\lesssim}
\def\om{\omega}
\def\Om{\Omega}
\def\Orw{O(r_1^{\f{1}{100}})}
\def\Ort{O(r_2^{\f{1}{100}})}
\def\Or{O(r^{\f{1}{100}})}

\newcommand{\e}{\epsilon}
\newcommand{\et} {\frac{\epsilon}{2}}
\newcommand{\ef} {\frac{\epsilon}{4}}
\newcommand{\LH} {L^2(H_u)}
\newcommand{\LHb} {L^2(\underline{H}_{\underline{u}})}
\newcommand{\M} {\mathcal}
\newcommand{\TM} {\tilde{\mathcal}}
\newcommand{\p}{\psi\hspace{1pt}}
\newcommand{\q}{\underline{\psi}\hspace{1pt}}
\newcommand{\Li}{_{L^{\infty}(S_{u,\underline{u}})}}
\newcommand{\Lt}{_{L^{2}(S)}}
\newcommand{\da}{\delta^{-\frac{\epsilon}{2}}}
\newcommand{\db}{\delta^{1-\frac{\epsilon}{2}}}
\newcommand{\D}{\Delta}
\newcommand{\snabla}{\slashed{\nabla}}
\newcommand{\sg}{\slashed{g}}
\newcommand{\sD}{\slashed{{\Delta}}}
\newcommand{\R}{\mathbb{R}}
\newcommand{\s}{\mathbb{S}}
\newcommand{\C}{\mathbb{C}}
\newcommand{\Q}{\mathbb{Q}}
\newcommand{\Z}{\mathbb{Z}}
\newcommand{\N}{\mathbb{N}}
\newcommand{\T}{\mathbf{T}}


\renewcommand{\div}{\mbox{div }}
\newcommand{\curl}{\mbox{curl }}
\newcommand{\trchb}{\mbox{tr} \chib}
\def\trch{\mbox{tr}\chi}
\newcommand{\tr}{\mbox{tr}}

\newcommand{\Ls}{{\mathcal L} \mkern-10mu /\,}
\newcommand{\eps}{{\epsilon} \mkern-8mu /\,}

\newcommand{\xib}{\underline{\xi}}
\newcommand{\psib}{\underline{\psi}}
\newcommand{\rhob}{\underline{\rho}}
\newcommand{\thetab}{\underline{\theta}}
\newcommand{\gammab}{\underline{\gamma}}
\newcommand{\nub}{\underline{\nu}}
\newcommand{\lb}{\underline{l}}
\newcommand{\mub}{\underline{\mu}}
\newcommand{\Xib}{\underline{\Xi}}
\newcommand{\Thetab}{\underline{\Theta}}
\newcommand{\Lambdab}{\underline{\Lambda}}
\newcommand{\vphb}{\underline{\varphi}}

\newcommand{\ih}{\hat{i}}

\newcommand{\tcL}{\widetilde{\mathscr{L}}}

\newcommand{\sRic}{Ric\mkern-19mu /\,\,\,\,}
\newcommand{\sL}{{\cal L}\mkern-10mu /}
\newcommand{\sLh}{\hat{\sL}}
\newcommand{\seps}{\epsilon\mkern-8mu /}
\newcommand{\sd}{d\mkern-10mu /}
\newcommand{\sR}{R\mkern-10mu /}
\newcommand{\snab}{\nabla\mkern-13mu /}
\newcommand{\sdiv}{\mbox{div}\mkern-19mu /\,\,\,\,}
\newcommand{\scurl}{\mbox{curl}\mkern-19mu /\,\,\,\,}
\newcommand{\slap}{\mbox{$\triangle  \mkern-13mu / \,$}}
\newcommand{\sGamma}{\Gamma\mkern-10mu /}
\newcommand{\somega}{\omega\mkern-10mu /}
\newcommand{\somb}{\omb\mkern-10mu /}
\newcommand{\spi}{\pi\mkern-10mu /}
\newcommand{\sJ}{J\mkern-10mu /}
\renewcommand{\sp}{p\mkern-9mu /}
\newcommand{\su}{u\mkern-8mu /}
\tikzset{every picture/.style={line width=0.75pt}}

\def\be{\begin{equation}}
\def\ee{\end{equation}}
\def\bes{\begin{equation*}}
\def\ees{\end{equation*}}
\newcommand\ba{\begin{align}}
\newcommand\ea{\end{align}}
\newcommand\bas{\begin{align*}}
\newcommand\eas{\end{align*}}
\def\Orw{O\big((M^{-1}r_1)^{\f{1}{100}}\big)}
\def\Ort{O\big((M^{-1}r_2)^{\f{1}{100}}\big)}
\def\Or{O\big((M^{-1}r)^{\f{1}{100}}\big)}
\def\ba{\begin{align}}
\def\ea{\end{align}}
\def\bas{\begin{align*}}
\def\eas{\end{align*}}

\begin{abstract}
Under spherical symmetry, with double-null coordinates $(u,v)$, we study the gravitational collapse of the Einstein--scalar field system with a positive cosmological constant. The spacetime singularities arise when area radius $r$ vanishes and they are spacelike. We derive new quantitative estimates, obtain polynomial blow-up rates $O(1/r^N)$ for various quantities, and extend the results in \cite{AZ} by the first author and Zhang and the arguments in \cite{AG} by the first author and Gajic to the cosmological settings. In particular, we sharpen the estimates of $r\partial_u r$ and $r\partial_v r$ in \cite{AZ} and prove that the spacelike singularities where $r(u,v)=0$ are $C^{1,1/3}$ in $(u,v)$ coordinates. As an application, these estimates also give quantitative blow-up upper bounds of fluid velocity and density for the hard-phase model of the Einstein-Euler system under irrotational assumption. Near the timelike infinity, we also generalize the theorems in \cite{AG} by linking the precise blow-up rates of the Kretschmann scalar to the exponential Price's law along the event horizon. In cosmological settings, this further reveals the mass-inflation phenomena along the spacelike singularities for the first time.
\end{abstract}

\maketitle

\tableofcontents

\section{Introduction}
In this paper, we study gravitational collapse and its associated spacetime singularities for the Einstein-scalar field system with a positive cosmological constant $\Lambda$. In the presence of this $\Lambda$, the Einstein-scalar field system takes the following form:
\begin{equation}\label{ESlam}
\begin{split}
&\mbox{Ric}_{\mu\nu}-\f12Rg_{\mu\nu}+\Lambda g_{\mu\nu}=2T_{\mu\nu},\\
&T_{\mu\nu}=\partial_{\mu}\phi \partial_{\nu}\phi-\f12g_{\mu\nu}\partial^{\sigma}\phi \partial_{\sigma}\phi,
\end{split}
\end{equation}

\noindent In literatures of astrophysics, the positive constant $\Lambda$ helps to build the Lambda cold dark matter ($\Lambda$CDM) model, which is widely used to characterize the accelerated expanding universe. The cosmological constant is intimately related to the concept of dark energy, which contributes to around $68\%$ of the mass-energy density in our universe. The simplest black-hole spacetime allowing a positive cosmological constant is the static, spherically symmetric Schwarzschild-de Sitter spacetime \cite{kottler,weyl}, which solves the Einstein vacuum equations
\begin{equation*}
\mbox{Ric}_{\mu\nu}-\f12Rg_{\mu\nu}+\Lambda g_{\mu\nu}=0.
\end{equation*}
The explicit expression of this solution will be given in Section \ref{pre}.

The system \eqref{ESlam} with positive $\Lambda$ can also be alternatively interpreted as a hard phase model of the Einstein-Euler system introduced by Christodoulou \cite{DC95}. There he introduced the following two-phase model for Einstein-Euler system
\begin{equation} \label{einstein euler normal}
\begin{split}
&\mbox{Ric}_{\mu\nu}-\f12Rg_{\mu\nu}+\Lambda g_{\mu\nu}=2T_{\mu\nu},\\
&T_{\mu\nu}=(\rho+p)\vartheta_\mu \vartheta_\nu+p g_{\mu\nu},
\end{split}
\end{equation}
where we use $\rho$ to denote the fluid density, $\vartheta^\mu$ to denote the fluid velocity, and $p$ to represent the pressure. In \cite{DC95}, Chirstodoulou requires
\begin{equation} \label{2 phase}
    p=
    \begin{cases}
    0 \quad \text{if}\ \rho\leq\rho_0\ (\text{soft phase}),\\
    \rho-\rho_0 \quad \text{if}\ \rho>\rho_0 \ (\text{hard phase})
    \end{cases}
\end{equation}
with the positive constant $\rho_0$ being the nuclear saturation density. Under assumption \eqref{2 phase} and irrotational requirement, this hard phase model can also be deduced to be
\begin{equation} \label{hard phase}
\begin{split}
&\mbox{Ric}_{\mu\nu}-\f12Rg_{\mu\nu}=2T_{\mu\nu},\\
&T_{\mu\nu}=\partial_\mu \phi \partial_\nu \phi+\frac12(\partial^\sigma \phi \partial_\sigma \phi-\rho_0)g_{\mu\nu},
\end{split}
\end{equation}
\noindent where $\phi$ is also called the velocity potential function. Noting that the above Einstein equation is equivalent to
\begin{equation*} 
\mbox{Ric}_{\mu\nu}-\f12Rg_{\mu\nu}+\rho_0 g_{\mu\nu}=2\partial_\mu \phi \partial_\nu \phi+\partial^\sigma \phi \partial_\sigma \phi g_{\mu\nu}.
\end{equation*}
Here, $\rho_0$ could be considered as the positive ``cosmological constant" $\Lambda$.

In the present article, for \eqref{ESlam}: the Einstein-scalar field system with positive constant $\Lambda$, we derive the \emph{quantitative} estimates for solutions near the spacetime singularities. We also obtain $C^{1,1/3}$ regularity for spacelike boundary in $(u,v)$ coordinates. This is optimal with respect to our method. Moreover, assuming the exponential Price's law along the event horizon, in black hole interior, we prove detailed asymptotic blow-up behaviours of solutions towards the timelike infinity. In cosmological settings, we give the first result of this kind on proving mass inflation along the spacelike singularities. We also exhibit the connection between these sharp quantitative blow-up rates and the \emph{exponential} Price's law. 
\subsection{Background of the $\Lambda=0$ scenario}
\begin{figure}[H]
\begin{center}
\begin{minipage}[!t]{0.4\textwidth}
\begin{tikzpicture}[scale=0.75]
\draw [white](-1, -2.5)-- node[midway, sloped, above,black]{$\Gamma$}(0, -2.5);
\draw [white](0.5, 0)-- node[midway, sloped, above,black]{$\mathcal{S}$}(4, 0);
\node[above] at (2.25, -0.75){$\mathcal{T}$};
\draw [white](-1, 0)-- node[midway, sloped, above,black]{$\mathcal{S}_0$}(1, 0);
\draw [white](5.5, 0.2)-- node[midway, sloped, above,black]{$i^+$}(7, 0.2);
\draw [white](10, -4.8)-- node[midway, sloped, above,black]{$i^0$}(12.5, -4.8);
\draw (0.07,0) to [out=-5, in=195] (5.43, 0.5);
\draw (0.05,-0.05) to [out=-40, in=215] (5.45, 0.45);
\draw [white](0, -3)-- node[midway, sloped, above,black]{$\mathcal{H}$}(7, -3);
\draw [white](7, -2)-- node[midway, sloped, above,black]{$\mathcal{I^+}$}(10, -2);
\node[below]at (2.2, -0.8){$\mathcal{A}$};
\draw [thick] (0, -5)--(0,-0.08);
\draw [thick] (5.45, 0.45)--(0,-5);
\draw (0,0) circle [radius=0.08];
\draw (5.5, 0.5) circle [radius=0.08];
\draw (11, -5) circle [radius=0.08];
\draw [thick] (5.55, 0.45)--(10.95,-4.95);
\draw [thick] (0,-5) to [out=5, in=175] (10.93, -5);
\end{tikzpicture}
\end{minipage}
\begin{minipage}[!t]{0.6\textwidth}
\end{minipage}
\hspace{0.05\textwidth}
\end{center}
\vspace{-0.2cm}
\caption{\textbf{Penrose diagram of the dynamical black hole spacetime for system \eqref{ES0}}. Here, $\Gamma$ denotes the center (invariant under $SO(3)$ group action) with the endpoint $\mathcal{S}_0$ being the first singularity along $\Gamma$; $\mathcal{S}$ is the spacelike singularity where the area radius vanishes; $\mathcal{I}^+$ denotes the future null infinity; $\mathcal{H}$ is the event horizon; the point $i^+$ represents the future timelike infinity; $\mathcal{T}$ denotes the trapped region; the hypersurface $\mathcal{A}$ denotes the apparent horizon.}
	\label{fig:chrBHs}
\end{figure}
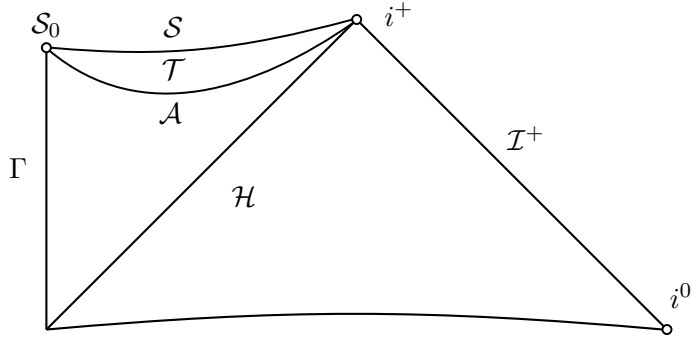

In \cite{DC91, DC93, DC94, DC99}, Christodoulou initiated the study of spherically symmetric gravitational collapse for the Einstein-scalar field system:
\begin{equation}\label{ES0}
\begin{split}
&\mbox{Ric}_{\mu\nu}-\f12Rg_{\mu\nu}=2T_{\mu\nu},\\
&T_{\mu\nu}=\partial_{\mu}\phi \partial_{\nu}\phi-\f12g_{\mu\nu}\partial^{\sigma}\phi \partial_{\sigma}\phi.
\end{split}
\end{equation}
In these works, Christodoulou established the almost-scale-critical trapped surface formation criterion and proved that the trapped region admits a spacelike $C^1$ future boundary, denoted as $\mathcal{S}$, where the area radius vanishes and the Kretschmann scalar $R_{\alpha \beta \gamma \delta}R^{\alpha \beta \gamma \delta}$ blows up. See Figure \ref{fig:chrBHs} for the Penrose diagram of such black hole spacetimes.

In this paper, with $\Lambda$ being a positive constant, under spherical symmetry, we foliate the spacetime with double null coordinates. The metric $g$ takes the following form:
\begin{equation}\label{metric}
g=-\Omega^2(u,v)dudv+r^2(u,v)\big(d\theta^2+\sin^2\theta d\phi^2\big).
\end{equation}
The level sets of $u$ and $v$ correspond to outgoing and ingoing null hypersurfaces, respectively.

Toward understanding the strength of the spacelike singularities, in \cite{AZ}, under spherical symmetry, the first author and Zhang proved that along $\mathcal{S}\backslash\mathcal{S}_0$ (where $r=0$) the Kretschmann scalar always obeys a polynomial blow-up rate in the trapped region:
\begin{equation} \label{an poly blowup rate}
r^{-6}\lesssim R^{\alpha\beta\gamma\delta}R_{\alpha\beta\gamma\delta}\lesssim r^{-N}
\end{equation}
with $N$ being a positive constant depending on the initial data. They also showed that the scalar field satisfies the following sharp bounds:
 \begin{equation*}
|\partial_u \phi|\lesssim r^{-2}, \quad \quad |\partial_v \phi|\lesssim r^{-2}.
\end{equation*}
A crucial ingredient in \cite{AZ} is the improved estimates for $r\partial_u r$ and $r\partial_v r$. As is shown by Christodoulou in \cite{DC93}, $-r\partial_u r$ and $-r\partial_v r$ admit positive limits on $\mathcal{S}$. In \cite{AZ}, the authors establish new quantitative estimates with a $r^{\frac{1}{100}}$ remainder, i.e., 
\begin{equation} \label{rk sharp dr} 
|r\partial_u r+C_1|(u,v)\lesssim r(u,v)^{\frac{1}{100}}, \quad |r\partial_v r+C_2|(u,v)\lesssim r(u,v)^{\f{1}{100}}
\end{equation}
with $C_1, C_2$ being the positive limits of $-r\partial_u r$ and $-r\partial_v r$ on $\mathcal{S}$, respectively. In addition, if prescribing polynomially decaying initial data (respecting the Price's law) along the event horizon $\mathcal{H}$:
\begin{equation*}
D_1v^{-q} \leq \partial_v\phi|_{\mathcal{H}} \leq D_2 v^{-p},
\end{equation*}
close to timelike infinity, the first author and Gajic in \cite{AG} further calculated $N$ in \eqref{an poly blowup rate} and specified the blow-up rate of $R^{\a\b\mu\nu}R_{\a\b\mu\nu}$ as $r^{-6-C v^{-2p} }$ with $C$ being a constant depending on $D_1$ and $D_2$. They also showed that in the trapped region the dynamical spacetime converges to the Schwarzschild black hole in a novel singular way as it approaches the timelike infinity $i^+$ along $\mathcal{S}$. The blow-up rate of the Kretschmann scalar in \cite{AG} varies at different points along $\mathcal{S}$, which also indicates a new PDE-type blow-up mechanism.

The results in \cite{AG} relates the polynomial decaying Price's law along the event horizon $\mathcal{H}$ with the strength of curvature singularities at the spacetime singularity $\mathcal{S}$, equipped with precise quantitative estimates. Along the event horizon, the derivatives of the scalar field are conjectured to decay in a certain polynomial rate. This is the so-called ``Price's law'' in the literature which was first raised in heuristic work by Price \cite{Price1972}. In \cite{aagprice}, it was shown that in the \emph{linearized setting} the scalar field has ``polynomial tails'' along the event horizon, i.e., when $\Lambda=0$, the derivatives of the scalar field decay polynomially. For the Einstein--Maxwell--scalar field system, Dafermos and Rodnianski proved the upper bound estimate in \cite{MDIR05}. As for the lower bound, Luk and Oh provides an $L^2$-version estimate in \eqref{eq:priceassm} for generic initial data in \cite{Luk2016b} in the same setting.

\subsection{Main results}
In this section, we summarize our main theorems.
\subsubsection{Dynamics along the spacelike singularities} 
\begin{figure}[H]
\begin{center}
\begin{minipage}[!t]{0.4\textwidth}
\begin{tikzpicture}[scale=0.75]
\draw [white](-1, -2.5)-- node[midway, sloped, above,black]{$\Gamma$}(0, -2.5);
\draw [white](0, 0)-- node[midway, sloped, above,black]{$\mathcal{S}$}(4, 0);
\draw [white](-1, 0)-- node[midway, sloped, above,black]{$\mathcal{S}_0$}(1, 0);
\draw [white](5.5, 0.2)-- node[midway, sloped, above,black]{$i^+$}(7, 0.2);
\draw (0.07,0) to [out=-5, in=195] (5.43, 0.5);
\draw (0.05,-0.05) to [out=-40, in=215] (5.45, 0.45);
\draw [white](0, -3)-- node[midway, sloped, above,black]{$\mathcal{H}$}(7, -3);
\node[below]at (2.2, -0.8){$\mathcal{A}$};
\draw [thick] (0, -5)--(0,-0.08);
\draw [thick] (5.45, 0.45)--(0,-5);
\draw (0,0) circle [radius=0.08];
\draw (5.5, 0.5) circle [radius=0.08];
\node[above] at (2.25, -0.75){$\mathcal{T}$};
\end{tikzpicture}
\end{minipage}
\begin{minipage}[!t]{0.6\textwidth}
\end{minipage}
\hspace{0.05\textwidth}
\end{center}
\vspace{-0.2cm}
\caption{\textbf{Penrose diagram of the cosmological spacelike singularity}.}
\label{pdiag lambda }
\end{figure}
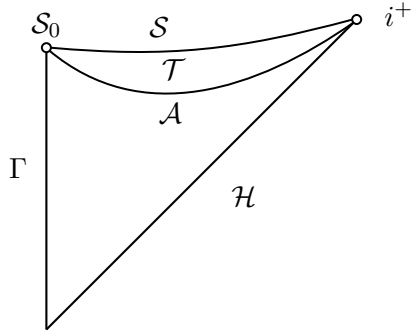

We first prove that the $r(u,v)=0$ singularities $\mathcal{S}$ are spacelike. We further generalize the arguments in \cite{AZ} to the Einstein-scalar field with positive $\Lambda$ case. In addition, we improve the remainder estimate in \eqref{rk sharp dr} from $r^\frac{1}{100}$ to $r^{\f23}$. 
\begin{theorem} \label{local version}
The spherically symmetric dynamical black hole spacetime solving (\ref{ESlam}) is bounded to the future by a spacelike singularity $\mathcal{S}$ where the area radius $r=0$. Let $(u_0,v_0)$ denotes a point along the spacelike singularity $\mathcal{S}$, different from the first singularity $\mathcal{S}_0$. Then, there exists a sufficiently small $r_0>0$ such that, for all $(u,v)$ lie both in the causal past of $(u_0,v_0)$ and in the causal future of $\{r=r_0\}$ we have
\begin{enumerate}[(i)]
\item For each $(u_0,v_0)\in\mathcal{S}\backslash\mathcal{S}_0$, there exist positive numbers $C_1$ and $C_2$, such that for any $(u,v)$ close to $(u_0,v_0)$, it holds
\begin{equation} \label{main sharp dr}
|r\partial_u r+C_1|(u,v)\lesssim r(u,v)^{\f23}, \quad |r\partial_v r+C_2|(u,v)\lesssim r(u,v)^{\f23}.
\end{equation}
\item The derivatives of $\phi$ obey
 \begin{equation}
\label{eq:dphinearr0}
r^2(u,v)|\partial_u \phi|(u,v)\lesssim1, \quad \quad r^2(u,v)|\partial_v \phi|(u,v)\lesssim1.
\end{equation}
\item The Krestchmann scalar satisfies the following lower and upper bound estimates:
\begin{equation}
\label{eq:upboundanzhang}
r(u,v)^{-6}\lesssim R^{\alpha\beta\gamma\delta}R_{\alpha\beta\gamma\delta}(u,v)\lesssim r(u,v)^{-N_{u_0,v_0}},
\end{equation}
where $N_{u_0,v_0}$ is a positive number and $(u,v)$ is close to $(u_0,v_0)\in\mathcal{S}\backslash\mathcal{S}_0$.
\end{enumerate}
\end{theorem}
\begin{remark}
Note that the values of the constants $C_1$, $C_2$ and $N_{u_0,v_0}$ depend on the location of $(u_0,v_0)$ along $\mathcal{S}$. The notation $\lesssim$ here means $\leq C_{u_0,v_0}\cdot$ with $C_{u_0,v_0}$ being a positive constant depending on the position of $(u_0,v_0)$. In \eqref{eq:upboundanzhang} we prove inverse polynomial blow-up rate for the Kretschmann scalar. Furthermore, employing \eqref{eq:dphinearr0}, we also have $|\phi|\lesssim |\ln r|$.
\end{remark}
\begin{remark}
By applying the generalized extension principle in \cite{kommemiphd} to our system, we have that the spacetime singularities ocuur at $r=0$. In this article, we further obtain \emph{quantitative} estimates of the solution's singular behaviours when it is close to and approaching $r=0$.
\end{remark}
In \cite{DC93}, for the $\Lambda=0$ case, Christodoulou proved that the spacelike singularity $\mathcal{S}$ is of $C^1$ regularity. Here, for both $\Lambda=0$ and $\Lambda>0$ cases, we improve it to the optimal regularity $C^{1,1/3}$ with respect to our method:
\begin{theorem} \label{sing holder}
The spacelike singularities $\mathcal{S}\backslash\mathcal{S}_0$, as in Figure \ref{pdiag lambda }, is $C^{1,1/3}$ in $(u,v)$ coordinates. In particular, assuming that $\mathcal{S}\backslash\mathcal{S}_0$ is parameterized\footnote{One can also parameterize $\mathcal{S}$ by $v$. A similar argument leads to the same $C^{1,1/3}$ regularity.} as a curve\footnote{Literally, the singularities $\mathcal{S}\backslash\mathcal{S}_0$ could be viewed as a three dimensional hypersurface. Within spherical symmetry, we consider $\mathcal{S}\backslash\mathcal{S}_0$ as a curve in the quotient space, i.e., the $(u,v)$ plane.} by $(u,v^*(u))$ with $r(u,v^*(u))=0$, then $v^*(u)$ as a function of $u$ is $C^{1,1/3}$.
\end{theorem}
Recall that the irrotational hard-phase model of Einstein-Euler system \eqref{einstein euler normal} can be recast as the Einstein-scalar field system with a positive cosmological constant $\Lambda=\rho_0$, i.e., \eqref{hard phase}. Our estimates thus can also be interpreted as the \emph{polynomial} blow-up upper bounds for the hard phase model. More precisely, we obtain
\begin{theorem} \label{hard phase thm}
Within spherical symmetry, we consider the hard-phase model of the Einstein-Euler system \eqref{einstein euler normal}. Under the irrotational condition, we have that if there is a spacetime singularity, it is along a spacelike curve $\mathcal{S}$. And for each interior point $(u_0,v_0)\in \mathcal{S}$, there exists a positive constant $C_{u_0,v_0}$, such that for all $(u,v)$ near $(u_0,v_0)$, the following estimates hold for the fluid velocity $\vartheta^\mu$ and density $\rho$:
\begin{equation*} 
    |\vartheta^\mu|(u,v)\lesssim r^{-2-2C_{u_0,v_0}}(u,v),\quad\quad |\rho|(u,v)\lesssim r^{-4-2C_{u_0,v_0}}(u,v).
\end{equation*}
\end{theorem}

\subsubsection{Late-time tails tied to the exponential Price's law}
 In the presence of the positive cosmological constant, it is widely expected that the the scalar field obeys exponential Price's law, i.e., inverse exponential decay rates along the event horizon. We refer the readers to linear results in \cite{dr4,dyatlov15,peter1,mavSdS1,vasydesitter}, nonlinear result in \cite{HV2016,mavSdS2} and references therein. In light of these results, in the current article, we further relate the curvature blow-up rates in Theorem \ref{local version} to the decay rates in the exponential Price's law.

With prescribed initial data along the event horizon $\mathcal{H}$, which converge to the Schwarzschild-de Sitter spacetime respecting \emph{exponential} Price's law, we construct spacetime up to spacelike singularities, and we compute the accurate blow-up rate of the Kretschmann scalar along $\mathcal{S}$. \textit{Here the blow-up rate is intimately related to the decay rate in the exponential Price's law. Furthermore, approaching the future timelike infinity $i^+$, various blow-up rates converge to the corresponding \emph{Schwarzschild-de Sitter} values, i.e,
\begin{equation} \label{SDS curv}
R_{\alpha \beta \gamma \delta}R^{\alpha \beta \gamma \delta}=24\left[\left(\frac{\Lambda}{3}\right)^4+ \frac{2M^2}{r^6}\right]
\end{equation}
with $M$ denoting the Schwarzschild-de Sitter mass.}

In this article, deriving the sharp curvature blow-up rates builds on a thorough analysis of the solution's dynamics in the black hole interior. We have
\begin{theorem}
\label{main thm}
Assume that along the event horizon $\mathcal{H}$, the derivatives of the scalar field satisfy the following exponential Price's law:
\begin{equation}
\label{eq:priceassm}
D_1 e^{-qv} \leq \partial_v\phi|_{\mathcal{H}} \leq D_2 e^{-pv},
\end{equation}
for $p,q$ satisfying $1<p\leq q<5p$.

Then, in the trapped region, when $v$ is sufficiently large (close to $i^+$) there exist positive constants $0<\rho<\sigma$, with $\rho$  and $\sigma$ depending on the constants in \eqref{eq:priceassm}, such that the Hawking mass satisfies
\begin{equation} \label{est mass infla}
     r^{-\rho e^{-2qv}}(u,v) \lesssim m(u,v) \lesssim  r^{-\sigma e^{-2pv}}(u,v),
\end{equation}
which indicates the ``mass inflation" phenomena along $r(u,v)=0$, and the Kretschmann scalar obeys
\begin{equation} \label{rough scalar curv}
 r^{-6-\rho e^{-2qv} }(u,v)\lesssim R^{\a\b\mu\nu}R_{\a\b\mu\nu}(u,v)
\lesssim r^{-6-\sigma e^{-2pv}}(u,v).
\end{equation}
\end{theorem}

\begin{remark}
Along $\mathcal{S}$, the geometric area radius $r(u,v)$ shrinks to $0$. Various geometric quantities blow up there at inverse polynomial rate of $r(u,v)$ as shown in \eqref{rough scalar curv}. Compared with Theorem \ref{local version}, Theorem \ref{main thm} demonstrates a more precise blow-up rate of the Kretschmann scalar. In particular, the blow-up rate for curvature is associated with exponential tails arising from the Price's law \eqref{eq:priceassm}. And it is stronger than, and moreover is convergent to the corresponding rate of the Schwarzschild-de Sitter spacetime in \eqref{SDS curv}. The readers are referred to Theorem \ref{thm:precisemain} for a more precise version of this main result.
\end{remark}

\subsection{Comparison with the $\Lambda=0$ case}
In this section, we compare the present result with the $\Lambda=0$ case in \cite{AG,AZ} and sketch the main steps in the proof. 
\begin{itemize}
    \item In a local region near the spacelike singularities $\mathcal{S}$, as is stated in \eqref{rk sharp dr} and Theorem \ref{sing holder}, we improve the estimates of $r\partial_u r$, $r\partial_v r$ and raise the obtained regularity of $\mathcal{S}$ in $(u,v)$ plane from $C^1$ in \cite{AZ} to $C^{1,1/3}$. In particular, for estimates of $r\partial_u r$ and $r\partial_v r$, we sharpen the remainder estimate from $r^\frac{1}{100}$ to $r^\f23$. Based on this improvement, we further demonstrate that the spacelike singularities $\mathcal{S}$, as a curve in $(u,v)$ plane, is of H\"{o}lder regularity $C^{1,1/3}$. Technically, there are also modifications such as in Proposition \ref{Prop 4.1}. In \cite{AZ}, this argument involves behaviour of the solution on the apparent horizon, which is however different when $\Lambda$ is present. Moreover, the results in the current paper can be applied to the hard-phase model of the Einstein-Euler system, where the parameter $\rho_0$ (nuclear saturation density) can be regarded as a positive ``cosmological constant''. We derive blow-up upper bounds for the corresponding fluid variables as well.\\
    
\item For a global region near the timelike infinity, when it is a bit away from $\mathcal{S}$, compared with \cite{AG} there is a key difference in the proof of Theorem \ref{main thm}. Here, the proof is further divided into two parts. We consider $$D_{v_{\infty},r_0}=[0,U_0]_U\times [v_0,v_{\infty}]\cap\{r\geq r_0\},$$
where $r_0$ is an arbitrarily small positive number, and we establish $L^{\infty}$ estimates for the quantities $(r,\Omega,\phi)$ via a bootstrap argument. In contrast with the $\Lambda=0$ case in \cite{AG}, here the corresponding bootstrap assumptions need to be imposed separately within two subsets of $D_{v_{\infty},r_0}$, i.e., the red-shift region and the no-shift region. While in \cite{AG}, the bootstrap assumptions are set for the whole $D_{v_{\infty},r_0}$. In this current paper, we first close the bootstrap argument in the red-shift region. Then, based on obtained estimates in the red-shift region, we improve the bootstrap assumptions in the no-shift region. Within this process, in the presence of the positive $\Lambda$, we need to deal with additional terms carrying the cosmological constant. With the upper bound estimates at our disposal, in $D_{v_{\infty},r_0}$ we also derive lower bound estimates in $\{r\geq r_0\}$ for the derivatives of $\phi$, which is essentially used to obtain the sharp curvature blow-up. In particular, the crucial upper bound estimates in the red-shift region and no-shift region are respectively proceeded in the following manners:
\begin{enumerate}

\item{\textit{Red-shift region $\mathcal{R}_\delta$.}}
We start with a red-shift region $\mathcal{R}_\delta$ where $v-|u|<-\delta^{-1}$, with $\delta>0$ suitably small. In the presence of positive $\Lambda$, we prescribe initial data satisfying the \underline{exponential} Price's law along the event horizon, while the Price's law for the case $\Lambda=0$ is polynomial. Here we need to use the red-shift effect to establish smallness and exponential decay of the difference quantities. While in \cite{AG} proving polynomial decay is enough.

\item{\textit{No-shift region $\mathcal{N}_{\delta,r_0}$.}}
The next region we consider is a \emph{no-shift} region $\mathcal{N}_{\delta,r_0}$ where $v-|u|>-\delta^{-1}$ but $r\geq r_0$. In this region, we derive estimates to identify the leading-order behaviour of (derivatives of) $(r,\Omega^2,\phi)$ in $r$ with error terms that decay in $e^{-v}$ and $e^{-|u|}$. However, unlike the $\Lambda=0$ case in \cite{AG}, here in order to derive estimates uniform in $r_0$ within the no-shift region, we encounter an obstruction which arises from the exponential tails. Specifically, in Proposition \ref{prop:baimpphinoshift}, we have to control the term $e^{p(v-|u|)}$ which depends on $r_0$. While its counterpart in the $\Lambda=0$ case is $\left(\frac{v}{|u|}\right)^{p}$, and it is uniformly close to $1$ when $v$ is sufficiently large. In order to overcome this difficulty, we first refine the estimates of the double null coordinates in Lemma \ref{lm:relatuvconstr}, which plays a \emph{key} role in our proof for the exponential-tail scenario. Then, fixing a uniform, sufficiently small $r_0$, which depends only on $M$ and the initial data, we show that all the $r_0$-relevant estimates are whence uniform. And the solution propagates in a desired way in the no-shift region.\\

\end{enumerate}

\item We also extend the estimates of $(r,\Omega,\phi)$ to the $r=0$ singularities for a global region $D_{\infty,0}=[0,U_0]_U\times [v_0,v_{\infty}]\cap\{r>0\}$. In the spirit of proof for Theorem \ref{local version}, we invoke the obtained estimates at $\{r=r_0\}$ in the last step as the initial state, and deduce upper and lower bound estimates of $(r,\Omega,\phi)$ in $D_{\infty,0}$. In this step, we also derive an optimal $r^\f23$ remainder for estimates of $r\partial_u r$ and $r\partial_v r$, which sharpens the result in \cite{AG}. Employing an algebraic calculation of the Kretschmann scalar $R^{\a\b\mu\nu}R_{\a\b\mu\nu}(u,v)$ in Theorem \ref{local version} and the above estimates of $\O^2(u,v)$, we then derive sharp lower and upper bound estimates of the Kretschmann scalar. The blow-up rate is \underline{different} from but \underline{approaching} the Schwarzschild de-Sitter value.
\end{itemize}

\subsection{Previous work}
In this section, we refer to some other related previous work on spacetime singularities in black hole interiors.

Prescribe the spherical symmetry. In \cite{DC91}\cite{DC99}, Christodoulou proved the strong and weak cosmic censorship conjecture for the Einstein-scalar field system. In the presence of a positive cosmological constant, in \cite{costacqg} Costa showed a trapped surface formation criterion. For Einstein--Maxwell--(real) scalar field system, we refer to Dafermos \cite{MD03, MD05c, MD12}, Dafermos--Rodnianski \cite{MDIR05}, Luk--Oh \cite{Luk2016a,Luk2016b} for studies on spherically symmetric dynamical black hole spacetimes. For the case of charged scalar field, we refer to An-Lim \cite{anlim}, Gajic--Luk \cite{gajluk}, Kommemi \cite{kommemiphd}, Moortel \cite{moortel18, moortel19, moortel20}. Adding a cosmological constant term to the Einstein equations, there can be a range of stability and instability phenomena at the Cauchy horizon, see Costa-Gir{\~a}o-Nat{\'a}rio-Silva \cite{Costa2015, Costa2014, Costa2014a,cgns}.

Outside of the spherical symmetry, in polarized axisymmetry, there is a recent result of Alexakis and Fournodavlos \cite{fournoalex} on stability for the Schwarzschild singularity. In the general setting with no symmetry, the linear theory is firstly developed. See for example Alho-Furnodavlos-Franzen \cite{alho19}, Dafermos--Shlapentokh-Rothman \cite{dafshl16}, Fournodavlos--Sbierski \cite{fournosbier19}, Franzen \cite{franzen14,franzen19}, Gajic \cite{gaj17a,gaj17b}, Hintz \cite{peter2}, Luk--Oh \cite{lukoh17}, Luk--Sbierski \cite{luksbier16}, Ringstr\"{o}m \cite{ring17}. For dynamics of nonlinear systems, we refer to Christodoulou \cite{DC09}, Dafermos--Luk \cite{dl-scc}, Fournodavlos \cite{fourno16}, Rodnianski-Speck \cite{rodspeck18} and references therein. In the cosmological setting, we refer to results of Cardoso-Costa-Destounis-Hintz-Jansen \cite{hprl18} and Costa-Franzen \cite{costfranz16}.

\subsection{Acknowledgements}
XA is supported by NUS startup grant R-146-000-269-133, and MOE Tier 1 grants R-146-000-321-144 and R-146-000-335-114. HC acknowledges the support of MOE Tier 1 grant R-146-000-335-114 and NSFC (Grant No. 12171097).

\section{Preliminaries} \label{pre}
In this section, with double null coordinates we derive the equations to be used and also exhibit the geometry of the Schwarzschild-de Sitter spacetime.

\subsection{The Einstein--scalar field system in double null coordinates} \label{pre equations}
Under spherical symmetry, we adopt the following ansatz for the Lorentzian metric
\begin{equation*}
g=-\Omega^2(u,v)dudv+r^2(u,v)\big(d\theta^2+\sin^2\theta d\phi^2\big).
\end{equation*}
Here, the area radius $r(u,v)$ and $\Omega^2(u,v)$ are strictly positive functions. In the presence of the positive cosmological constant $\Lambda>0$, the Einstein--scalar field system \eqref{ESlam} is then reduced to the 1+1-dimensional system below for $(r,\Omega^2,\phi)$ with variables $(u,v)\in \mathbb{R}^2$. 
\begin{enumerate}[1)]
\item \textit{Propagation equations for $r$ and $\Omega^2$}:
\begin{align}
\label{propeq:r1}
\partial_u(r\partial_v r)=&-\frac{1}{4}\Omega^2(1-\Lambda r^2),\\
\label{propeq:r2}
\partial_v(r\partial_u r)=&-\frac{1}{4}\Omega^2(1-\Lambda r^2),\\
\label{propeq:Omega}
\partial_u\partial_v\log \Omega^2=&-2\partial_u\phi\partial_v\phi+\frac{1}{2}r^{-2}\Omega^2+2r^{-2}\partial_ur \partial_v r.
\end{align}
A combination of \eqref{propeq:r1} and \eqref{propeq:Omega} also yields the following equations for $r\Omega^2$ and $r^2\Omega^2$:
\begin{align}
\label{propeq:Omegarescale1}
\partial_u\partial_v\log (r\Omega^2)=&\:\frac{1}{4}r^{-2}\Omega^2(1+\Lambda r^2)-2\partial_u\phi\partial_v\phi,\\
\label{propeq:Omegarescale2}
\partial_u\partial_v\log (r^2\Omega^2)=&-2r^{-2}\partial_ur \partial_v r-2\partial_u\phi\partial_v\phi+\frac{\Lambda}{2}\Omega^2.
\end{align}
\item \textit{Propagation equations for the scalar field $\phi$}:
\begin{equation}\label{eqn phi}
r\partial_u \partial_v \phi=-\partial_u r \partial_v \phi-\partial_v r \partial_u \phi,
\end{equation}
which can also be written as
\begin{align}
\label{propeq:phi1}
\partial_u(r\partial_v \phi)=&-\partial_vr \partial_u \phi,\\
\label{propeq:phi2}
\partial_v(r\partial_u \phi)=&-\partial_ur \partial_v \phi.
\end{align}

\noindent Denote
\begin{equation}
\label{eq:defY}
Y=\frac{1}{-\partial_u r}\partial_u.
\end{equation}

\noindent Combining \eqref{propeq:r2} and \eqref{propeq:phi2}, we also have the following equation for $Y\phi$:
\begin{equation}
\label{propeq:Yphi}
\partial_v(r^2 Y\phi)=\left[-\frac{1}{4}r^{-1}\left(\frac{\Omega^2}{-\partial_u r}\right)(1-\Lambda r^2)+2r^{-1}\partial_vr\right] r^2 Y\phi+r\partial_v\phi.
\end{equation}

\item \textit{Constraint equations (also called the ``Raychaudhuri equations'')}:
\begin{align}
\label{conseq:r1}
\partial_u(\Omega^{-2}\partial_u r)=&-r \Omega^{-2}(\partial_u \phi)^2,\\
\label{conseq:r2}
\partial_v(\Omega^{-2}\partial_v r)=&-r \Omega^{-2}(\partial_v \phi)^2.
\end{align}

\end{enumerate}
To understand the dynamics better, we further define the Hawking mass:
\begin{equation}\label{Hawking mass}
m(u,v):=\f{r}{2}(1+4\Omega^{-2}\partial_u r \partial_v r),
\end{equation}
the dimensionless mass ratio $\mu:=\f{2m}{r}$ and the modified Hawking mass
\begin{equation}\label{ modified Hawking mass}
\varpi(u,v):=\f{r}{2}(1+4\Omega^{-2}\partial_u r \partial_v r)-\frac{\Lambda}{6}r^3.
\end{equation}

Regarding the value of $\mu$, we divide the spacetime into:
\begin{itemize}
\item \emph{The regular region $\mathcal{R}$}, where $\partial_v r>0, \partial_u r<0$ and it holds
$$1-\f{2m}{r}>0, \quad \mu<1, \quad 2m<r.\\$$
    \item \emph{The apparent horizon $\mathcal{A}$}, where  $\partial_v r=0$ and hence
$$1-\f{2m}{r}=0, \quad \mu=1, \quad 2m=r.\\$$
\item \emph{The trapped region $\mathcal{T}$}, where $\partial_v r<0, \partial_u r<0$ and it holds
$$1-\f{2m}{r}<0, \quad \mu>1, \quad 2m>r.$$
\end{itemize}

\noindent With $m$ and $\mu$, a straightforward calculation gives
\begin{equation}\label{r v}
\partial_u (\partial_v r)=\f{\mu-\Lambda r^2}{(1-\mu)r}\partial_v r \partial_u r,
\end{equation}

\begin{equation}\label{r u}
\partial_v (\partial_u r)=\f{\mu-\Lambda r^2}{(1-\mu)r}\partial_v r \partial_u r,
\end{equation}

\begin{equation}\label{m u}
 \partial_u m=\frac{1-\mu}{2\partial_u r}r^2 (\partial_u \phi)^2+\frac{\Lambda r^2}{2} \partial_u r,
\end{equation}

\begin{equation}\label{m v}
 \partial_v m=\frac{1-\mu}{2\partial_v r}r^2 (\partial_v \phi)^2+\frac{\Lambda r^2}{2} \partial_v r.\\
\end{equation}
The modified Hawking mass $\varpi$ also satisfies
\begin{equation}\label{modm u}
 \partial_u \varpi=\frac{1-\mu}{2\partial_u r}r^2 (\partial_u \phi)^2,
\end{equation}

\begin{equation}\label{modm v}
\partial_v \varpi=\frac{1-\mu}{2\partial_v r}r^2 (\partial_v \phi)^2.\\
\end{equation}
\subsection{Geometry of Schwarzschild-de Sitter black holes}
With constant $M>0$ and cosmological constant $\Lambda>0$, we consider the Lorentzian manifold $(\mathcal{M}_{\rm int},g_{SdS})$, with $\mathcal{M}_{\rm int}=(\R_u\times \R_v\cap\{v+u<0\})\times \s^2$ and
\begin{equation} \label{sph metric}
g_{SdS}=-\Omega^2_{SdS}(r_{SdS}(u,v))dudv+r_{SdS}^2(u,v) (d\theta^2+\sin^2\theta d\varphi^2),
\end{equation}
where
\begin{equation*}
\Omega^2_{SdS}=\frac{\Lambda}{3} r_{SdS}^2+\frac{2M}{r_{SdS}}-1 .
\end{equation*}
One can check that \eqref{sph metric} solves \eqref{ESlam} when $T_{\mu\nu}\equiv 0$. And we call $g_{SdS}$ in \eqref{sph metric} as the Schwarzschild-de Sitter metric. Define $P(r_{SdS}):=\frac{3}{\Lambda}r_{SdS} \O^2_{SdS}$. Then, it holds
\begin{align*}
 P(r_{SdS}) &=  r_{SdS}^3+\frac{6M}{\Lambda}-\frac{3}{\Lambda}r_{SdS}\\
 &=(r_S-r_{SdS})(r_D-r_{SdS})(r_{SdS}-r_-),
\end{align*}
where $r_S,r_D,r_-$ are the three roots of $P(r_{SdS})=0$. When $3\sqrt{\Lambda}M<1$, these roots are all real. In particular, two of them are positive satisfying $r_S<r_D$. And
\begin{equation*}
r_D=\frac{2}{\sqrt{\Lambda}} \cos \frac{\theta}{3},\quad r_S=\frac{2}{\sqrt{\Lambda}} \cos \left( \frac{\theta}{3}+\frac{4\pi}{3} \right)\quad \text{with}\  \cos{\theta}=-3\sqrt{\Lambda}M.
\end{equation*}
\noindent The third root $r_-=-(r_S+r_D)$ is negative. And the event horizon corresponds to $r_{SdS}=r_S$;  the cosmological horizon corresponds to $r_{SdS}=r_D$ .

The black hole interior, i.e., $0<r_{SdS}<r_S$, is our focus of analysis. There, it holds 
\begin{align*}
1-\Lambda r_{SdS}^2\geq& 1-\Lambda r_S^2=-\frac{\Lambda}{3} P'(r_S)=\frac{\Lambda}{3}(r_D-r_S)(2r_S+r_D)>0.
\end{align*}
Now we introduce the tortoise coordinate $r_*$: 
\begin{equation} \label{tort coord}
r_*(r_{SdS}):=\int_0^{r_{SdS}} -\frac{1}{\Omega^2_{SdS}(r)} dr.
\end{equation}
Invoking the aforementioned expression of $\O^2_{SdS}$, by direct calculation, we have that
\begin{equation*}
r_*(r_{SdS})=-\beta_D \log (r_D-r_{SdS})+\beta_S \log (r_S-r_{SdS})+\beta_- \log (r_{SdS}-r_-)-C^*_0
\end{equation*}
with the constant $C^*_0=-\beta_D \log (r_D)+\beta_S \log (r_S)+\beta_- \log (-r_-)$ and
\begin{equation*}
\beta_D=\frac{ 3r_D}{\Lambda(r_D-r_S)(2r_D+r_S)},
\end{equation*}
\begin{equation*}
\beta_S=\frac{ 3r_S}{\Lambda(r_D+2r_S)(r_D-r_S)},
\end{equation*}
\begin{equation*}
\beta_-=\frac{3(r_S+r_D)}{\Lambda (r_D+2r_S)(2r_D+r_S)}.
\end{equation*}
Moreover, the double null coordinates $(u,v)$ are chosen to be the so-called Eddington-Finkelstein coordinates satisfying
\begin{equation*}
\frac{1}{2}(v+u)=r_*(r_{SdS}).
\end{equation*}
Together with the definition of $r_*(r_{SdS})$ in \eqref{tort coord}, we also have
\begin{equation*}
\partial_ur_{SdS}=\partial_vr_{SdS}=-\frac{1}{2}\Omega^2_{SdS}.
\end{equation*}
\begin{figure}[H]
\begin{tikzpicture}[x=0.75pt,y=0.75pt,yscale=-1,xscale=1]

\draw [fill={rgb, 255:red, 155; green, 155; blue, 155 }  ,fill opacity=1 ]   (161.32,41.64) -- (209.87,89.67) -- (258.7,137.97) ;
\draw [fill={rgb, 255:red, 155; green, 155; blue, 155 }  ,fill opacity=1 ]   (260.58,138.95) -- (310.12,89.69) -- (358.82,41.39) ;
\draw [shift={(259.87,139.67)}, rotate = 315.16] [color={rgb, 255:red, 0; green, 0; blue, 0 }  ][line width=0.75]      (0, 0) circle [x radius= 2.01, y radius= 2.01]   ;
\draw [fill={rgb, 255:red, 155; green, 155; blue, 155 }  ,fill opacity=1 ]   (209.87,90) -- (258.09,41.91) ;
\draw [fill={rgb, 255:red, 155; green, 155; blue, 155 }  ,fill opacity=1 ]   (261.71,41.81) -- (310.12,90.03) ;
\draw    (209.87,90) -- (161.45,138.47) ;
\draw  [dash pattern={on 0.84pt off 2.51pt}]  (157.87,40.33) -- (60.68,40.1) ;
\draw [shift={(59.67,40.1)}, rotate = 180.14] [color={rgb, 255:red, 0; green, 0; blue, 0 }  ][line width=0.75]      (0, 0) circle [x radius= 2.01, y radius= 2.01]   ;
\draw  [dash pattern={on 0.84pt off 2.51pt}]  (160.68,140.1) -- (258.45,139.97) ;
\draw [shift={(159.67,140.1)}, rotate = 359.93] [color={rgb, 255:red, 0; green, 0; blue, 0 }  ][line width=0.75]      (0, 0) circle [x radius= 2.01, y radius= 2.01]   ;
\draw    (60.47,41.1) -- (109.62,90.05) -- (158.87,139.1) ;
\draw    (159,42.77) -- (109.67,90.1) ;
\draw  [dash pattern={on 0.84pt off 2.51pt}]  (160.88,40.13) -- (257.64,40.22) ;
\draw [shift={(159.87,40.13)}, rotate = 0.05] [color={rgb, 255:red, 0; green, 0; blue, 0 }  ][line width=0.75]      (0, 0) circle [x radius= 2.01, y radius= 2.01]   ;
\draw  [dash pattern={on 0.84pt off 2.51pt}]  (260.88,40.06) -- (261.45,40.06) -- (261.45,40.06) -- (358.81,40.22) ;
\draw [shift={(359.82,40.22)}, rotate = 0.09] [color={rgb, 255:red, 0; green, 0; blue, 0 }  ][line width=0.75]      (0, 0) circle [x radius= 2.01, y radius= 2.01]   ;
\draw [shift={(259.87,40.05)}, rotate = 0.33] [color={rgb, 255:red, 0; green, 0; blue, 0 }  ][line width=0.75]      (0, 0) circle [x radius= 2.01, y radius= 2.01]   ;
\draw    (109.62,90.05) ;
\draw [shift={(109.62,90.05)}, rotate = 0] [color={rgb, 255:red, 0; green, 0; blue, 0 }  ][fill={rgb, 255:red, 0; green, 0; blue, 0 }  ][line width=0.75]      (0, 0) circle [x radius= 1.34, y radius= 1.34]   ;
\draw    (209.87,90) -- (209.87,89.67) ;
\draw [shift={(209.87,90)}, rotate = 270] [color={rgb, 255:red, 0; green, 0; blue, 0 }  ][fill={rgb, 255:red, 0; green, 0; blue, 0 }  ][line width=0.75]      (0, 0) circle [x radius= 1.34, y radius= 1.34]   ;
\draw    (310.12,90.03) ;
\draw [shift={(310.12,90.03)}, rotate = 0] [color={rgb, 255:red, 0; green, 0; blue, 0 }  ][fill={rgb, 255:red, 0; green, 0; blue, 0 }  ][line width=0.75]      (0, 0) circle [x radius= 1.34, y radius= 1.34]   ;

\draw (199.87,27.9) node [anchor=north west][inner sep=0.75pt]  [font=\tiny]  {$r=0$};
\draw (302.03,27.23) node [anchor=north west][inner sep=0.75pt]  [font=\tiny]  {$r=\infty $};
\draw (216.76,66.28) node [anchor=north west][inner sep=0.75pt]  [font=\tiny,rotate=-315]  {$r=r_{S}$};
\draw (284.64,46.07) node [anchor=north west][inner sep=0.75pt]  [font=\tiny,rotate=-45]  {$r=r_{D}$};
\draw (175.64,59.74) node [anchor=north west][inner sep=0.75pt]  [font=\tiny,rotate=-45]  {$r=r_{S}$};
\draw (332.1,76.62) node [anchor=north west][inner sep=0.75pt]  [font=\tiny,rotate=-315]  {$r=r_{D}$};
\draw (172.33,46) node [anchor=north west][inner sep=0.75pt]  [font=\tiny]  {$u\rightarrow \infty $};
\draw (365.57,42.78) node [anchor=north west][inner sep=0.75pt]  [font=\tiny]  {$v\rightarrow \infty $};
\draw (94.03,28.57) node [anchor=north west][inner sep=0.75pt]  [font=\tiny]  {$r=\infty $};
\draw (222.64,107.07) node [anchor=north west][inner sep=0.75pt]  [font=\tiny,rotate=-45]  {$r=r_{S}$};
\draw (281.43,123.28) node [anchor=north west][inner sep=0.75pt]  [font=\tiny,rotate=-315]  {$r=r_{D}$};
\draw (119.64,108.07) node [anchor=north west][inner sep=0.75pt]  [font=\tiny,rotate=-45]  {$r=r_{D}$};
\draw (171.43,113.95) node [anchor=north west][inner sep=0.75pt]  [font=\tiny,rotate=-315]  {$r=r_{S}$};
\draw (117.43,67.28) node [anchor=north west][inner sep=0.75pt]  [font=\tiny,rotate=-315]  {$r=r_{D}$};
\draw (72.64,61.07) node [anchor=north west][inner sep=0.75pt]  [font=\tiny,rotate=-45]  {$r=r_{D}$};
\draw (199.87,145.57) node [anchor=north west][inner sep=0.75pt]  [font=\tiny]  {$r=0$};

\end{tikzpicture}
\vspace{-0.2cm}
\caption{A Penrose diagram for the Schwarzschild-de Sitter black hole.}
	\label{fig:schw}
\end{figure}
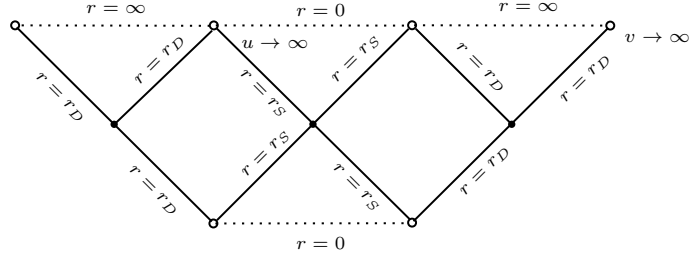
\noindent Here we also list some calculations to be used later. Letting $\alpha_S=\frac{1}{2\beta_S}$, inside the black hole region we have
\begin{equation} \label{reform OSdS1}
\begin{split}
e^{\alpha_S(v-|u|)}&=e^{2\alpha_S r_*(r_{SdS})}\\
&= e^{-2\alpha_S C^*_0}{(r_D-r_{SdS})}^{-2\alpha_S\beta_D} {(r_S-r_{SdS})} {(r_{SdS}-r_-)}^{2\alpha_S\beta_-}\\
&\geq c (r_S-r_{SdS}),
\end{split}
\end{equation}
and
\begin{equation} \label{reform OSdS2}
\frac{3r_{SdS}\Omega^2_{SdS}(u,v)}{\Lambda}=(r_D-r_{SdS})(r_S-r_{SdS})(r_{SdS}-r_-)\leq C e^{\alpha_S(v-|u|)},
\end{equation}
with $c$ and $C$ being positive constants. These imply
\begin{equation}
\label{eq:durschwest}
|\partial_ur_{SdS}|=|\partial_vr_{SdS}|=\frac{1}{2}\Omega^2_{SdS}\leq \frac{\Lambda C_{SdS}}{6r_{SdS}} e^{\alpha_S(v-|u|)}.
\end{equation}

\noindent We further define $U={(\alpha_S)}^{-1}e^{\alpha_S u}$. Then $g_{SdS}$ can also be expressed as
\begin{equation*}
g_{SdS}=-\Omega^2_{SdS}(r_{SdS}(u,v))e^{\alpha_S {|u|}}dUdv+r_{SdS}^2 (d\theta^2+\sin^2\theta d\varphi^2).
\end{equation*}
And $(\mathcal{M}_{\rm int},g_{SdS})$ can be smoothly extended to 
\begin{equation*}
\overline{\mathcal{M}_{\rm int}}=\left([0,\infty)_U \times  \R_v\cap\left\{v+{(\alpha_S)}^{-1}\log\left(\alpha_S{U}\right)\leq 0\right\} \right)\times \s^2.
\end{equation*}

\section{Local dynamics near the spacelike singularity (Theorem \ref{local version}-\ref{sing holder})} \label{pf of local bds}
In this section, we establish general estimates for spherically symmetric solutions to \eqref{ESlam} near $\mathcal{S}$. And we will prove the main results Theorem \ref{local version} and Theorem \ref{sing holder}.

\subsection{Estimates for $\partial_u r$ and $\partial_v r$} We begin with the following proposition for $\partial_u r$ and $\partial_v r$ with the presence of the cosmological constant $\Lambda$.
\begin{proposition}\label{Prop 4.1}
Given a spherically symmetric solution to the Einstein-scalar field system \eqref{ESlam} up to a future singular boundary $\mathcal{S}$, where the area radius $r=0$, then in double null coordinates for each $b_0=(u^*(v), v)\in \mathcal{S}$, there exists a positive continuous function $E$ of $v$ such that
$$-(r\partial_v r)(u,v)\rightarrow E(v)\quad \text{as}\quad u\rightarrow {u^*(v)^-}.$$
Similarly, for $b_0\in \mathcal{S}$ with coordinate $(u, v^*(u))$, the following limit holds
$$-(r\partial_u r)(u,v)\rightarrow E^{*}(u)\quad \text{as} \quad v\rightarrow {v^*(u)^-},$$ 
where $E^*$ is a positive continuous function of $u$. Moreover, $\mathcal{S}\backslash\mathcal{S}_0$, the non-central component of $\mathcal{S}$ is a $C^1$ spacelike curve.
\end{proposition}
\begin{proof}
Inside the trapped region $\mathcal{T}$, by (\ref{m u}), we have 
$$\partial_u m \geq \frac{1}{2}\Lambda r^2 \partial_u r.$$ 
For each $v$, consider a 2-sphere $b_1=(u_\sigma(v),v)\in\mathcal{T}$ with radius $r(u_\sigma(v),v)=\sigma$. Then, choosing $\sigma$ to be sufficiently small, we have that for any $(u,v)\in \mathcal{T}\cap \{r(u,v)\leq \sigma\}$, the estimate below holds
\begin{eqnarray*}
2m(u,v)&\geq & 2m(u_\sigma(v),v)+\int_{u_\sigma(v)}^{u} \Lambda r^2(u',v) \partial_u r du'\\
&>&\sigma-\frac{\Lambda}{3}\sigma^3+\frac{\Lambda}{3}r^3(u,v) \\
&\geq& \sigma(1-\frac{\Lambda}{3}\sigma^2) \geq \frac{\sigma}{2}.
\end{eqnarray*}
Hence, in $\mathcal{T}\cap \{r(u,v)\leq \sigma\}$, it follows that
\begin{equation} \label{est mu}
\mu(u,v)> \f{\sigma}{2r(u,v)}.
\end{equation}
Now we rewrite (\ref{r v}) as
$$\f{-\partial \log |\partial_v r|/\partial u}{\partial \log r/\partial u}=\f{\mu-\Lambda r^2}{\mu-1}.$$
Applying \eqref{est mu} to the right hand side (RHS) of this equation in the region $\mathcal{T}\cap \{r(u,v)\leq \frac{\sigma}{2}\}$, we have
\begin{equation} \label{low up est}
1<\f{-\partial \log |\partial_v r|/\partial u}{\partial \log r/\partial u}\leq \f{\sigma}{\sigma-2r}.
\end{equation}
The first inequality in \eqref{low up est} implies 
$\f{\partial \log r|\partial_v r|}{\partial u}>0$, and hence
\begin{equation} \label{low lim}
r|\partial_v r|(u,v)>r|\partial_v r|(u_1,v),
\end{equation}
where $u_{\frac{\sigma}{2}}(v)<u_1\leq u \leq u^*(v)$ and $(u^*(v), v)\in \mathcal{S}$.

Next, using the second inequality in \eqref{low up est}, we deduce that
$$\f{\partial \log r|\partial_v r|}{\partial u}\leq -\f{\partial \log r}{\partial u}\cdot \f{2r}{\sigma-2r}=-2\f{\partial r}{\partial u}\cdot \f{1}{\sigma-2r}=\f{\partial \log (\sigma-2r)}{\partial u}.$$
Integrating this inequality with respect to $u$, we derive
\begin{equation} \label{up lim}
\f{r|\partial_v r|(u,v)}{r|\partial_v r|(u_1, v)}\leq \f{\sigma-2r(u,v)}{\sigma-2r(u_1,v)}\leq \f{\sigma}{\sigma-2r(u_1,v)}.
\end{equation}
Combing \eqref{low lim} with \eqref{up lim}, we obtain
\begin{equation} \label{lim est}
\f{\sigma}{\sigma-2r(u_1,v)} (r\partial_v r)(u_1, v)\leq (r\partial_v r)(u,v)< (r\partial_v r)(u_1, v).
\end{equation} 
A direct calculation further yields
$$\limsup_{u\rightarrow u^*(v)^-}\big(-(r\partial_v r)(u,v)\big)-\liminf_{u\rightarrow u^*(v)^-}\big(-(r\partial_v r)(u,v)\big)\leq \f{2r(u_1,v)}{\sigma-2r(u_1,v)}\big(-(r\partial_v r)(u_1,v)\big).$$
Noting that $r(u_1, v)\rightarrow 0$ as $u_1\rightarrow u^*(v)^-$, the above estimate whence indicates that $-(r\partial_v r)(u,v)$ tends to a positive limit $E(v)$ as $u\rightarrow u^*(v)^-.$ In the same fashion, we can also prove that $-(r\partial_u r)(u,v)$ tends to a positive limit $E^*(u)$ as $v\rightarrow v^*(u)^-.$ Moreover, since the convergence is uniform in a small neighborhood, the limit functions $E(v)$ and $E^*(u)$ are both continuous.

Next, employing
$$\f{\partial r^2}{\partial v}=2r \partial_v r, \quad  \quad \quad \f{\partial r^2}{\partial u}=2r \partial_u r,$$
we further conclude that the function $r^2(u,v)$ is $C^1$ in $\mathcal{Q}\cup(\mathcal{S}\backslash \mathcal{S}_0)$. Since $\mathcal{S}\backslash\mathcal{S}_0$ can be characterized by a curve $\{r^2(u,v)=0\}$, it is also a $C^1$ curve.

Next, we show that $\mathcal{S}\backslash \mathcal{S}_0$ is spacelike. We first fix a point $b_0=(u^*(v), v)\in \mathcal{S}$. Along the past incoming null cone, for any $(u,v)$ close enough to $b_0$, there holds
\begin{equation*}
   r(u,v)^2=\int_u^{u^*(v)} -2r\partial_u r(u',v) du'\gtrsim u^*(v)-u>0.
\end{equation*}
Thus, the singular boundary $\mathcal{S}$ does not contain the incoming null piece. We then exclude the outgoing null piece. Consider the following shadowed rectangle region:
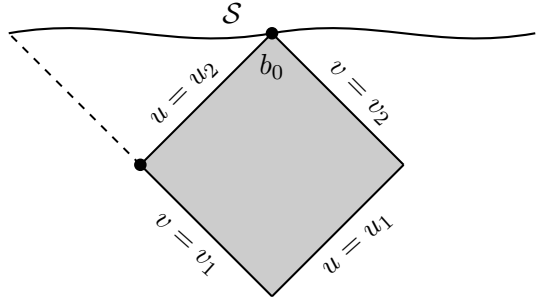
\begin{center}
\begin{tikzpicture}[scale=0.7]
\filldraw[white, fill=gray!40] (0,0)--(2.5,-2.5)--(0,-5)--(-2.5,-2.5)--(0,0);
\node [below] at(0,-0.2){$b_0$};
\draw [white](-3, 0)-- node[midway, sloped, above,black]{$\mathcal{S}$}(1.5, 0);
\draw [white](0, -5)-- node[midway, sloped, below,black]{$u=u_1$}(2.8, -2.5);
\draw [white](0, -5)-- node[midway, sloped, below,black]{$v=v_1$}(-2.8, -2.5);
\draw [white](0, 0)-- node[midway, sloped, above, black]{$u=u_2$}(-2.8,-2.8);
\draw [white](0, 0)-- node[midway, sloped, above, black]{$v=v_2$}(2.8,-2.8);
\draw[thick] (0,0)--(-2.5,-2.5);
\draw[thick] (0,0)--(2.5, -2.5);
\draw [thick] (-5,0) to [out=10,in=-170] (0,0);
\draw [thick] (0,0) to [out=10,in=-170] (5,0);
\draw[dashed] (-2.5,-2.5)--(-5,0);
\draw [thick] (-2.5,-2.5)--(0,-5);
\draw [thick] (2.5,-2.5)--(0,-5);
\draw[fill] (0,0) circle [radius=0.1];
\draw[fill] (-2.5,-2.5) circle [radius=0.1];
\end{tikzpicture}
\end{center}
Let $b_0=(u^*(v_2), v_2)\in \mathcal{S}$. Assume that the shadowed region is sufficiently small, i.e., $u_1$ is sufficiently close to $u_2$ and $v_1$ is sufficiently close to $v_2$. Following \eqref{r v}, it holds that $\partial_u\partial_v r<0$ and
\begin{equation*}
    \partial_v r(u_2,v)<\partial_v r(u_1,v),
\end{equation*}
where $v_1\leq v\leq v_2$. Integrating the above inequality on the interval $[v_1,v_2]$, we have
\begin{equation*}
    r(u_2,v_1)>r(u_1,v_1)-r(u_1,v_2)>0.
\end{equation*}
Therefore, $(u_2,v_1)$ does not reach the singularity $\mathcal{S}$ where $r(u,v)=0$. We thus falsify the existence of outgoing null piece in $\mathcal{S}$ and we conclude that the singularity $\mathcal{S}\backslash\mathcal{S}_0$ is spacelike.

\end{proof}
Employing the monotonicity of $\O^{-2}\partial_u r$ guaranteed by (\ref{conseq:r1}) and applying Proposition \ref{Prop 4.1}, we have
\begin{equation}\label{Omega upper}
\O^2(u,v)\leq c_0^{{\color{black}-1}}\cdot [-\partial_u r (u,v)]\leq \f{D}{r(u,v)}
\end{equation}
with $D$ a locally uniform constant depending on initial data.

To prove the desired upper bounds of $\partial \phi$ as stated in (ii) of Theorem \ref{local version}, we first derive a preliminary estimate for it. We operate in the trapped diamond region below. The inner rectangular region $D_0$ is chosen to be small enough. 
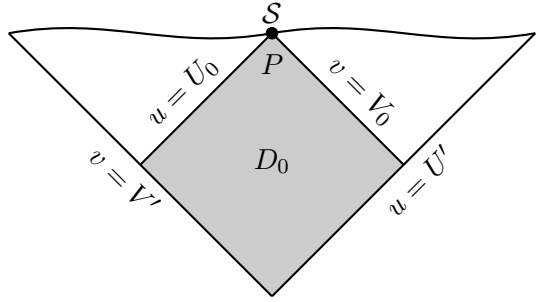
\begin{center}
\begin{tikzpicture}[scale=0.7]
\filldraw[white, fill=gray!40] (0,0)--(2.5,-2.5)--(0,-5)--(-2.5,-2.5)--(0,0);
\node[below] at(0, -0.2){$P$};
\node[below] at(0, -2){$D_0$};
\draw [white](-3, 0)-- node[midway, sloped, above,black]{$\mathcal{S}$}(3, 0);
\draw [white](0, -5)-- node[midway, sloped, below,black]{$u=U'$}(5, 0);
\draw [white](0, -5)-- node[midway, sloped, below,black]{$v=V'$}(-5, 0);
\draw [white](0, 0)-- node[midway, sloped, above, black]{$u=U_0$}(-2.8,-2.8);
\draw [white](0, 0)-- node[midway, sloped, above, black]{$v=V_0$}(2.8,-2.8);
\draw[thick] (0,0)--(-2.5,-2.5);
\draw[thick] (0,0)--(2.5, -2.5);
\draw [thick] (-5,0) to [out=10,in=-170] (0,0);
\draw [thick] (0,0) to [out=10,in=-170] (5,0);
\draw [thick] (-5, 0)--(0,-5);
\draw [thick] (5, 0)--(0,-5);
\draw[fill] (0,0) circle [radius=0.10];
\end{tikzpicture}
\end{center}
\begin{proposition}\label{preliminary}
For $0<\alpha\leq 1$, in $D_0$, the following estimates hold
\begin{equation} \label{est prelim}
|\partial_u \phi(u,v)|\leq \f{I_0}{r^{3+\alpha}(u,v)} \mbox{ and }\ |\partial_v \phi(u,v)|\leq \f{I_0}{r^{3+\alpha}(u,v)},
\end{equation}
where $I_0$ is a locally uniform number depending on initial data.
\end{proposition}
\begin{proof}
Note that equation (\ref{eqn phi}) for the scalar field does not contain the cosmological constant. Hence \eqref{est prelim} can be obtained in a similar way to the proof of Propostion 4.1 in \cite{AZ}. For an outline, employing  (\ref{eqn phi}) and Proposition \ref{Prop 4.1}, we have:
\begin{equation} \label{lim const}
r\partial_v r+C_2=o(1), \quad \quad r\partial_u r+C_1=o(1).
\end{equation}
For $0<\alpha\leq 1$, employing \eqref{lim const}, we can deduce that
\begin{equation*}
\int_{v=V'}^{v=V_0}C_1\cdot(r^{2+\alpha}\partial_v  \phi)^2 (U_0, v)dv+\int_{u=U'}^{u=U_0}C_2\cdot(r^{2+\alpha}\partial_u  \phi)^2 (u, V_0)du \leq I_0^2.
\end{equation*}
Based on $P=(U_0, V_0)$, we integrate \eqref{propeq:phi1}. Applying the above bound, we have
\begin{equation}
|r\partial_v \phi (U_0, V_0)|\lesssim \text{const}+\frac{I_0}{r^{2+\alpha}(U_0,V_0)}.
\end{equation}
This implies $|\partial_v \phi|\leq \f{I_0}{r^{3+\alpha}}$ for $0<\alpha<1$. Likewise we also obtain $|\partial_u \phi|\leq \f{I_0}{r^{3+\alpha}}$.
\end{proof}

\subsection{Refined Estimates of $r\partial_u r$ and $r\partial_v r$}
To obtain sharp blow-up rates for $\partial_u \phi$ and $\partial_v \phi$, here we improve estimates for $r\partial_u r$ and $r\partial_v r$.

To study the singular boundary $\mathcal{S}$ (where $r=0$), we still consider the diamond region $D_0$ (in $\mathcal{T}$) as in Proposition \ref{preliminary}. Without loss of generality, we zoom in $D_0$ and assume $U_0=0, V_0=0$ and $\tilde{U}, \tilde{V}, U', V' < 0$ as portrayed in the picture below: 
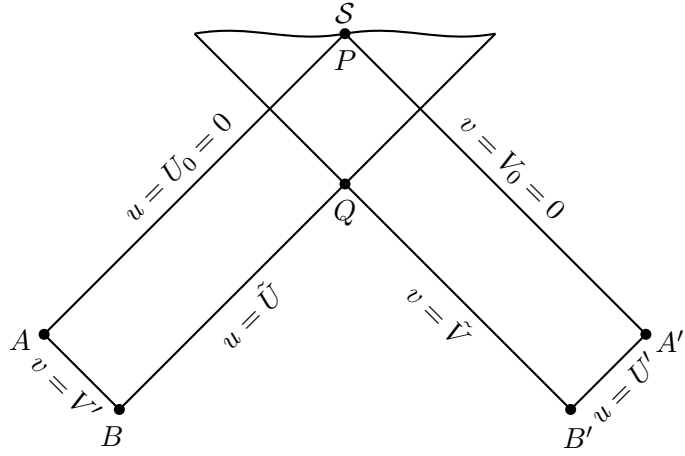
\begin{center}
\begin{tikzpicture}[scale=0.4]
\draw [white](-3, -0.2)-- node[midway, sloped, below,black]{$P$}(3, -0.2);
\draw [white](-3, -5.2)-- node[midway, sloped, below,black]{$Q$}(3, -5.2);
\draw [white](-14.6, -9.5)-- node[midway, sloped, below,black]{$A$}(-7, -9.5);
\draw [white](-6, -12.7)-- node[midway, sloped, below,black]{$B$}(-9.5, -12.7);
\draw [white](14.6, -9.5)-- node[midway, sloped, below,black]{$A'$}(7, -9.5);
\draw [white](6, -12.7)-- node[midway, sloped, below,black]{$B'$}(9.5, -12.7);
\draw [white](-3, 0)-- node[midway, sloped, above,black]{$\mathcal{S}$}(3, 0);

\draw [white](0, -5)-- node[midway, sloped, below,black]{$u=\tilde{U}$}(-7.5, -12.5);
\draw [white](0, -5)-- node[midway, sloped, below,black]{$v=\tilde{V}$}(7.5, -12.5);
\draw [white](7.5,-12.5)-- node[midway, sloped, below,black]{$u=U'$}(10, -10);
\draw [white](-7.5, -12.5)-- node[midway, sloped, below,black]{$v=V'$}(-10, -10);
\draw [white](0, 0)-- node[midway, sloped, above, black]{$u=U_0=0$}(-10,-10);
\draw [white](0, 0)-- node[midway, sloped, above, black]{$v=V_0=0$}(10,-10);
\draw[thick] (0,0)--(-10,-10);
\draw[thick] (0,0)--(10, -10);
\draw[thick] (0,-5)--(-7.5,-12.5);
\draw[thick] (0,-5)--(7.5,-12.5);
\draw[thick] (-10, -10)--(-7.5, -12.5);
\draw[thick] (10, -10)--(7.5, -12.5);

\draw [thick] (-5,0) to [out=10,in=-170] (0,0);
\draw [thick] (0,0) to [out=10,in=-170] (5,0);

\draw [thick] (-5, 0)--(0,-5);
\draw [thick] (5, 0)--(0,-5);
\draw[fill] (0,0) circle [radius=0.15];
\draw[fill] (0,-5) circle [radius=0.15];
\draw[fill] (-10,  -10) circle [radius=0.15];
\draw[fill] (-7.5,-12.5) circle [radius=0.15];
\draw[fill] (10,  -10) circle [radius=0.15];
\draw[fill] (7.5,-12.5) circle [radius=0.15];
\end{tikzpicture}
\end{center}

Given $P\in \mathcal{S}$ and $Q\in D_0$ close to $P$. Denoting $r(Q)=r_0$, along $u=0$, one can find $A$ (in the past of $P$) satisfying
$$r(A)=r(Q)^{\f{1}{8}}=r_0^{\f{1}{8}}.$$
After fixing $P$, $Q$ and $A$, we then define the points $B$, $A'$ and $B'$ as points of intersection in the above picture. Within $D_0$, we are now ready to prove improved estimates for $r\partial_u r$ and $r \partial_v r$.
\begin{proposition}\label{improved ru rv}
For any $Q\in D_0$ sufficiently close to $P\in \mathcal{S}$, the following improved estimates hold
\begin{equation}
|r\partial_u r+C_1|(Q)\leq 2r(Q)^{\f{1}{8}}, \quad |r\partial_v r+C_2|(Q)\leq 2r(Q)^{\f{1}{8}}.
\end{equation}
\end{proposition}
\begin{proof}
Using the second equation in \eqref{lim const}, we get the following estimate for $r^2$:
\begin{equation*}
\begin{split}
r^2(u,0)=&r^2(0,0)+\int_0^u \partial_u (r^2(u',0))du'=\int_0^u (2r\partial_u r) (u',0)du'\\
=&\int_0^u[-2C_1+o(1)]du'=[-2C_1+o(1)]u.
\end{split}
\end{equation*}
Utilizing the first equation in \eqref{lim const}, we further derive
\begin{equation}\label{asympupperbdofrsq}
\begin{split}
r^2(u,v)=&r^2(u,0)+\int_0^v \partial_v (r^2(u, v'))dv'=r^2(u,0)+\int_0^v (2r\partial_v r) (u,v')dv'\\
=&[-2C_1+o(1)]u+[-2C_2+o(1)]v.
\end{split}
\end{equation}
In particular, at $Q$, we have $u=\tilde{U}, v=\tilde{V}$. Hence the above estimate \eqref{asympupperbdofrsq} reads
$$[-2C_1+o(1)]\tilde{U}+[-2C_2+o(1)]\tilde{V}=r^2_0.$$
This implies that
\begin{equation} \label{est uvQ}
|\tilde{U}|\leq \f{r^2_0}{C_1}, \quad \quad \quad \quad |\tilde{V}|\leq \f{r^2_0}{C_2}.
\end{equation}
Along $AP$, noting that $r \ll 1$ and taking (\ref{Omega upper}) into account, we then integrate equation \eqref{propeq:r2} and obtain
\begin{equation}\label{PA}
\begin{split}
|(r\partial_u r)(P)-(r\partial_u r)(A)|=&\int_{V'}^0\f14\O^2(1-\Lambda r^2)(0,v')dv'\lesssim \int_{V'}^0\f{D}{r(0,v')}dv'
\\
\lesssim& \int_{V'}^0 -\partial_v r(0,v')dv' = r(A)=r_0^{\f{1}{8}}.
\end{split}
\end{equation}
\noindent Along $BA$, we rewrite (\ref{conseq:r1}) as
\begin{equation}\label{ruu}
\partial_u (r\partial_u r)=-r^2(\partial_u \phi)^2+2\partial_u \log\O\cdot r\cdot \partial_u r+\partial_u r \partial_u r.
\end{equation}
We then bound every term on the RHS. First, employing Proposition \ref{Prop 4.1}, there holds $|\partial_u r\partial_v r|\lesssim 1/r^2$. Then, by Proposition \ref{preliminary}, we have $r^2(\partial_u \phi)^2\leq {I^2_0}/{r^{4+2\a}}.$ To estimate $\partial_u\log\O$, integrating (\ref{propeq:Omega}) and using (\ref{Omega upper}), (\ref{asympupperbdofrsq}) and Proposition \ref{preliminary}, we deduce $$|\partial_u\log\O|\leq I^2_0/r^{4+2\a},$$
and hence
$$|2\partial_u\log\O\cdot r\partial_u r|\leq I^2_0/r^{4+2\a}.$$
With these estimates, for $0<\alpha<1$, we integrate (\ref{ruu}) and obtain the following estimate
\begin{equation}\label{AB}
\begin{split}
|(r\partial_u r)(A)-(r\partial_u r)(B)|\leq& \int_{\tilde{U}}^0 \f{I^2_0}{r^{{\color{black} 4}+2\a}}(u',V')du'\lesssim \f{1}{r^{4+2\a}(A)}\cdot |\tilde{U}|\\
\leq&\f{1}{r_0^{\f{4+2\a}{8}}}\cdot \f{r_0^2}{C_1}\leq r_0.
\end{split}
\end{equation}
Combining (\ref{PA}) and (\ref{AB}), we then derive
\begin{equation}\label{PB}
|(r\partial_u r)(P)-(r\partial_u r)(B)|\leq r_0^{\f{1}{8}}.
\end{equation}
\noindent Along $AB$, since $r\partial_u r\approx C_1$, it holds that
$$|\partial_u r|\lesssim \f{1}{r(A)}=\f{1}{r_0^{ \f{1}{8}}}.$$
This then implies
$$|r(B)-r(A)|\leq \int_{\tilde{U}}^0 |\partial_u r(u', V')|du'\leq \f{1}{r_0^{\f{1}{8}}}\cdot \f{r_0^2}{C_1}\leq r_0.$$
Hence, for $r(B)$, we obtain the following lower and upper bounds,
$$r_0^{\f{1}{8}}\leq r(A) \leq r(B) \leq r(A)+r_0\leq 2 r_0^{\f{1}{8}}.$$
Analogously, along $BQ$, via integrating equation \eqref{propeq:r2}, we derive the following estimate
\begin{equation}\label{BQ}
\begin{split}
|(r\partial_u r)(B)-(r\partial_u r)(Q)|=&\int_{V'}^{\tilde{V}}\f14\O^2(1-\Lambda r^2)(\tilde{U},v')dv'\\
\lesssim& \int_{V'}^{\tilde{V}} -\partial_v r(\tilde{U},v')dv' = r(B)-r(Q)\leq \frac{3}{2}r_0^{\f{1}{8}}.
\end{split}
\end{equation}
\noindent A combination of (\ref{PB}) and (\ref{BQ}) then gives
\begin{equation}\label{PQ}
|(r\partial_u r)(P)-(r\partial_u r)(Q)|\leq 2r_0^{\f{1}{8}}.
\end{equation}
which is equivalent to
$$|r\partial_u r+C_1|(Q)\leq 2r(Q)^{\f{1}{8}}.$$
Similarly, by analyzing $A'$ and $B'$, we deduce $|r\partial_v r+C_2|(Q)\leq 2r(Q)^{\f{1}{8}}.$ This concludes the proof of the proposition.
\end{proof}

\subsection{Sharp polynomial upper bounds for $\partial_u \phi$ and $\partial_v \phi$}\label{section sharp phi}

In this subsection, we will prove statement (ii) of Theorem \ref{local version}: At any point $(u,v)\in\mathcal{T}$ near $\mathcal{S}$, there exists positive number $F_1$ and $F_2$ (depending on the initial data), such that \eqref{eq:dphinearr0} holds, i.e.,
$$|\partial_u \phi(u,v)|\leq \f{F_1}{r(u,v)^2}, \quad \quad |\partial_v \phi(u,v)|\leq \f{F_2}{r(u,v)^2}.$$
Within the spacetime region $D_0$, we consider constant $r$-level sets $\{L_r\}$. Define
$$\Psi(r):=\max\{\sup_{P\in L_r} |C_2\cdot r\partial_u \phi|(P), \sup_{Q\in L_r} |C_1\cdot r\partial_v \phi|(Q)\}.$$
Then using equation (\ref{eqn phi}) for the scalar field  (which does not contain $\Lambda$), applying \eqref{lim const} and Proposition \ref{improved ru rv}, for any small enough $\tilde{r}>0$ one can obtain that
$$\Psi(\tilde{r})\leq I.D.\times e^{\int_{\tilde{r}}^{2^{-l+1}} \f{1+O(r^{\f{1}{8}})}{r}\cdot} dr=I.D.\times e^{-\ln \tilde{r}+O(1)}\leq\f{C}{\tilde{r}}.$$
Here, $C$ is a locally uniform constant depending only on initial data. This immediately leads to \eqref{eq:dphinearr0}, and hence completes the proof of (ii) of Theorem \ref{local version}. Interested readers are referred to Section 6 in \cite{AZ} for more details.

As a corollary, with the above estimates at our disposal, we derive lower bound estimate for $\O^2$. Integrating (\ref{propeq:Omega}) twice and employing \eqref{eq:dphinearr0}, \eqref{lim const}, (\ref{Omega upper}), for all $(u,v)$ near $\mathcal{S}$ we deduce that
$$\log\Omega(u,v)\geq \log r^{\tilde{C}}(u,v),$$
where $\tilde{C}$ is a (locally) uniform positive constant depending on initial data. Together with (\ref{Omega upper}) we have
\begin{equation} \label{rough omega lowerbd}
r^{2\tilde{C}}(u,v)   \leq \Omega^2(u,v)\leq  \f{D}{r(u,v)}, \quad \f{r(u,v)}{D}\leq \Omega^{-2}(u,v)\leq r^{-2\tilde{C}}(u,v).
\end{equation}

\subsection{Optimal estimates of $r\partial_u r$ and $r\partial_v r$} \label{subsec sharp dr}
Now the sharp estimates for $\partial_u \phi$ and $\partial_v \phi$ have been derived, we can further refine the estimates for $r\partial_u r$ and $r\partial_v r$ stated in Proposition \ref{improved ru rv} and obtain our optimal\footnote{Using our aforementioned method, the remainder estimate in this section is optimal.} estimates. We still consider the following diamond region: 
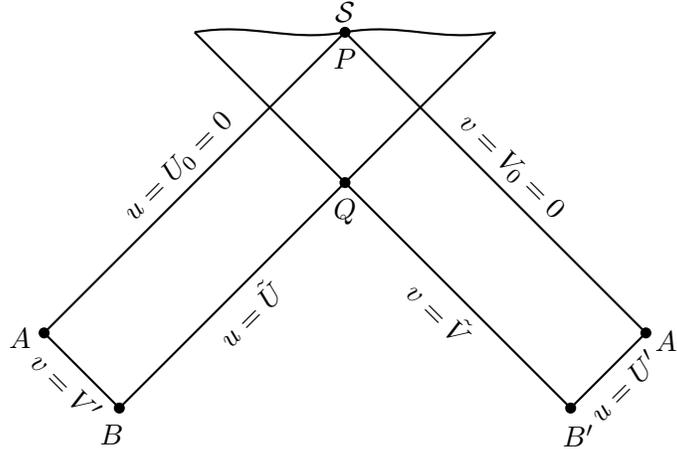
\begin{center}
\begin{tikzpicture}[scale=0.4]
\draw [white](-3, -0.2)-- node[midway, sloped, below,black]{$P$}(3, -0.2);
\draw [white](-3, -5.2)-- node[midway, sloped, below,black]{$Q$}(3, -5.2);
\draw [white](-14.6, -9.5)-- node[midway, sloped, below,black]{$A$}(-7, -9.5);
\draw [white](-6, -12.7)-- node[midway, sloped, below,black]{$B$}(-9.5, -12.7);
\draw [white](14.6, -9.5)-- node[midway, sloped, below,black]{$A'$}(7, -9.5);
\draw [white](6, -12.7)-- node[midway, sloped, below,black]{$B'$}(9.5, -12.7);
\draw [white](-3, 0)-- node[midway, sloped, above,black]{$\mathcal{S}$}(3, 0);

\draw [white](0, -5)-- node[midway, sloped, below,black]{$u=\tilde{U}$}(-7.5, -12.5);
\draw [white](0, -5)-- node[midway, sloped, below,black]{$v=\tilde{V}$}(7.5, -12.5);
\draw [white](7.5,-12.5)-- node[midway, sloped, below,black]{$u=U'$}(10, -10);
\draw [white](-7.5, -12.5)-- node[midway, sloped, below,black]{$v=V'$}(-10, -10);
\draw [white](0, 0)-- node[midway, sloped, above, black]{$u=U_0=0$}(-10,-10);
\draw [white](0, 0)-- node[midway, sloped, above, black]{$v=V_0=0$}(10,-10);
\draw[thick] (0,0)--(-10,-10);
\draw[thick] (0,0)--(10, -10);
\draw[thick] (0,-5)--(-7.5,-12.5);
\draw[thick] (0,-5)--(7.5,-12.5);
\draw[thick] (-10, -10)--(-7.5, -12.5);
\draw[thick] (10, -10)--(7.5, -12.5);

\draw [thick] (-5,0) to [out=10,in=-170] (0,0);
\draw [thick] (0,0) to [out=10,in=-170] (5,0);

\draw [thick] (-5, 0)--(0,-5);
\draw [thick] (5, 0)--(0,-5);
\draw[fill] (0,0) circle [radius=0.15];
\draw[fill] (0,-5) circle [radius=0.15];
\draw[fill] (-10,  -10) circle [radius=0.15];
\draw[fill] (-7.5,-12.5) circle [radius=0.15];
\draw[fill] (10,  -10) circle [radius=0.15];
\draw[fill] (7.5,-12.5) circle [radius=0.15];
\end{tikzpicture}
\end{center}
Given $P\in\mathcal{S}$ and $Q\in D_0$ close to $P$. Denoting $r(Q)=r_0$, along $u=0$, we pick $A$ in the past of $P$ satisfying $r(A)=r(Q)^{\beta}=r_0^{\beta}$ with $\beta$ to be determined. Similarly to the proof of Proposition \ref{improved ru rv}, along $AP$ we also have that
\begin{equation}\label{PAsharp}
|(r\partial_u r)(P)-(r\partial_u r)(A)|\lesssim r_0^{\beta}.
\end{equation}
Then along $BA$, employing the sharp estimates for $\partial_u \phi$ and $\partial_v \phi$ obtained in Section \ref{section sharp phi}, repeating \eqref{AB}, we get
\begin{equation}\label{ABsharp}
|(r\partial_u r)(A)-(r\partial_u r)(B)|\lesssim r_0^{2-2\beta}.
\end{equation}
Comparing (\ref{PAsharp}) and (\ref{ABsharp}), one can see that choosing $\beta=2-2\beta$, i.e., $\beta=\f23$ optimizes the upper bound. Now we have
\begin{equation}\label{PBsharp}
|(r\partial_u r)(P)-(r\partial_u r)(B)|\leq r_0^{\beta}=r_0^{\f23}.
\end{equation}
The remaining part of this proof is a replication of Proposition \ref{improved ru rv}. We hence prove the following optimal estimates for $r\partial_u r$ and $r\partial_v r$:
\begin{equation} \label{sharp ru rv}
|r\partial_u r+C_1|(Q)\lesssim r(Q)^{\f23}, \quad |r\partial_v r+C_2|(Q)\lesssim r(Q)^{\f23}.
\end{equation}
This gives the statement (i) of Theorem \ref{local version}. As a corollary, with \eqref{sharp ru rv}, we further improve the regularity of curve $r(u,v)=0$ in $(u,v)$ plane. By Proposition \ref{Prop 4.1}, we know that this curve is $C^1$. Now we improve it to $C^{1,1/3}_{loc}$.
\begin{proof}[Proof of Theorem \ref{sing holder}]
Consider a point $P=(u,v^*(u))$ where $\partial_u r^2(P)=-2C_1$ and $\partial_v r^2(P)=-2C_2$. Note that along $\mathcal{S}$ there holds
\begin{equation*}
    \frac{dr}{du}=\partial_u r+\frac{dv^*}{du} \partial_v r=0,
\end{equation*}
which implies
\begin{equation} \label{curve}
\frac{dv^*}{du}=-\frac{\partial_u r}{\partial_v r}.
\end{equation}
Immediately, for $\bar{P}=(\bar{u},v^*(\bar{u}))$ being a singular point near $P$ and $Q=(\bar{u},v^*(u))$, we have
\begin{equation}  \label{holder triangle0}
\Big|\frac{dv^*}{du}(P)-\frac{dv^*}{du}(\bar{P})\Big| \leq  \Big|\frac{\partial_u r}{\partial_v r}(P)-\frac{\partial_u r}{\partial_v r}(Q)\Big|+\Big|\frac{\partial_u r}{\partial_v r}(Q)-\frac{\partial_u r}{\partial_v r}(\bar{P})\Big|. 
\end{equation}
See the following figure.
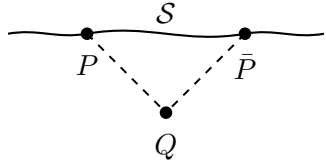
\begin{center}
\begin{tikzpicture}[scale=0.7]
\draw [white](-3, -0.2)-- node[midway, sloped, below,black]{$P$}(0, -0.2);
\draw [white](0, -0.2)-- node[midway, sloped, below,black]{$\bar{P}$}(3, -0.2);
\draw [white](-3, -1.7)-- node[midway, sloped, below,black]{$Q$}(3, -1.7);
\draw [white](-3, 0)-- node[midway, sloped, above,black]{$\mathcal{S}$}(3, 0);
\draw [thick] (-3,0) to [out=10,in=-170] (-1.5,0);
\draw [thick] (-1.5,0) to [out=10,in=-170] (1.5,0);
\draw [thick] (1.5,0) to [out=10,in=-170] (3,0);
\draw [dashed] (-1.5,0)--(0,-1.5);
\draw [dashed] (1.5,0)--(0,-1.5);
\draw[fill] (-1.5,0) circle [radius=0.1];
\draw[fill] (1.5,0) circle [radius=0.1];
\draw[fill] (0,-1.5) circle [radius=0.1];
\end{tikzpicture}
\end{center}
By \eqref{sharp ru rv}, we deduce that
\begin{equation*}
\begin{split}
 \Big|\frac{\partial_u r}{\partial_v r}(P)-\frac{\partial_u r}{\partial_v r}(Q)\Big| =&  \Big|\frac{C_1}{C_2}-\frac{r\partial_u r}{r\partial_v r}(Q)\Big|\\
 =&\Big|\frac{C_1(r\partial_v r+C_2)-C_2(C_1+r\partial_u r)}{C_2r\partial_v r}(Q)\Big|\\
 \lesssim& (r(Q)^2)^\frac{1}{3}.
 \end{split}
\end{equation*}
A similar estimate based on $\bar{P}$ can be obtained in the same fashion, i.e.,
\begin{equation*}
 \Big|\frac{\partial_u r}{\partial_v r}(Q)-\frac{\partial_u r}{\partial_v r}(\bar{P})\Big| \lesssim (r(Q)^2)^\frac{1}{3}.
\end{equation*}
Back to \eqref{holder triangle0}, with the above inequalities, we obtain
\begin{equation} \label{holder triangle}
\Big|\frac{dv^*}{du}(P)-\frac{dv^*}{du}(\bar{P})\Big| \lesssim (r(Q)^2)^\frac{1}{3}.  
\end{equation}
For $r(Q)^2$, it also holds 
\begin{align*}
    r(Q)^2&=\frac12(r^2(\bar{u},v^*(u))-r^2(\bar{u},v^*({\bar{u}}))+\frac12(r^2(\bar{u},v^*(u))-r^2(u,v^*({u}))\\
    &\leq 2C_2\cdot|v^*(u)-v^*(\bar{u})|+2C_1\cdot|u-\bar{u}|.
\end{align*}
For the first term on the RHS, by mean value theorem, we have
\begin{equation*}
    |v^*(u)-v^*(\bar{u})| \leq \sup \left|\frac{dv^*}{du}\right|\cdot |u-\bar{u}|.
\end{equation*}
Invoking \eqref{curve}, we then obtain
\begin{equation*}
    |v^*(u)-v^*(\bar{u})| \leq \sup \left|\frac{\partial_u r}{\partial_v r}\right|\cdot |u-\bar{u}| \leq 4 \frac{C_1}{C_2}\cdot |u-\bar{u}|.
\end{equation*}
This leads to the following estimate of $r(Q)^2$:
\begin{equation*}
    r(Q)^2 \leq 10C_1 \cdot |u-\bar{u}|.
\end{equation*}
Back to \eqref{holder triangle}, we hence prove the (local) H\"{o}lder continuity of curve $r(u,v)=0$ in $(u,v)$ plane:
\begin{equation*}
\Big|\frac{dv^*}{du}(P)-\frac{dv^*}{du}(\bar{P})\Big|  \lesssim |u-\bar{u}|^\frac{1}{3}.
\end{equation*}
\end{proof}

\subsection{Higher Order Estimates} \label{high order est}
To estimate the Kretschmann scalar, we also need estimates for higher order derivatives of $\O$, $r$ and $\phi$. We summarize these bounds in the following proposition. 
\begin{proposition}
For the first order derivatives of $\O^2(u,v)$, we have
\begin{equation}\label{1stderivofOmega}
\begin{aligned}
&|\partial_u\log\O|(u,v)\lesssim\f{1}{r^2(u,v)}, \quad  |\partial_v\log\O|(u,v)\lesssim\f{1}{r^2(u,v)},\\
&|\partial_u (\O^2(u,v))|\lesssim \f{1}{r^3(u,v)}, \quad |\partial_v(\O^2(u,v))(u,v)|\lesssim \f{1}{r^3(u,v)}.
\end{aligned}
\end{equation}
For the second order derivatives of $r(u,v)$, there hold
\begin{equation}\label{2ndorderr}
\begin{aligned}
&|\partial_u \partial_v {\color{black}(r^2)}|{\color{black}\lesssim \f{1}{r}\lesssim} \f{1}{r^2}, \quad|\partial_u \partial_u {\color{black}(r^2)}|{\color{black}\lesssim} \f{1}{r^2}, \quad |\partial_v \partial_v {\color{black}(r^2)}|{\color{black}\lesssim} \f{1}{r^2},\\
&|\partial_u \partial_v r|{\color{black}\lesssim} \f{1}{r^3}, \quad|\partial_u \partial_u r|{\color{black}\lesssim} \f{1}{r^3}, \quad |\partial_v \partial_v r|{\color{black}\lesssim} \f{1}{r^3}.
\end{aligned}
\end{equation}
The second order derivatives of $\phi(u,v)$ satisfy
\begin{equation}\label{2ndderivofphi}
 |\partial_u \partial_v \phi|{\color{black}\lesssim} \f{1}{r^4}, \quad |\partial_u \partial_u \phi|{\color{black}\lesssim} \f{1}{r^4}, \quad |\partial_v \partial_v \phi|{\color{black}\lesssim} \f{1}{r^4}.   
\end{equation}
For second order derivatives of $\O^2(u,v)$, we have
\begin{equation}\label{2ndderivofOmega}
\begin{aligned}
&|\partial_u\partial_v \log{\color{black}(\O^2)}|{\color{black}\lesssim} \f{1}{r^4}, \quad |\partial_u\partial_u \log{\color{black}(\O^2)}|{\color{black}\lesssim} \f{1}{r^4}, \quad |\partial_v\partial_v \log{\color{black}(\O^2)}|{\color{black}\lesssim} \f{1}{r^4},\\
&|\partial_v\partial_u {\color{black}(\O^2)}|{\color{black}\lesssim}\f{1}{r^5}, \quad |\partial_u \partial_u {\color{black}(\O^2)}|{\color{black}\lesssim} \f{1}{r^5}, \quad |\partial_v \partial_v {\color{black}(\O^2)}|{\color{black}\lesssim} \f{1}{r^5}.
\end{aligned}
\end{equation}
\end{proposition}
\begin{proof}
For the first order derivatives of $\O^2(u,v)$, integrating (\ref{propeq:Omega}), by \eqref{eq:dphinearr0} and (\ref{Omega upper}), one can obtain the desired estimates \eqref{1stderivofOmega}. For $\partial_u(\partial_v {\color{black}(r^2)})$, we first rewrite (\ref{propeq:r1})
$$\partial_u(\partial_v {\color{black}(r^2)})=-\f12\O^2(1-\Lambda r^2).$$
The estimate of $\partial_u(\partial_v {(r^2)})$ then follows from $\O^2(1-\Lambda r^2)\ls1/r\ls 1/r^2$. Then, by employing the constraint equations \eqref{conseq:r1}\eqref{conseq:r2}, the estimates \eqref{eq:dphinearr0}, \eqref{Omega upper} and \eqref{1stderivofOmega}, the estimates of $\partial_u(\partial_u (r^2))$ and $|\partial_v(\partial_v r^2)|$ can be derived in the same manner as in \cite{AZ}. For second order derivatives of the scalar field and $\Omega$, employing the equation (\ref{eqn phi}) and \eqref{eq:dphinearr0}, one can obtain \eqref{2ndderivofphi}. Using the equation of (\ref{propeq:Omega}) and the above derived estimates, one can further derive \eqref{2ndderivofOmega}. We refer the readers to Section 7 in \cite{AZ} for more details on how to prove this proposition.
\end{proof}

\subsection{Estimates of Kretschmann scalar}\label{Kretschmann}
As calculated in \cite{AZ}, there holds the following expression for the Kretschmann scalar:
\begin{equation}\label{Kretschmann2}
\begin{split}
&R^{\alpha\beta\rho\sigma}R_{\alpha\beta\rho\sigma}\\
=&\f{4}{r^4 \O^8}\bigg(16\cdot(\f{\partial^2 r}{\partial u \partial v})^2\cdot r^2\cdot\O^4+16\cdot \f{\partial^2 r}{\partial u^2}\cdot\f{\partial^2 r}{\partial v^2} \cdot r^2\cdot\O^4 \bigg)\\
&+\f{4}{r^4 \O^8}\bigg(-32\cdot \f{\partial^2 r}{\partial u^2}\cdot\partial_v r \cdot r^2 \cdot \O^4\cdot \partial_v \log\O -32\cdot\f{\partial^2 r}{\partial v^2}\cdot r^2\cdot\partial_u r \cdot \O^4\cdot \partial_u \log\O \bigg)\\
&+\f{4}{r^4 \O^8}\bigg( 16\cdot (\partial_v r)^2\cdot (\partial_u r)^2\cdot \O^4+64\cdot \partial_v r\cdot r^2\cdot \partial_u r\cdot \O^4\cdot \partial_u \log\O\cdot \partial_v\log\O+8\cdot\partial_v r\cdot \partial_u r\cdot \O^6\bigg)\\
&+\f{4}{r^4 \O^8}\bigg(16\cdot r^4\cdot (\f{\partial^2 \O}{\partial v \partial u})^2\cdot \O^2-32\cdot r^4\cdot \f{\partial^2 \O}{\partial v \partial u}\cdot \O^3 \cdot \partial_v \log\O \cdot \partial_u \log\O \bigg)\\
&+\f{4}{r^4 \O^8}\bigg( 16\cdot r^4\cdot \O^4\cdot (\partial_v \log\O)^2\cdot (\partial_u \log\O)^2+\O^8 \bigg).\\
\end{split}
\end{equation}

We first derive a lower bound for the Kretschmann scalar. The following three properties holds:
\begin{itemize}
\item[1)]The Kretschmann scalar obeys a polynomial lower bound,  \begin{equation}\label{mass inequality}
R_{\a\b\gamma\delta}R^{\a\b\gamma\delta}\geq \f{16\, m^2}{r^6}\geq \f{16\, \varpi^2}{r^6}.
\end{equation}
As in \cite{DC91}, we choose a null frame $(e_1,e_2,e_3,e_4)$ with $e_A$ (where $A=1,2$) being orthonormal on the 2-sphere and $e_3,e_4$ being the corresponding null vectors satisfying $\langle e_3,e_4 \rangle=-2$. The Kretschmann scalar can now be expressed as
$$
\f14 R_{\a\b\gamma\delta}R^{\a\b\gamma\delta}=\f14 R_{ABCD}R_{ABCD}+\f12(R_{A3B3}R_{A4B4}+R_{A3B4}R_{A4B3})+\bigg(\f14 R_{3434}\bigg)^2.
$$
To get the lower bound, for the RHS, the last term can be neglected since it is non-negative. The first term is equal to $\f{4\, m^2}{r^6}$, which is the desired lower bound in \eqref{mass inequality}. It then suffices to show that the second term is also non-negative. In fact, one can verify that $R_{A3B4}=R_{A4B3}$, $R_{A3B3}=(D_3 \phi)^2$ and $R_{A4B4}=(D_4 \phi)^2$. Hence the second term is also non-negative and then \eqref{mass inequality} follows.

\item [2)] By \eqref{modm u}, the modified Hawking mass $\varpi$ is monotonic in the trapped region $\mathcal{T}$, i.e., $\partial_u \varpi(u,v)\geq 0$.
\item [3)] The apparent horizon $\mathcal{A}$ consists of points $(u_{\mathcal{A}} (v),v)$ such that $\partial_v r(u_{\mathcal{A}} (v),v)=0$ and hence we have $m(u_{\mathcal{A}} (v),v)=\frac{1}{2} r(u_{\mathcal{A}} (v),v)$.
\end{itemize}

These properties together imply the following uniform lower bound for the Kretschmann scalar in $\mathcal{T}$:
\begin{equation}
\label{eq:prelimlowboundkretsch}
R_{\a\b\gamma\delta}R^{\a\b\gamma\delta}(u,v) \geq 4\frac{(r-\Lambda r^3/3)^2(u_{\mathcal{A}}(v),v)}{r^6(u,v)}.
\end{equation}

In \eqref{eq:prelimlowboundkretsch}, we can see that the Kretschmann scalar blows up \emph{at least} as fast as $r^{-6}$ when $r(u,v)\to0$. Note that this $r^{-6}$ lower bound is also consistent with its blow-up rate in Schwarzschild-de Sitter black holes.
~\\~

Now we deduce the upper bound estimate for the Kretschmann scalar. In the following, we outline how to estimate the terms on the RHS of \eqref{Kretschmann2}. We refer the readers to Section 8 in \cite{AZ} for more details. With Proposition \ref{Prop 4.1} and \eqref{2ndorderr}, we obtain polynomial upper bounds for $|\partial_u r|, |\partial_v r|, |\partial_u \partial_v r|, |\partial_u \partial_u r|, |\partial_v \partial_v r|$. Via (\ref{Omega upper}), \eqref{rough omega lowerbd} and \eqref{1stderivofOmega}, we can control $|\O|$, $|\partial_u \log\O|$, $|\partial_v \log\O|$, $\O^{-2}$. For $\partial_v \partial_u \O$, we use the following equation
\begin{equation}\label{Omega 3}
\O\cdot\partial_v \partial_u \O=\f12\cdot \partial_u (\O^2)\cdot \partial_v \log(\O^2)-\O^2\cdot\partial_v \log\O\cdot \partial_u \log\O+\f12\cdot \O^2\cdot\partial_v\partial_u \log{\color{black}(\O^2)}.
\end{equation}
Invoking (\ref{Omega upper}), \eqref{1stderivofOmega} and \eqref{2ndderivofOmega} to this equation, we then obtain polynomial upper bound for $|\O\cdot\partial_v\partial_u\O|$. Combining all the above estimates, we conclude the proof of statement (iii) of Theorem \ref{local version}, i.e., for all $(u,v)$ close to $(u_0,v_0)\in \mathcal{S}$, it holds
\begin{equation*}
r(u,v)^{-6}\lesssim R^{\alpha\beta\gamma\delta}R_{\alpha\beta\gamma\delta}(u,v)\lesssim r(u,v)^{-N_{u_0,v_0}},
\end{equation*}
where $N_{u_0,v_0}$ is a positive number depending on the initial data.

\subsection{Application to hard phase model}
We further apply the above derived estimates to the irrotational hard-phase Einstein-Euler system and prove Theorem \ref{hard phase thm}. Consider the hard phase model of Einstein-Euler system \eqref{einstein euler normal} 
\begin{equation*}
\begin{split}
&\mbox{Ric}_{\mu\nu}-\f12Rg_{\mu\nu}+\Lambda g_{\mu\nu}=2T_{\mu\nu},\\
&T_{\mu\nu}=(\rho+p)\vartheta_\mu \vartheta_\nu+p g_{\mu\nu},
\end{split}
\end{equation*}
where $\rho$ denotes the fluid density, $\vartheta^\mu$ is the fluid velocity, and $p$ represents the pressure. In particular, for the hard phase model, the pressure satisfies the equation of state $$p=\rho-\rho_0\quad \text{for}\quad \rho>\rho_0$$
with the constant $\rho_0$ being the nuclear saturation density.

Now we define the future timelike vector field 
$$V=\|V\|\vartheta$$ 
with
\begin{eqnarray} \label{renorm vel}
 \nonumber   \|V\|&=&\sqrt{-g(V,V)}:=\sqrt{\frac{p+\rho}{\rho_0}}\\
 &=&\sqrt{\frac{2\rho}{\rho_0}-1} \geq 1.
\end{eqnarray}
The fluid flow is irrotational if there exists a potential function $\phi$ such that
\begin{equation} \label{irr}
V^\mu=-\nabla^\mu \phi.
\end{equation} 
With this velocity potential function $\phi$, we can recast the Einstein-Euler system \eqref{einstein euler normal} as 
\begin{equation*} 
\begin{split}
&\mbox{Ric}_{\mu\nu}-\f12Rg_{\mu\nu}=2T_{\mu\nu},\\
&T_{\mu\nu}=\partial_\mu \phi \partial_\nu \phi+\frac12(\partial^\sigma \phi \partial_\sigma \phi-\rho_0)g_{\mu\nu},
\end{split}
\end{equation*}
which is \eqref{hard phase}.

 According to Theorem \ref{local version}, the derivatives of the velocity potential satisfy
\begin{equation} \label{est pot}
    |\partial_u \phi|\lesssim r^{-2},\quad\quad |\partial_v \phi| \lesssim r^{-2}.
\end{equation}
We then aim at deducing the quantitative estimates for the fluid velocity $\vartheta^\mu$ and density $\rho$ from \eqref{est pot}. It follows from \eqref{rough omega lowerbd}, \eqref{renorm vel}, \eqref{irr} and \eqref{est pot} that 
\begin{align*}
|\vartheta^\mu| &=\frac{|V^\mu|}{\|V\|}\leq |\nabla^\mu \phi|\leq 4\O^{-2}|\partial \phi|
\lesssim r^{-2-2\tilde{C}}.
\end{align*}
Next, it follows from \eqref{irr} that $\|V\|=\|d\phi\|$. Hence, together with \eqref{renorm vel}, this implies
$$
\rho=\frac{\rho_0}{2} (1+\|d\phi\|^2)=\frac{\rho_0}{2} (1+4\O^{-2}|\partial_u\phi \partial_v \phi|).
$$
Employing \eqref{rough omega lowerbd} and \eqref{est pot}, we deduce that
$$
|\rho| \lesssim r^{-4-2\tilde{C}}.
$$
This concludes the proof of Theorem \ref{hard phase thm}.

\section{Initial data respecting exponential Price's law}
\label{sec:initialdata}
In this section, we set up the characteristic initial value problem in Theorem \ref{main thm}. In particular, under $(U,v)$ coordinates, for the solution $(r,\widehat{\Omega}^2, \phi)$ to the spherically symmetric Einstein-scalar field system \eqref{ESlam}, we prescribe initial data satisfying exponential Price's law on the following two hypersurfaces:
\begin{align*}
H_0:=&\:\{(U,v)\in \R^2\,|\, U=0, v_0\leq v<\infty\},\\
\underline{H}_0:=&\:\{(U,v)\in \R^2\,|\, 0\leq U\leq U_0, v=v_0\}.
\end{align*}
We first fix the double-null coordinates $(U,v)$ by imposing the following gauge condition along $H_0$:
\begin{align}
\label{gaugeeq:Omega1}
\widehat{\Omega}^2(0,v)&=\widehat{\Omega}_{SdS}^2(0,v){\underset{\eqref{reform OSdS2}}{\overset{\eqref{reform OSdS1}}{=}}}e^{2\alpha_S C_0^\ast}\frac{\Lambda}{3r_S}(r_D-r_S)^{1+2\alpha_S\beta_D}(2r_S+r_D)^{1-2\alpha_S\beta_-}e^{\alpha_S v} \nonumber\\
&=: A_S e^{\alpha_S v} \quad \textnormal{for all $v\geq v_0$},
\end{align}
and the gauge condition along $\underline{H}_0$:
\begin{align}
\label{gaugeeq:Omega2}
\widehat{\Omega}^2(U,v_0)=&\widehat{\Omega}_{SdS}^2(U,v_0)\nonumber\\=&\:e^{2\alpha_S C_0^\ast}\frac{\Lambda}{3r_{SdS}(U, v_0)}(r_D-r_{SdS}(U,v_0))^{1+2\alpha_s\beta_D}\nonumber\\&\times(r_{SdS}(U,v_0)+r_S+r_D)^{1-2\alpha_S\beta_-}e^{\alpha_S v_0}
 \quad \textnormal{for all $0\leq U\leq U_0$.}
\end{align}
We set that $r(0,v)$ and $\phi(0,v)$ approach the Schwarzschild-de-Sitter values, i.e.,
\begin{equation*}
\lim_{v \to \infty} r(0,v)=r_S \quad \text{and}\quad \lim_{v \to \infty} \phi(0,v)=0.
\end{equation*}
Moreover, for $\partial_v\phi$ along $H_0$, we require it to satisfy the exponential Price's law: for $0<p\leq q<5p<\alpha_S/2$, it holds that
\begin{equation}
\label{eq:estdataphi}
D_1  e^{-qv}\leq r\partial_v\phi(0,v)\leq D_2 e^{-pv},
\end{equation}
where $D_{1}$ and $D_2$ are positive dimensionless constants.

\noindent Along the incoming cone $v=v_0$, we also set that
\begin{equation}
\label{eq:estdataphi2}
\sup_{0\leq U\leq U_0}  |rY\phi|(U,v_0)+|\partial_Ur|(0,v_0)\leq D_3\quad  \text{and} \quad rY\phi(U,v_0)>0
\end{equation}
with $D_3$ being a positive dimensionless constant.

 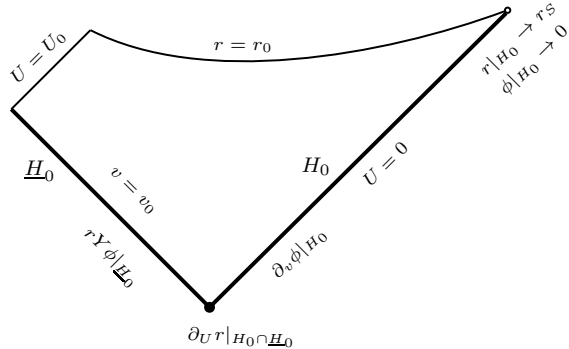
\begin{figure}[H]
	\begin{center}

\begin{tikzpicture}[x=0.75pt,y=0.75pt,yscale=-1,xscale=1]

\draw [line width=1.5]    (100,100.2) -- (200,200.2) ;
\draw [line width=1.5]    (200,200.2) -- (349.6,51.73) ;
\draw [shift={(200,200.2)}, rotate = 315.22] [color={rgb, 255:red, 0; green, 0; blue, 0 }  ][fill={rgb, 255:red, 0; green, 0; blue, 0 }  ][line width=1.5]      (0, 0) circle [x radius= 1.74, y radius= 1.74]   ;
\draw    (100,100.2) -- (139.94,60.29) ;
\draw    (139.94,60.29) .. controls (196.51,89.14) and (286.81,73.68) .. (348.86,50.74) ;
\draw [color={rgb, 255:red, 0; green, 0; blue, 0 }  ,draw opacity=1 ][fill={rgb, 255:red, 0; green, 0; blue, 0 }  ,fill opacity=1 ][line width=0.75]    (350.34,50.26) ;
\draw [shift={(350.34,50.26)}, rotate = 0] [color={rgb, 255:red, 0; green, 0; blue, 0 }  ,draw opacity=1 ][line width=0.75]      (0, 0) circle [x radius= 1.34, y radius= 1.34]   ;

\draw (200.57,64.69) node [anchor=north west][inner sep=0.75pt]  [font=\tiny]  {$r=r_{0}$};
\draw (96.65,83.14) node [anchor=north west][inner sep=0.75pt]  [font=\tiny,rotate=-315]  {$U=U_{0}$};
\draw (105.14,124.6) node [anchor=north west][inner sep=0.75pt]  [font=\scriptsize]  {$\underline{H}_{0}$};
\draw (141.29,156.02) node [anchor=north west][inner sep=0.75pt]  [font=\tiny,rotate=-45]  {$rY\phi |_{\underline{H}_{0}}$};
\draw (244.85,123.17) node [anchor=north west][inner sep=0.75pt]  [font=\scriptsize]  {$H_{0}$};
\draw (227.51,181.2) node [anchor=north west][inner sep=0.75pt]  [font=\tiny,rotate=-315]  {$\partial _{v} \phi |_{H_{0}}$};
\draw (342.36,87.96) node [anchor=north west][inner sep=0.75pt]  [font=\tiny,rotate=-315]  {$\phi |_{H_{0}}\rightarrow 0$};
\draw (332.65,78.53) node [anchor=north west][inner sep=0.75pt]  [font=\tiny,rotate=-315]  {$r|_{H_{0}}\rightarrow r_{S}$};
\draw (187.43,208.17) node [anchor=north west][inner sep=0.75pt]  [font=\tiny]  {$\partial _{U} r|_{H_{0} \cap \underline{H}_{0}}{}$};
\draw (152.29,127.02) node [anchor=north west][inner sep=0.75pt]  [font=\tiny,rotate=-45]  {$v=v_{0}$};
\draw (274.51,135.2) node [anchor=north west][inner sep=0.75pt]  [font=\tiny,rotate=-315]  {$U=0$};

\end{tikzpicture}
\end{center}
\vspace{-0.2cm}
\caption{The initial data.}
	\label{fig:contour}
\end{figure}
Subsequently, we derive the following estimates that the initial data obey along $H_0$:
\begin{lemma}
\label{lm:rdata}
There exist positive constants $c$ and $C$, such that along ${H}_0$ it hold:
\begin{align}
\label{eq:estdatadvr}
 c D_1^2 e^{-2qv}\leq \partial_vr(0,v)\leq&\: C D_2^2 e^{-2pv},\\
 \label{eq:estdatar}
 \frac{c D_1^2}{2q}e^{-2qv}\leq r_S-r(0,v)\leq &\:\frac{C D_2^2}{2p}e^{-2pv},\\
 \label{eq:estdatadUr}
\left|\frac{\partial_Ur}{\widehat{\Omega}^2}(0,v)+\frac{1}{2}\right|\leq&\: CD_2^2 e^{-2pv}+{C D_3(1+D_1^2)}e^{-\alpha_S v}.
\end{align}
Moreover, for any arbitrarily small $\epsilon>0$, there exists a $v_1$ (depending on $D_1$, $D_2$, $D_3$) greater than $v_0$ such that for all $v\geq v_1$ we have
\begin{equation}
\label{eq:idlowboundYphi}
r^2Y\phi(0,v)\geq \frac{D_1}{\alpha_S-q}(1-\epsilon)e^{-qv}.
\end{equation}
\end{lemma}
\begin{proof}[Proof of Lemma \ref{lm:rdata}]
We appeal to the constraint equations \eqref{conseq:r1} and \eqref{conseq:r2} for proving the above estimates. First, integrating \eqref{conseq:r2} along $H_0$, by \eqref{eq:estdataphi}, we get
\begin{equation}
    \big|A_S^{-1} e^{-\a_S v} \partial_v r\Big|^{v}_{v_0}\big|=\int^{v}_{v_0} r A_S^{-1} e^{-\a_S v'}(\partial_v \phi)^2 dv'\lesssim \int^{v}_{v_0}  e^{-(\a_S+2p) v'} dv' <\infty.
\end{equation}
Therefore, $A_S^{-1} e^{-\alpha_S v}\partial_vr(0,v)$ is bouned on $H_0$. Moreover, noting that $\partial_vr(0,v)$ attains a finite limit as $v\to \infty$ since $r(0,v)\to r_S$, it hence holds that $A_S^{-1} e^{-\alpha_S v}\partial_vr(0,v)$ tends to $0$ as $v\to \infty$. Furthermore, by the monotonicity in \eqref{conseq:r2}, it follows that $\partial_v r(0,v)\geq 0$ and hence $r(0,v)\leq r_S$. Thus, by integrating \eqref{conseq:r2} and using \eqref{eq:estdataphi}, we have for all $v\geq v_0$ it holds that
\begin{equation*}
\begin{split}
 c D_1^2 e^{-\a_S v} e^{-2qv}\leq e^{-\a_S v}\partial_vr(0,v)\leq &\: C D_2^2 e^{-\a_S v} e^{-2pv}.
\end{split}
\end{equation*}
This implies \eqref{eq:estdatadvr}. And \eqref{eq:estdatar} follows directly from integrating \eqref{eq:estdatadvr} in $v$.

Next, to estimate $\partial_Ur(0,v)$, we integrate \eqref{propeq:r2} along $H_0$ and obtain
\begin{equation*}
r\partial_U r(0,v)=\:r\partial_U r(0,v_0)-\int_{v_0}^{v}\frac{1}{4} A_S e^{\a_S v}{(1-\Lambda r^2)}\,dv'.
\end{equation*}
Using \eqref{eq:estdatar}, we deduce
\begin{equation*}
\begin{split}
     &|r\partial_U r(0,v)+\f{1}{4}\a_S^{-1}A_S{(1-\Lambda r_S^2)}e^{\a_S v}|\\ \leq& r\partial_U r(0,v_0)+\f{1}{4}\a_S^{-1}A_S{(1-\Lambda r_S^2)}e^{\a_S v_0}+\f{1}{12} C \Lambda D_2^2A_S \int_{v_0}^{v}e^{(\a_S-2p) v}\,dv',
\end{split}
\end{equation*}
which gives that 
\begin{equation*}
\left|r\partial_U r(0,v)+\f{1}{4}\a_S^{-1}A_S{(1-\Lambda r_S^2)}e^{\a_S v}\right|\leq {C D_3(1+D_1^2)} +CD_2^2 e^{(\alpha_S-2p)v}.
\end{equation*}
Together with \eqref{eq:estdatar}, this implies 
\begin{equation*}
\left|\frac{\partial_Ur}{\widehat{\Omega}^2}(0,v)+\f{1}{4r_S}\a_S^{-1}{(1-\Lambda r_S^2)}\right|\leq {C D_3(1+D_1^2)} e^{-\alpha_S}+CD_2^2 e^{-2pv}.
\end{equation*}
Observing that $\beta_S=\frac{3r_S}{-\Lambda P'(r_S)}=\frac{r_S}{1-\Lambda r_S^2 }$ and $\alpha_S^{-1}=2\beta_S$, we thus have $\f{1}{4r_S}\a_S^{-1}{(1-\Lambda r_S^2)}=\frac{1}{2}$ and this gives (\ref{eq:estdatadUr}).

Finally, we bound $r^2Y\phi(0,v)$. Employing \eqref{propeq:Yphi}, \eqref{eq:estdatadvr}, \eqref{eq:estdatar} and \eqref{eq:estdatadUr}, along $H_0$ we deduce that
\begin{equation}
\label{eq:YphialongH0}
|\partial_v(e^{\alpha_S v}r^2Y\phi)-e^{\alpha_S v}r\partial_v\phi|\leq C[{D_3(1+D_1^2)} e^{-\alpha_S v}+D_2^2 e^{-2pv}]|e^{\alpha_S v}r^2Y\phi|.
\end{equation}
Integrating this inequality along $H_0$ and applying \eqref{eq:estdataphi}, we get
\begin{equation*}
\begin{split}
|e^{\alpha_S v}r^2Y\phi|(0,v)\leq &\: |e^{\alpha_S v_0}r^2Y\phi|(0,v_0)+ \int_{v_0}^v D_2e^{(\alpha_S-p) v'}\,dv'\\
&+C\int_{v_0}^v[D_2^2 e^{-2pv'}+{D_3(1+D_1^2)}e^{-\alpha_S v'}]|e^{\alpha_S v'}r^2Y\phi|(0,v')\,dv'\\
\leq &\:  |e^{\alpha_S v_0}r^2Y\phi|(0,v_0)+ CD_2 e^{(\alpha_S-p) v}\\
&+C\int_{v_0}^v[D_2^2 e^{-2pv'}+{D_3(1+D_1^2)}e^{-\alpha_S v'}]|e^{\alpha_S v'}r^2Y\phi|(0,v')\,dv'
\end{split}
\end{equation*}
Via using Gr\"onwall's inequality, we hence obtain
\begin{equation}
\label{eq:upboundYphialongH0}
|e^{\alpha_S v}r^2Y\phi|(0,v) \leq C(D_1,D_2,D_3,v_0,p,q) (|e^{\alpha_S v_0}r^2Y\phi|(0,v_0)+CD_2 e^{(\alpha_S-p) v}).
\end{equation}
Combining \eqref{eq:estdataphi}, \eqref{eq:estdataphi2}, \eqref{eq:YphialongH0}, and \eqref{eq:upboundYphialongH0}, we further derive:
\begin{equation*}
\begin{split}
r^2Y\phi(0,v)\geq&\: e^{-\alpha_S(v-v_0)}r^2Y\phi(0,v_0)+ e^{-\alpha_S v}\int_{v_0}^v D_1e^{(\alpha_S-q) v'}\,dv'\\
&- C e^{-\alpha_S v}\int_{v_0}^v [D_2^2 e^{-2pv'}+D_3(1+D_1^2)e^{-\alpha_S v'}]|e^{\alpha_S v}r^2Y\phi|(0,v')\,dv' \\
\geq&\: e^{-\alpha_S(v-v_0)}r^2Y\phi(0,v_0)+ e^{-\alpha_S v}\int_{v_0}^v D_1e^{(\alpha_S-q) v'}\,dv'\\
&- C e^{-\alpha_S v}\int_{v_0}^v [D_2^2 e^{-2pv'}+D_3(1+D_1^2)e^{-\alpha_S v'}]\\
&C(D_1,D_2,D_3,v_0,p, q) (|e^{\alpha_S v_0}r^2Y\phi|(0,v_0)+CD_2 e^{(\alpha_S-p) v'})\,dv'\\
\geq &\: e^{-\alpha_S v} \frac{D_1}{\alpha_S-q}(e^{(\alpha_S-q)v}-e^{(\alpha_S-q)v_0})+ e^{-\alpha_S(v-v_0)}r^2Y\phi(0,v_0)\\
&-e^{-\alpha_S v} C(D_1,D_2,D_3,v_0,p, q)(\sum_{i=1}^4 \frac{e^{k_iv}-e^{k_iv_0}}{k_i})\\
\geq &\: e^{-\alpha_S v} \frac{D_1}{\alpha_S-q}(e^{(\alpha_S-q)v}-e^{(\alpha_S-q)v_0})\\
&-e^{-\alpha_S v} C(D_1,D_2,D_3,v_0,p, q)(\sum_{i=1}^4 \frac{e^{k_iv}-e^{k_iv_0}}{k_i}).
\end{split}
\end{equation*}
Here $k_1=-\alpha_S$, $k_2=-p$, $k_3=-2p$, $k_4=\alpha_S-3p$. Noting that $k_1,k_2,k_3\leq0$ and $k_4\geq0$, we rearrange the above inequality as
\begin{equation*}
\begin{split}
r^2Y\phi(0,v)
\geq& \frac{D_1}{\alpha_S-q}e^{-qv}\Big[1-e^{(q-\alpha_S) v} C(D_1,D_2,D_3,v_0,p,q)\\
&-e^{(q-\alpha_S )v} C(D_1,D_2,D_3,v_0,p,q) \frac{e^{k_4v}-e^{k_4v_0}}{k_4}\Big]\\
\geq &\:  \frac{D_1}{\alpha_S-q}e^{-qv}\Big[1-e^{(q-\alpha_S) v} C(D_1,D_2,D_3,v_0,p,q)\left(1+\frac{e^{k_4v}}{k_4}\right)\Big].
\end{split}
\end{equation*}
Employing the facts $q-\alpha_S<0$ and $q-\alpha_S+k_4<0$, we hence conclude the proof of \eqref{eq:idlowboundYphi}.
\end{proof}

\section{Estimates away from the singularity}\label{Section7}
In this section, we investigate the solution's dynamics in the region away from the spacelike singularity, i.e., in $\{r\geq r_0\}$ with $r_0$ a positive small parameter. To carry out the estimates, we will further divide this region into two parts: the red-shift region and the no-shift region.

\subsection{Basic setup}
To begin with, we decompose the domains and list the bootstrap assumptions. Let $r_0$ be a sufficiently small positive constant satisfying $r_0< \inf_{0\leq U\leq U_0}r(U,v_0)$. With initial data prescribed in Section \ref{sec:initialdata}, in the region
\begin{align*}
D_{v_{\infty},r_0}&=[0,U_0)\times [v_0,v_{\infty})\cap\{r\geq r_0\}\\
&=[0,U_0)\times [v_0,v_{\infty})\cap J^-(\gamma_{r_0})\quad \text{for some}\ v_{\infty}>v_0,
\end{align*}  
a smooth spherically symmetric solution $(r,\hat{\Omega},\phi)$ to the Einstein-scalar field system \eqref{ESlam} is guaranteed to exist by a standard local existence argument. Then, applying the same argument as in Lemma 7.1 of \cite{AG}, we have that, for $U_0$ suitably small, the level set
\begin{equation*}
\gamma_{r_0}=\{r=r_0\}\cap D_{v_{\infty},r_0}
\end{equation*}
is either empty or a spacelike curve intersecting $\{U=U_0\}$ at $v>v_0$.

In the following, in the region $D_{v_{\infty},r_0}\cap \{U>0\}$, instead of $U$, we will use the Eddington--Finkelstein-type coordinate $u$, which is defined via
\begin{equation}
\label{eq:reluU}
u:=-{(\alpha_S)^{-1}} \log \left(\frac{{(\alpha_S)}^{-1}}{U}\right)
\end{equation}
and we denote $u_0:=u(U_0)$. Accommodating the $(u,v)$ coordinates, the metric component $\O^2$ respects the following transformation
\begin{equation}
\label{eq:relhatomega}
\Omega^2(u(U),v):=\widehat{\Omega}^2(U,v)\frac{dU}{du}=\widehat{\Omega}^2(U(u),v)e^{-\alpha_S |u|(U)}.
\end{equation}
In the rest of this section, we will use the $(u,v)$ coordinates. 

Now we divide the region $D_{v_{\infty},r_0}$ into two parts. For $\delta>0$, we define the red-shift region
\begin{equation*}
\mathcal{R}_{\delta}:=\:D_{v_{\infty},r_0}\cap \{v-|u|\leq -\delta^{-1}\},
\end{equation*}
which is close to the event horizon, and we set the no-shift region to be
\begin{equation*}
\mathcal{N}_{\delta,r_0}:=\: D_{v_{\infty},r_0}\setminus \mathcal{R}_{\delta}.
\end{equation*} 
We also denote the hypersurface of intersection as
\begin{equation*}
\gamma_{\delta}= D_{v_{\infty},r_0}\cap\{v-|u|= -\delta^{-1}\}.
\end{equation*}
 \begin{figure}[H]
	\begin{center}
\begin{tikzpicture}[x=0.75pt,y=0.75pt,yscale=-1,xscale=1]

\draw [line width=0.75]    (100,100.2) -- (200,200.2) ;
\draw [line width=0.75]    (200,200.2) -- (320.8,80.8) ;
\draw    (100,100.2) -- (120.14,80.08) -- (139.94,60.29) ;
\draw [color={rgb, 255:red, 0; green, 0; blue, 0 }  ,draw opacity=1 ][fill={rgb, 255:red, 0; green, 0; blue, 0 }  ,fill opacity=1 ][line width=0.75]    (350.34,50.26) ;
\draw [shift={(350.34,50.26)}, rotate = 0] [color={rgb, 255:red, 0; green, 0; blue, 0 }  ,draw opacity=1 ][line width=0.75]      (0, 0) circle [x radius= 1.34, y radius= 1.34]   ;
\draw    (304.4,64.72) -- (320.8,80.8) ;
\draw    (139.94,60.29) .. controls (180.06,80) and (236.63,81.43) .. (304.4,64.72) ;
\draw    (119.97,80.24) .. controls (166.06,110.57) and (264.34,101.43) .. (313.41,73.84) ;
\draw    (304.4,64.72) -- (313.41,73.84) ;
\draw    (139.94,60.29) -- (119.97,80.24) ;

\draw  [dash pattern={on 0.84pt off 2.51pt}]  (304.4,64.72) .. controls (330.34,57.71) and (326.91,58.29) .. (350.34,50.26) ;
\draw  [dash pattern={on 0.84pt off 2.51pt}]  (320.8,80.8) -- (349.6,51.73) ;
\draw [fill={rgb, 255:red, 155; green, 155; blue, 155 }  ,fill opacity=1 ] [dash pattern={on 0.84pt off 2.51pt}]  (313.41,73.84) -- (349.6,51.73) ;

\draw [fill={rgb, 255:red, 155; green, 155; blue, 155 }  ,fill opacity=1 ]   
 (139.94,60.29) .. controls (180.06,80) and (236.63,81.43) .. (304.4,64.72) --  (313.41,73.84) .. controls (264.34,101.43) and (166.06,110.57) .. (119.97,80.24) --  (139.94,60.29) ;

\draw (200.57,64.69) node [anchor=north west][inner sep=0.75pt]  [font=\tiny]  {$r=r_{0}$};
\draw (96.65,83.14) node [anchor=north west][inner sep=0.75pt]  [font=\tiny,rotate=-315]  {$u=u_{0}$};
\draw (105.14,124.6) node [anchor=north west][inner sep=0.75pt]  [font=\scriptsize]  {$\underline{H}_{0}$};
\draw (141.29,156.02) node [anchor=north west][inner sep=0.75pt]  [font=\tiny,rotate=-45]  {$v=v_{0}$};
\draw (272.05,137.57) node [anchor=north west][inner sep=0.75pt]  [font=\scriptsize]  {$H_{0}$};
\draw (318.97,84.83) node [anchor=north west][inner sep=0.75pt]  [font=\tiny]  {$v=v_{\infty }$};
\draw (200.86,102.69) node [anchor=north west][inner sep=0.75pt]  [font=\tiny]  {$\gamma _{\delta }$};

\end{tikzpicture}
\end{center}
\vspace{-0.2cm}
\caption{The red-shift region and the no-shift region.}
	\label{fig:contour1}
\end{figure}
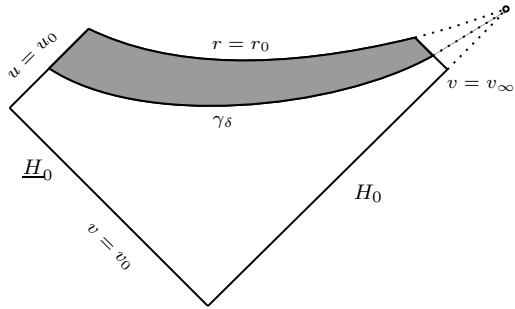
The upper bound estimates in this section are acquired via a bootstrap argument. We first list the following \textbf{bootstrap assumptions} for $(r,\Omega,\phi)$:
\begin{align}
\label{ba:dvphi}
r| \partial_v\phi|(u,v)\leq &\: {\Delta_1} e^{-pv}\quad \textnormal{for all $(u,v)\in \mathcal{R}_{\delta}$},\\
\label{ba:Yphi}
|Y\phi|(u,v)\leq &\: {\Delta_1} e^{-pv} \quad \textnormal{for all $(u,v)\in \mathcal{R}_{\delta}$},\\
\label{ba:Omega}
\left| \partial_u \log \left(\frac{\Omega^2}{\Omega^2_{SdS}}\right)\right|\leq&\: \Delta_2 e^{\alpha_S(v-|u|)}e^{-2pv}\quad \textnormal{for all $(u,v)\in \mathcal{R}_{\delta}$},\\
\label{ba:dvphi2}
r| \partial_v\phi|(u,v)\leq &\: {\Delta_1'} e^{-pv}\quad \textnormal{for all $(u,v)\in \mathcal{N}_{\delta,r_0}$},\\
\label{ba:Yphi2}
|Y\phi|(u,v)\leq &\: {\Delta_1'} e^{-pv} \quad \textnormal{for all $(u,v)\in \mathcal{N}_{\delta,r_0}$}.
\end{align}
with $\Delta_1,\Delta_2,\Delta_1'$ being sufficiently large constants to be determined later.

\subsection{Estimates in $\mathcal{R}_{\delta}$}
\label{estimate in R delta}
We first work in the red-shift region $\mathcal{R}_\delta$. We start with deriving the a priori estimates for $r$ and $\Omega^2$.
\begin{proposition}
\label{prop:metricest}
Under the bootstrap assumptions \eqref{ba:dvphi} and \eqref{ba:Yphi}, for all $(u,v)\in \mathcal{R}_\delta$, we have that
\begin{align}
\label{eq:durest1prop}
\left|- \Omega^{-2}\partial_u r-\frac{1}{2}\right|(u,v)\leq&\:C(\Delta_1,D_1,D_2,D_3,v_0,p) e^{-2pv},\\
\label{eq:dvrestprop}
\left|r \partial_vr +\frac{\Lambda}{6}P(r)\right|(u,v)\leq&\: C(\Delta_1,D_1,D_2,D_3,v_0,p) e^{-2pv},
\end{align}
where $C(\Delta_1,D_1,D_2,D_3,v_0,p)>0$ is a constant depending on $\Delta_1,D_1,D_2,D_3,v_0$ and $p$.
\end{proposition}
\begin{proof}
First, we rewrite \eqref{conseq:r1} as below
\begin{equation*}
\partial_u\log(- \Omega^{-2}\partial_u r)=\frac{r}{-\partial_ur}(\partial_u\phi)^2=r(-\partial_ur)(Y\phi)^2.
\end{equation*}
Note that from the Raychaudhuri equation \eqref{conseq:r1} it follows that $\partial_u r<0$. Hence, together with \eqref{eq:estdatar}, one can bound $r$ by a positive number from above. Integrating the above equation, by \eqref{ba:Yphi}, we get
\begin{equation*}
\log(- \Omega^{-2}\partial_u r)(u,v)-\log(- \Omega^{-2}\partial_u r)(-\infty,v)=\int_{r(u,v)}^{r_{\mathcal{H}^+}(v)} r(Y\phi)^2\,dr'\leq C {\Delta_1}^2 e^{-2pv}.
\end{equation*}
Noticing that $\Omega^{-2}\partial_ur =\widehat{\Omega}^{-2} \partial_Ur$ and using \eqref{eq:estdatadUr}, we thus obtain
\begin{equation*}
\begin{split}
\left|- \Omega^{-2}\partial_u r(u,v)-\frac{1}{2}\right|\leq &\:C(D_1^2+D_2^2) e^{-2pv}+C(D_1^2+D_3) e^{-\alpha_S v}+(e^{C {\Delta_1}^2 e^{-2pv}}-1).
\end{split}
\end{equation*}
Choosing $C=C(\Delta_1,D_1,D_2,D_3,v_0,p)>0$ to be sufficiently large, this estimate implies \eqref{eq:durest1prop}.

\noindent To show \eqref{eq:dvrestprop}, we integrate \eqref{propeq:r1} and obtain
\begin{equation*}
\begin{split}
r\partial_vr(u,v)=&\:r\partial_vr(-\infty,v)+\int_{-\infty}^u\frac{-\frac{1}{4}\Omega^2}{\partial_u r} \partial_u r (1-\Lambda r'^2) du'\\
=&\:r\partial_vr(-\infty,v)-\frac{1}{2}(r_{\mathcal{H}^+}(v)-r(u,v))+\frac{\Lambda}{6}(r_{\mathcal{H}^+}^3(v)-r^3(u,v))\\
&-\int^{r_{\mathcal{H}^+}(v)}_r \left[-\frac{1}{4}{\Omega^2}({\partial_u r})^{-1}-\frac{1}{2}\right]\cdot (1-\Lambda {r'}^2)\,dr'.
\end{split}
\end{equation*}
Thus, it holds that
\begin{equation*}
\begin{split}
\left|r \partial_vr+\frac{\Lambda}{6}P(r)\right|\leq & r\partial_vr(-\infty,v)+ \frac{1}{2}(r_S-r_{\mathcal{H}^+}(v))+\frac{\Lambda}{6}(r_{\mathcal{H}^+}^3(v)-r_S^3)\\
&+\int^{r_{\mathcal{H}^+}(v)}_r \left|-\frac{1}{4}{\Omega^2}({\partial_u r})^{-1}-\frac{1}{2}\right|\cdot (1-\Lambda {r'}^2)\,dr'.
\end{split}
\end{equation*}
Applying Lemma \ref{lm:rdata} and \eqref{eq:durest1prop}, the estimate \eqref{eq:dvrestprop} then follows.
\end{proof}

Next, we estimate the differences between our solution's value and the corresponding Schwarzschild-de-Sitter value. We study $r-r_{SdS}$ and $\frac{\Omega^2}{\Omega^2_{SdS}}$.
\begin{proposition}
	\label{lm:metricredshift}
	There exist a suitably large $C(\Delta_1,D_1,D_2,D_3,v_0,p,q)>0$, a sufficiently small $\epsilon>0$ and $\delta=\delta(\epsilon)>0$ such that the following estimates hold for all $(u,v)\in \mathcal{R}_{\delta}$:
	\begin{align}
	\label{eq:redshiftOmega}
	\left|\frac{\Omega^2}{\Omega^2_{SdS}}-1\right|\leq&\: C\epsilon  \Delta_2 e^{-2pv},\\
	\label{eq:redshiftdur}
	|\partial_u(r-r_{SdS})| \leq&\: C\epsilon \Delta_2 e^{\alpha_S(v-|u|)}  e^{-2pv}+C(\Delta_1,D_1,D_2,D_3,v_0)e^{\alpha_S(v-|u|)}e^{-2pv},\\
	\label{eq:redshiftr}
	|r-r_{SdS}| \leq&\:  C\epsilon \Delta_2 e^{\alpha_S(v-|u|)} e^{-2pv}+C(\Delta_1,D_1,D_2,D_3,v_0,p,q) e^{-2pv},\\
	\label{eq:redshiftdvr}
	|\partial_v(r-r_{SdS})| \leq&\:  C\epsilon \Delta_2 e^{\alpha_S(v-|u|)} e^{-2pv}+C(\Delta_1,D_1,D_2,D_3,v_0,p)e^{-2pv}.
	\end{align}
Furthermore, we also have
	\begin{equation}
	\label{eq:redshiftOmega2}
	\left|\log \left(\frac{3r^2\Omega^2}{\Lambda rP(r)}\right)\right|\leq C\epsilon  \Delta_2 e^{\alpha_S(v-|u|)} e^{-2pv}+C(\Delta_1,D_1,D_2,D_3,v_0,p)e^{-2pv}\quad \textnormal{on $\gamma_{\delta}$}.
	\end{equation}
\end{proposition}
\begin{proof}
Note that in $\mathcal{R}_{\delta}$, there hold
	\begin{equation*}
 \frac{\Omega^2}{\Omega^{2}_{SdS}}(-\infty,v)=1\quad \text{and}\quad	e^{\alpha_S(v-|u|)}\le e^{-\frac{\alpha_S}{\delta}}.
	\end{equation*}
Integrating \eqref{ba:Omega} and using the above two facts, we obtain estimate \eqref{eq:redshiftOmega}. The estimate \eqref{eq:redshiftdur} follows from \eqref{eq:durest1prop} and \eqref{eq:redshiftOmega}. Employing \eqref{eq:estdatar}, one can obtain \eqref{eq:redshiftr} via integration of \eqref{eq:redshiftdur}. Finally, a combination of \eqref{eq:dvrestprop} and \eqref{eq:redshiftr} gives the estimate \eqref{eq:redshiftdvr}. Now denote the constant $r_{\delta}:=r_{SdS}(u,v_{\gamma_{\delta} (u)})$. As a corollary, together with \eqref{eq:redshiftOmega} and \eqref{eq:redshiftr}, for $r_\delta<r_S$, we have \eqref{eq:redshiftOmega2}.
\end{proof}
As defined in Section \ref{pre equations}, the apparent horizon $\mathcal{A}$ (within $D_{v_{\infty},r_0}$) is corresponding to
\begin{equation*}
\mathcal{A}=\{\partial_vr=0\}\cap D_{v_{\infty},r_0}.
\end{equation*}
It follows immediately from \eqref{propeq:r1} that $\mathcal{A}$ contains only (possible empty) spacelike or outgoing null segments. Moreover, we can show that the apparent horizon is close to the event horizon:
\begin{proposition}
	\label{prop:apparenthordecay}
	For sufficiently large $|u_0|$, we have $\mathcal{A}\subset \mathcal{R}_{\delta}$ and it holds
	\begin{align}
	\label{eq:rA1}
	|r_{\mathcal{A}}(v)-r_S|\leq &\:C(\Delta_1,D_1,D_2,D_3,v_0,p) e^{-2pv}.
	\end{align}
\end{proposition}
\begin{proof}
By definition $\partial_vr=0$ along $\mathcal{A}$. Invoking this in \eqref{eq:dvrestprop}, we hence derive \eqref{eq:rA1}. Note that $r_\delta<r_S$. Combining \eqref{eq:rA1} and \eqref{eq:redshiftr}, we deduce that $\mathcal{A}\subset \mathcal{R}_{\delta}$ with $\delta>0$ being suitably small.
\end{proof}

Now we are ready to derive upper bound estimates for $\partial_v \phi$ and $Y\phi$, and we will improve the bootstrap assumptions \eqref{ba:dvphi} and \eqref{ba:Yphi}.
\begin{proposition}
	\label{prop:baimpphired}
There exist a sufficiently large $\Delta_1>0$ and a sufficiently small $\delta>0$, such that for all $(u,v)\in\mathcal{R}_{\delta}$ , there hold
	\begin{align}
	\label{eq:dvphiredshift0}
	r|\partial_v\phi|(u,v)\leq &\: \frac{1}{4}{\Delta_1}e^{-pv},\\
	\label{eq:Yphiredshift0}
	|Y\phi|(u,v)\leq &\: \frac{1}{4}{\Delta_1}  e^{-pv}.
	\end{align}
\end{proposition}
\begin{proof}
Integrating \eqref{propeq:phi1}, by \eqref{eq:durschwest}, \eqref{lm:metricredshift}, \eqref{ba:Yphi} and  \eqref{eq:dvrestprop}, we obtain that
	\begin{equation}
	\label{eq:dvphiredshiftest}
	\begin{split}
	|r\partial_v\phi|(u,v)\leq&\: |r\partial_v\phi|(-\infty,v)+\int_{-\infty}^u |\partial_vr||\partial_ur||Y\phi|(u',v)\,du'\\
	\leq &\: C D_2e^{-pv}+C(\Delta_1,D_2,D_3,v_0)e^{-pv}e^{\alpha_S(v-|u|)}\\
	\leq &\: C D_2e^{-pv}+\epsilon C(\Delta_1,D_2,D_3,v_0) e^{-pv}.
	\end{split}
	\end{equation}
By taking $\Delta_1$ suitably large compared to $D_2$ and $\epsilon>0$ suitably small compared to $C(\Delta_1,D_2,D_3,v_0)$, the estimate \eqref{eq:dvphiredshift0} follows.

Next, for sufficiently small $\epsilon>0$ and $\delta>0$, we apply the estimates in Proposition \ref{lm:metricredshift} to equation \eqref{propeq:Yphi} and derive
	\begin{equation*}
	\partial_v(|r^2Y\phi|)\leq -\alpha_S (1-\epsilon) |r^2Y\phi|+r|\partial_v\phi|.
	\end{equation*}
This implies
	\begin{equation*}
	\partial_v(e^{\alpha_S(1-\epsilon)v}|r^2Y\phi|)\leq e^{\alpha_S (1-\epsilon)v}r|\partial_v\phi|.
	\end{equation*}
Integrating the above inequality and employing \eqref{eq:dvphiredshiftest}, we obtain
	\begin{equation}
	\label{eq:Yphiredshiftest}
	\begin{split}
    |r^2Y\phi|(u,v)\leq&\: e^{-\alpha_S(1-\epsilon)(v-v_0)} |r^2Y\phi|(u,v_0)+ e^{-\alpha_S(1-\epsilon)v} \int_{v_0}^v e^{\alpha_S(1-\epsilon)v'} r|\partial_v\phi|(u,v')\,dv'\\
	\leq &\: Ce^{-\alpha_S(1-\epsilon)v} D_{3}+e^{-\alpha_S(1-\epsilon)v} \int_{v_0}^v e^{\alpha_S(1-\epsilon)v'} r|\partial_v\phi|(u,v')\,dv\\
	\leq &\: Ce^{-\alpha_S(1-\epsilon)v} D_3+C D_2e^{-pv}+\epsilon C(\Delta_1,D_2,D_3,v_0)e^{-pv}.
	\end{split}
	\end{equation}
	In the above estimate, choosing $\epsilon>0$ to be suitably small (depending on $r_0$ and $C(\Delta_1,D_2,D_3,v_0)$) and $\Delta_1$ to be suitably large (depending on $D_2$ and $D_3$), we finish proving \eqref{eq:Yphiredshift0}.
\end{proof}
We wrap up this subsection with the following proposition, which improves the bootstrap assumption \eqref{ba:Omega}.
\begin{proposition}
	\label{prop:closingbaOmega}
	Let $\delta>0$ be suitably small. There exists a suitably large $\Delta_2>0$  such that for all $(u,v)\in \mathcal{R}_{\delta}$, we have
	\begin{equation*}
	\left| \partial_u \log \left(\frac{r^2\Omega^2}{r_{SdS}^2\Omega^2_{SdS}}\right)\right|\leq \frac{1}{2}\Delta_2 e^{\alpha_S(v-|u|)} e^{-2pv}.
	\end{equation*}
\end{proposition}
\begin{proof}
We first rearrange the terms on the RHS of \eqref{propeq:Omegarescale2} as follows:
	\begin{align*}
	-2r^{-2}\partial_ur\partial_vr=&-2r^{-2}[\partial_ur_{SdS}\partial_vr_{SdS}+\partial_u(r-r_{SdS})\partial_vr_{SdS}\\
	&+\partial_ur_{SdS}\partial_v(r-r_{SdS})+\partial_u(r-r_{SdS})\partial_v(r-r_{SdS})],\\
	\partial_u\phi\partial_v\phi=&-\partial_ur Y\phi\partial_v\phi=-\partial_u r_{SdS} Y\phi\partial_v\phi- \partial_u(r-r_{SdS}) Y\phi\partial_v\phi,
	\end{align*}
and recall the fact 
\begin{equation*}
    r^{-1}=r_{SdS}^{-1}\left(1-\frac{r_{SdS}-r}{r_{SdS}}\right)^{-1}.
\end{equation*}
Note that when $\delta>0$ being sufficiently small, it holds $r_{SdS}>\frac{r_S}{2}$ in $\mathcal{R}_\delta$. Hence, using the above expression for $r^{-1}$ and \eqref{eq:redshiftr}, for $v\geq v_1$, with $v_1$ suitably large (depending on $D_1$, $D_2$, $D_3$, $\Delta_1$), we obtain that $r>\frac{r_S}{4}$ in $\mathcal{R}_{\delta}\cap \{v\geq v_1\}$. Moreover, by taking $|u_0|$ to be sufficiently large (depending on $v_1$), in $\mathcal{R}_{\delta}\cap \{v\leq v_1\}$ we prove that
	\begin{equation*}
	r\geq \frac{1}{2} \inf_{\underline{H}_0} r|_{\underline{H}_0}.
	\end{equation*}
To prove the desired estimate, we integrate the equation \eqref{propeq:Omegarescale2}. Together with the above observations on lower bounds of $r$, and noting that
\begin{equation*}
    \partial_u\partial_v\log (r_{SdS}^2\Omega_{SdS}^2)=-2r_{SdS}^{-2}\partial_ur_{SdS} \partial_v r_{SdS}+\frac{\Lambda}{2}\Omega_{SdS}^2,
\end{equation*} 
it hence holds
	\begin{equation*}
	\begin{split}
	\left|\partial_v\partial_u\log\left(\frac{r^2\Omega^2}{r_{SdS}^2\Omega^2_{SdS}}\right)\right| \leq&\: C\Big[ |\partial_vr_{SdS}|( |\partial_u(r-r_{SdS})|+|\partial_v(r-r_{SdS})|+|Y\phi||\partial_v\phi|)\\
	&+|\partial_u(r-r_{SdS})||\partial_v(r-r_{SdS})|+|\partial_u(r-r_{SdS})||Y\phi||\partial_v\phi|\\
    &+ |\partial_u r_{SdS}| |\partial_v r_{SdS}| |r-r_{SdS}|+\Omega^2_{SdS}|\f{\Omega^2}{\Omega^2_{SdS}}-1|	\Big].
	\end{split}
	\end{equation*}
Then, applying estimate \eqref{eq:durschwest}, Proposition \ref{lm:metricredshift}, the bootstrap assumptions \eqref{ba:dvphi} and \eqref{ba:Yphi}, when $\epsilon\Delta_2\ll1$, we deduce that
	\begin{equation*}
	\left|\partial_v\partial_u\log\left(\frac{r^2\Omega^2}{r_{SdS}^2\Omega^2_{SdS}}\right)\right|\leq  C  e^{\alpha_S(v-|u|)} [\epsilon^2 \Delta_2 e^{-2pv}+C(\Delta_1,D_1,D_2,D_3,v_0,p) e^{-2pv}].
	\end{equation*}
Integrating the above inequality and choosing $\Delta_2$ suitably large compared to $C(\Delta_1,D_1,D_2,D_3,v_0,p)$, we arrive at
	\begin{equation*}
	\left| \partial_u \log \left(\frac{r^2\Omega^2}{r_{SdS}^2\Omega^2_{SdS}}\right)\right|\leq \frac{1}{2}\Delta_2 e^{\alpha_S(v-|u|)} e^{-2pv}\quad \textnormal{}.
	\end{equation*}
This concludes the proof of this proposition.
\end{proof}

\subsection{Estimates in $\mathcal{N}_{\delta,r_0}$}
\label{estimate in N delta r0}
Based on the obtained estimates in the red-shift region, we move to derive estimates in the no-shift region. Employing the bootstrap assumptions \eqref{ba:dvphi2} and \eqref{ba:Yphi2}, we first derive the following estimates for $r$ and $\Omega^2$ in $\mathcal{N}_{\delta,r_0}$ in the same manner as in Proposition \ref{prop:metricest}.
\begin{proposition}
\label{prop:metricest2}
Supposing that \eqref{ba:dvphi2} and \eqref{ba:Yphi2} are verified, then for all $(u,v)\in \mathcal{N}_{\delta,r_0}$, there hold 
\begin{align}
\label{eq:durest1prop2}
\left|- \Omega^{-2}\partial_u r-\frac{1}{2}\right|(u,v)\leq&\:C(\Delta_1',D_1,D_2,D_3,v_0,p) e^{-2pv},\\
\label{eq:dvrestprop2}
\left|r \partial_vr +\frac{\Lambda}{6}P(r)\right|(u,v)\leq&\: C(\Delta_1',D_1,D_2,D_3,v_0,p) e^{-2pv},
\end{align}
where $C(\Delta_1',D_1,D_2,D_3,v_0,p)>0$ is a constant depending on $\Delta_1',D_1,D_2,D_3,v_0$ and $p$.
\end{proposition}

In the following, we derive a useful lemma on estimating the double-null coordinates.
\begin{lemma}
	\label{lm:relatuvconstr}
	For all $(u,v)\in \mathcal{N}_{\delta,r_0}$, there holds
	\begin{equation}
	\label{eq:reluvN}
	\left|v+u-2r^\ast(r)\right|\leq  C_{\delta,r_0}C(\Delta_1', \Delta_1 ,D_1,D_2,D_3,v_0,p)(1+\epsilon^2 \Delta_2)e^{-2pv}.
	\end{equation}
\end{lemma}
\begin{proof}
First recall that
\begin{equation*}
    r^\ast(r):=\int_0^{r} -\frac{1}{\Omega^2_{SdS}(r')} dr'=\int_0^{r} -\frac{3r'}{\Lambda P(r')} dr'.
\end{equation*}
Then it follows
\begin{equation*}
    \f{dr^*}{dr}=-\f{3r}{\Lambda P(r)}.
\end{equation*}
Denoting $u_\delta (v)=-\delta^{-1}-v$, then we have that $(u_\delta (v), v)$ lies on $\gamma_\delta$. For any $(u, v)\in \mathcal{N}_{\delta,r_0}$, we observe that the following equality holds
\begin{equation*}
    \begin{split}
        \l v+u-2r^*(r)\r\Big|_{u_\delta (v)}^{u}=&\int_{u_\delta (v)}^{u} \partial_u \l v+u'-2r^*(r)\r du' \\
        =&\int_{r(u_\delta (v), v)}^{r(u ,v)} (1-2\f{dr^*}{dr} \partial_u r) (\partial_u r)^{-1} dr' \\
        =& \int_{r(u_\delta (v), v)}^{r(u ,v)} 2 (-\partial_u r)^{-1} \f{3r}{\Lambda P(r)} (-\partial_u r-\f{\Lambda P(r)}{6r})  dr'.
    \end{split}
\end{equation*}
 Applying \eqref{eq:durest2prop}, we then obtain 
 \begin{equation}\label{v+u dif}
     \Big| \big(v+u-2r^*(r)\big)\big| _{u_\delta (v)}^{u}\Big|\leq C_{\delta,r_0}C(\Delta_1', D_1,D_2,D_3,v_0,p)(1+\epsilon^2 \Delta_2)e^{-2pv}.
 \end{equation}
To evaluate $v+u-2^*(r)$ on $\gamma_\delta$, we appeal to the following fact. Note that via using \eqref{eq:redshiftr}, in $\mathcal{R}_\delta$ it holds that
\begin{equation*}
    |v+u-2r^*(r)|=2|r^*(r_{SdS})-r^*(r)|\leq  C\epsilon \Delta_2 e^{\alpha_S(v-|u|)} e^{-2pv}+C(\Delta_1,D_1,D_2,D_3,v_0,p,q) e^{-2pv}.
\end{equation*}
In particular, restricting on $\gamma_\delta$, we derive
\begin{equation}\label{v+u 2}
    |v+u_\delta (v)-2r^*(r(u_\delta (v), v))| \leq C(\Delta_1,D_1,D_2,D_3,v_0,p,q) e^{-2pv}.
\end{equation}
Combining \eqref{v+u dif} and \eqref{v+u 2}, we conclude \eqref{eq:reluvN}.
\end{proof}

In the no-shift region, the geometric quantities are not close to the corresponding value in the explicit Schwarzchild-de-Sitter solution since the decaying effect around the event horizon gets weaker here. However, the convergence of the quantities to their Schwarzchild-de-Sitter values still holds, but in a weaker way. In particular, we derive the following weak version estimates for the solution:
\begin{proposition}
\label{prop:durdvrr0}
In $\mathcal{N}_{\delta, r_0}$, under bootstrap assumptions \eqref{ba:dvphi2} and \eqref{ba:Yphi2}, the following estimates hold:
\begin{align}
\label{eq:rescaledOmegaestprop}
&\left| \frac{3r^2\Omega^2}{\Lambda rP(r)}-1\right|\leq\: C_{\delta,r_0}C(\Delta_1',D_1,D_2,D_3,v_0,p)(1+\epsilon^2 \Delta_2) e^{-2pv},\\
\label{eq:prelimestOmega}
&|\partial_v\log (r^2\Omega^2)(u,v)+\frac{2\Lambda}{3} r-r^{-1}+Mr^{-2}|(u,v)\leq\: C_{r_0}\cdot C(\Delta_1',D_1,D_2,D_3,v_0,p,q) e^{-2pv},\\
\label{eq:durest2prop}
&\left|- \partial_u r- \frac{\Lambda P(r)}{6r}\right|(u,v)\leq\: C_{\delta,r_0}C(\Delta_1',D_1,D_2,D_3,v_0,p)(1+\epsilon^2 \Delta_2) e^{-2pv},\\
\label{eq:ratiodurdvr}
&\left|\frac{\partial_u r}{\partial_vr}-1\right|(u,v)\leq\: C_{\delta,r_0}C(\Delta_1',D_1,D_2,D_3,v_0,p)(1+\epsilon^2 \Delta_2) e^{-2pv}.
\end{align}
\end{proposition}
\begin{proof}
Integrating \eqref{propeq:Omegarescale2}, we have
\begin{equation*}
\begin{split}
\partial_v\log (r^2\Omega^2)(u,v)=&\:\partial_v\log (r^2\Omega^2)(-\infty,v)+\int_{-\infty}^u[-2r^{-2}\partial_ur \partial_v r-2\partial_u\phi\partial_v\phi+\frac{\Lambda}{2}\Omega^2] (u',v)\,du'\\
=&\:\partial_v\log (r^2\Omega^2)(-\infty,v)+\int_{r_{\mathcal{H}^+}(v)}^{r(u,v)} [-2r^{-2} \partial_v r+2 Y \phi\partial_v\phi+\frac{\Lambda}{2}\Omega^2\cdot{(\partial_u r)}^{-1}]|_{v'=v}(r')\,dr'\\
=&\: \partial_v\log (r^2)(-\infty,v)+ \partial_v\log(\Omega^2)(-\infty,v)-\int_{r_{\mathcal{H}^+}(v)}^{r(u,v)}[r'^{-2}-2Mr'^{-3} ]\,dr'+\frac{2\Lambda}{3} (r_{\mathcal{H}^+}(v)-r)\\
&\:+\int_{r_{\mathcal{H}^+}(v)}^{r(u,v)} [-2r^{-2} (\partial_v r+\frac{\Lambda P(r')}{6r'})+2 Y \phi\partial_v\phi+\frac{\Lambda}{2}(\Omega^2\cdot{(\partial_u r')}^{-1}+2)]|_{v'=v}(r')\,dr'.
\end{split}
\end{equation*}
\noindent On the RHS of the above equality, the first two terms regarding the initial data obeys
\begin{equation*}
\left|\partial_v\log (r^2)(-\infty,v)+ \partial_v\log(\Omega^2)(-\infty,v)+\frac{2\Lambda}{3}r_S-r_S^{-1}+Mr_S^{-2}\right| \leq C(D_1,D_2,D_3) e^{-2pv}.
\end{equation*}
The last three terms can be bounded via using \eqref{ba:dvphi2}, \eqref{ba:Yphi2}, \eqref{eq:durest1prop2} and \eqref{eq:dvrestprop2} and they satisfy
\begin{equation*}
\left|\int_{r_{\mathcal{H}^+}(v)}^{r(u,v)} [2r^{-2} (\partial_v r+\frac{\Lambda P(r)}{6r})+2 Y \phi\partial_v\phi+\frac{\Lambda}{2}(\Omega^2\cdot{(\partial_u r')}^{-1}+2)]|_{v'=v}(r')\,dr'\right|\leq C_{r_0}\cdot C(\Delta_1',D_1,D_2,D_3,v_0,p)e^{-2pv}.
\end{equation*}
Combining the above estimates, we hence obtain \eqref{eq:prelimestOmega}:
\begin{equation*}
|\partial_v\log (r^2\Omega^2)(u,v)+\frac{2\Lambda}{3} r-r^{-1}+Mr^{-2}|(u,v)\leq C_{r_0}\cdot C(\Delta_1',D_1,D_2,D_3,v_0,p) e^{-2pv}.
\end{equation*}

To deduce \eqref{eq:rescaledOmegaestprop}, we write
\begin{equation} \label{d r2 O2}
\begin{split}
\partial_v\log\left(\frac{3r^2\Omega^2}{\Lambda rP(r)}\right)=&\: \partial_v\log (r^2\Omega^2)+\frac{2\Lambda}{3} r-r^{-1}+Mr^{-2}- \frac{2\Lambda r-3r^{-1}+3Mr^{-2}}{3P(r)}(\frac{6}{\Lambda} r\partial_vr+P(r)).
\end{split}
\end{equation}
It remains to control the last term on the RHS. Notice that $P(r)^{-1}\lesssim(r_S-r)^{-1}$ and
\begin{equation*}
\begin{split}
 (r_S-r_{SdS})^{-1}= (r_S-r)^{-1}\frac{r_S-r}{r_S-r_{SdS}}= (r_S-r)^{-1} (1+\frac{r_{SdS}-r}{r_S-r_{SdS}}).
\end{split}
\end{equation*}
To bound $(r_S-r)^{-1}$, based on the above observations, on $\gamma_{\delta}$ we first derive that
\begin{equation*}
(r_S-r)^{-1}(u,v)\leq \frac{1}{r_S-r_{\delta}} \frac{1}{1+\frac{(r_{SdS}-r)(u,v)}{r_S-r_\delta}}.
\end{equation*}
Applying \eqref{eq:redshiftr} and using the fact that $v-|u|= -\frac{1}{\delta}$ along $\gamma_\delta$, we deduce
\begin{equation} \label{rs-r inv}
(r_S-r)^{-1}(u_{\delta}(v),v)\leq C_{\delta} [1+\epsilon^2 \Delta_2 e^{-2pv}+ C(\Delta_1,D_1,D_2,D_3,v_0,p) e^{-2pv}].
\end{equation}
Thanks to the monotonic property $\partial_u r<0$, the estimate \eqref{rs-r inv} holds for all $(u,v)$ in $\mathcal{N}_{\delta,r_0}$.
Back to \eqref{d r2 O2}, together with \eqref{eq:dvrestprop2} and \eqref{eq:prelimestOmega}, this implies for all $(u,v)$ in $\mathcal{N}_{\delta,r_0}$,
\begin{equation}
\label{eq:dvrescaledOmegaest}
\begin{split}
    \left|\partial_v\log\left(\frac{3r^2\Omega^2}{\Lambda rP(r)}\right)\right|\leq&  C_{\delta,r_0}C(\Delta_1',D_1,D_2,D_3,v_0,p)(\epsilon^2\Delta_2+1) e^{-2pv}\\
    \leq & C_{\delta,r_0}C(\Delta_1',D_1,D_2,D_3,v_0,p)(\epsilon^2\Delta_2+1) e^{-2p|u|}.
\end{split}
\end{equation}
Integrating the above inequality with respect to $v$, by \eqref{eq:redshiftOmega2}, \eqref{eq:reluvN} and \eqref{eq:dvrescaledOmegaest}, we obtain \eqref{eq:rescaledOmegaestprop}
\begin{equation}
\label{eq:rescaledOmegaest}
\begin{split}
\left| \log \left(\frac{3r^2\Omega^2}{\Lambda rP(r)}\right)\right|\leq&\:  \left|\log \left(\frac{3r^2\Omega^2}{\Lambda rP(r)}\right)\right|(u,v_{\gamma_{\delta}}(u))\\
&+ C_{\delta,r_0}C(\Delta_1',D_1,D_2,D_3,v_0,p)(\epsilon^2\Delta_2+1) e^{-2p|u|}\\
\leq&\: C_{\delta,r_0}C(\Delta_1',D_1,D_2,D_3,v_0,p)(1+ \epsilon^2 \Delta_2 ) e^{-2pv}.
\end{split}
\end{equation}
Employing the above estimate and \eqref{eq:durest1prop2}, we then prove \eqref{eq:durest2prop}. Finally, a combination of \eqref{eq:dvrestprop2} and \eqref{eq:durest2prop} yields \eqref{eq:ratiodurdvr}. This concludes the proof of this proposition.
\end{proof}

Now we are in a good position to improve the bootstrap assumptions \eqref{ba:dvphi2} and \eqref{ba:Yphi2}:
\begin{proposition}
	\label{prop:baimpphinoshift}
Assume that \eqref{ba:dvphi2} and \eqref{ba:Yphi2} hold. Let $\delta>0$ be suitably small and $|u_0|$ be suitably large (depending on $\delta$, $\Delta_1$, $\Delta_2$, $D_2$, $D_3$, $v_0$, $r_0$). There exists a sufficiently large $\Delta_1'>0$, such that for all  $(u,v)\in\mathcal{N}_{\delta,r_0}$, we have the following estimates:
	\begin{align}
	\label{eq:dvphinoshift}
	r|\partial_v\phi|(u,v)\leq &\: \frac{1}{2} \Delta_1' e^{-pv},\\
	\label{eq:duphinoshift}
	r|\partial_u\phi|(u,v)\leq &\: \frac{1}{2} \Delta_1' e^{-pv},\\
	\label{eq:Yphirnoshift}
	|Y\phi|(u,v)\leq &\: \frac{1}{2} \Delta_1' e^{-pv}.
	\end{align}
\end{proposition}
\begin{proof}
	Define
	\begin{equation*}
	\overline{\Phi}(s):=\max \Big \{\sup_{\{r(u,v)= s\}\cap \{v\geq v_{\gamma_{\delta}}(u_0)\}} e^{pv}|r\partial_v\phi| , \sup_{\{r(u,v)= s\}\cap\{v\geq v_{\delta}(u_0)\}}  e^{-pu}|r\partial_u\phi|\Big \}
	\end{equation*}
	with $r_0\leq s\leq r_{\delta}+\eta$. Here, we choose $\eta>0$ to be suitably small and $|u_0|$ to be suitably large, such that $v_{\gamma_{\delta}}(u_0)>v_0$ and the curve $\{r=r_{\delta}+\eta\}\cap \{v\geq v_{\gamma_{\delta}}(u_0)\}\subset\mathcal{R}_{\delta}$. The region under our consideration is shown in the figure below. 
		\begin{figure}[H]
		\begin{center}
		\begin{tikzpicture}[x=0.75pt,y=0.75pt,yscale=-1,xscale=1]

\draw [fill={rgb, 255:red, 155; green, 155; blue, 155 }  ,fill opacity=1 ]   
 (139.94,60.29) .. controls (180.06,80) and (236.63,81.43) .. (304.4,64.72) --  (313.41,73.84) .. controls (264.34,101.43) and (166.06,110.57) .. (119.97,80.24) --  (139.94,60.29) ;

\draw [line width=0.75]    (100,100.2) -- (200,200.2) ;
\draw [line width=0.75]    (200,200.2) -- (320.8,80.8) ;
\draw    (100,100.2) -- (120.14,80.08) -- (139.94,60.29) ;
\draw [color={rgb, 255:red, 0; green, 0; blue, 0 }  ,draw opacity=1 ][fill={rgb, 255:red, 0; green, 0; blue, 0 }  ,fill opacity=1 ][line width=0.75]    (350.34,50.26) ;
\draw [shift={(350.34,50.26)}, rotate = 0] [color={rgb, 255:red, 0; green, 0; blue, 0 }  ,draw opacity=1 ][line width=0.75]      (0, 0) circle [x radius= 1.34, y radius= 1.34]   ;
\draw    (304.4,64.72) -- (320.8,80.8) ;
\draw    (140.23,60.29) .. controls (180.34,80) and (236.91,81.43) .. (304.69,64.72) ;
\draw    (120.26,80.24) .. controls (166.34,110.57) and (264.63,101.43) .. (313.7,73.84) ;
\draw    (304.69,64.72) -- (313.7,73.84) ;
\draw    (140.23,60.29) -- (120.26,80.24) ;
\draw  [dash pattern={on 0.84pt off 2.51pt}]  (304.4,64.72) .. controls (330.34,57.71) and (326.91,58.29) .. (350.34,50.26) ;
\draw  [dash pattern={on 0.84pt off 2.51pt}]  (320.8,80.8) -- (349.6,51.73) ;
\draw [fill={rgb, 255:red, 155; green, 155; blue, 155 }  ,fill opacity=1 ] [dash pattern={on 0.84pt off 2.51pt}]  (313.41,73.84) -- (349.6,51.73) ;
\draw  [dash pattern={on 0.84pt off 2.51pt}]  (110.07,90.14) -- (210.07,190.14) ;
\draw    (123.36,103) .. controls (166.4,121.52) and (277.6,107.92) .. (317.6,77.92) ;
\draw  [dash pattern={on 0.84pt off 2.51pt}]  (317.6,77.92) -- (349.6,51.73) ;
\draw    (200.56,110) -- (226.4,84.32) ;
\draw [shift={(200.56,110)}, rotate = 315.18] [color={rgb, 255:red, 0; green, 0; blue, 0 }  ][fill={rgb, 255:red, 0; green, 0; blue, 0 }  ][line width=0.75]      (0, 0) circle [x radius= 1.34, y radius= 1.34]   ;
\draw    (226.4,84.32) -- (246,104.32) ;
\draw [shift={(246,104.32)}, rotate = 45.58] [color={rgb, 255:red, 0; green, 0; blue, 0 }  ][fill={rgb, 255:red, 0; green, 0; blue, 0 }  ][line width=0.75]      (0, 0) circle [x radius= 1.34, y radius= 1.34]   ;
\draw [shift={(226.4,84.32)}, rotate = 45.58] [color={rgb, 255:red, 0; green, 0; blue, 0 }  ][fill={rgb, 255:red, 0; green, 0; blue, 0 }  ][line width=0.75]      (0, 0) circle [x radius= 1.34, y radius= 1.34]   ;

\draw (174,61.54) node [anchor=north west][inner sep=0.75pt]  [font=\tiny]  {$r=r_{0}$};
\draw (96.65,83.14) node [anchor=north west][inner sep=0.75pt]  [font=\tiny,rotate=-315]  {$u=u_{0}$};
\draw (105.14,124.6) node [anchor=north west][inner sep=0.75pt]  [font=\scriptsize]  {$\underline{H}_{0}$};
\draw (141.29,156.02) node [anchor=north west][inner sep=0.75pt]  [font=\tiny,rotate=-45]  {$v=v_{0}$};
\draw (272.05,137.57) node [anchor=north west][inner sep=0.75pt]  [font=\scriptsize]  {$H_{0}$};
\draw (318.97,84.83) node [anchor=north west][inner sep=0.75pt]  [font=\tiny]  {$v=v_{\infty }$};
\draw (159.26,84.74) node [anchor=north west][inner sep=0.75pt]  [font=\tiny]  {$\gamma _{\delta }$};
\draw (145.14,113.31) node [anchor=north west][inner sep=0.75pt]  [font=\tiny]  {$r=r_{\delta } +\eta $};
\draw (233.71,80.11) node [anchor=north west][inner sep=0.75pt]  [font=\tiny]  {$( u,\ v)$};
\draw (210.11,191.31) node [anchor=north west][inner sep=0.75pt]  [font=\tiny]  {$v=v_{\gamma _{\delta }}( u_{0})$};

\end{tikzpicture}
		\end{center}
		\vspace{-0.2cm}
		\caption{Relevant region for the upper bound estimate.}
		\label{fig:gronwall1}
	\end{figure}
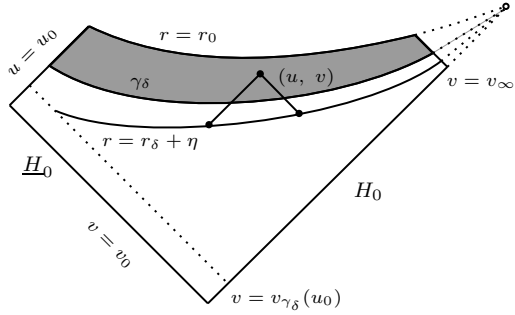
We first integrate \eqref{propeq:phi1} in $u$ and \eqref{propeq:phi2} in $v$, respectively. These two integrals together lead to the following estimate:
	\begin{equation*}
	\begin{split}
	\overline{\Phi}(r(u,v))\leq &\: \overline{\Phi}(r_{\delta}+\eta) +\max\Bigg \{ \int_{u_{r_{\delta}+\eta}(v)}^u r^{-1}|\partial_vr| e^{pv}|r\partial_u\phi|(u',v)\,du' ,\\
	&\int_{v_{r_{\delta}+\eta}(u)}^v r^{-1}|\partial_ur|e^{-pu} |r\partial_v\phi|(u,v')\,dv'\Bigg\}\\
	= &\: \overline{\Phi}(r_{\delta}+\eta) +\max\Bigg \{ \int_{u_{r_{\delta}+\eta}(v)}^u r^{-1}\f{|\partial_vr|}{|\partial_ur|}\cdot |\partial_ur|e^{p(v+u)}\cdot e^{-pu}|r\partial_u\phi|(u',v)\,du' ,\\
	&\int_{v_{r_{\delta}+\eta}(u)}^v r^{-1}\f{|\partial_ur|}{|\partial_vr|}\cdot |\partial_vr|e^{-p(v+u)}\cdot e^{pv} |r\partial_v\phi|(u,v')\,dv'\Bigg\}\\
	\leq &\: \overline{\Phi}(r_{\delta}+\eta) +\int^{r_{\delta}+\eta}_{r(u,v)} e^{p|v+u|}\max\left\{\f{|\partial_vr|}{|\partial_ur|},\f{|\partial_ur|}{|\partial_vr|}\right\} r'^{-1} \overline{\Phi}(r')\,dr'.
	\end{split}
	\end{equation*}
By Lemma \ref{lm:relatuvconstr}, it holds
	\begin{equation} \label{gronwall cond1}
		e^{p|v+u|}\leq e^{pC_{\delta,r_0}+pC_{\delta,r_0}(C(\Delta_1',\Delta_1, D_1,D_2,D_3,v_0,p)+\epsilon^2 \Delta_2)e^{-pv}}.
		\end{equation}
Then, using the estimate for the ratio of $\partial_u r$ and $\partial_v r$ in \eqref{eq:ratiodurdvr}, we deduce
	\begin{equation} \label{gronwall cond2}
\max\left\{\f{|\partial_vr|}{|\partial_ur|},\f{|\partial_ur|}{|\partial_vr|}\right\}\leq 1+C_{\delta,r_0}C(\Delta_1',D_1,D_2,D_3,v_0,p,q)(1+\epsilon^2 \Delta_2)e^{-2pv}.
	\end{equation}
Note that, in $\mathcal{N}_{\delta,r_0}$, one can pick $|u_0|$ to be sufficiently large (depending on $\delta$) such that
		\begin{equation} \label{gronwall cond3}
		v\geq |u|-\frac{1}{\delta}>\frac{1}{2}|u_0|.
		\end{equation} 
Gathering estimates \eqref{gronwall cond1}, \eqref{gronwall cond2}, \eqref{gronwall cond3}, for $|u_0|$ being sufficiently large (depending on $\Delta_2$, $\Delta_1'$, $\delta$, $C(\Delta_1,D_2,D_3,v_0)$), we have
\begin{equation*}
  e^{p|v+u|}\max\left\{\f{|\partial_vr|}{|\partial_ur|},\f{|\partial_ur|}{|\partial_vr|}\right\}\leq 4e^{pC_{\delta, r_0}} .
\end{equation*}
Now we take $\Delta_1'\geq\left(\frac{r_S}{r_0}\right)^{4e^{pC_{\delta,r_0}}}\Delta_1$. Employing the standard Gr\"onwall's inequality, we obtain
	\begin{equation} \label{gronwall upper est}
	\begin{split}
	\overline{\Phi}(r(u,v))\leq&\: \overline{\Phi}(r_{\delta}+\eta)(\frac{r_{\delta}+\eta}{r})^{4e^{pC_{\delta, r_0}}} \\
	\leq&\frac{1}{2} \Delta_1  \left(\frac{r_S}{r_0}\right)^{4e^{pC_{\delta,r_0}}}\leq \frac{1}{2}\Delta_1',
	\end{split}
	\end{equation}
where $\overline{\Phi}(r_{\delta}+\eta)$ is controlled via \eqref{ba:dvphi} and \eqref{ba:Yphi}. This establishes \eqref{eq:dvphinoshift} and \eqref{eq:duphinoshift}. And \eqref{eq:Yphirnoshift} follows directly by applying \eqref{eq:durest2prop} to \eqref{eq:duphinoshift}.

\end{proof}
Since we already improve the estimates in bootstrap assumptions \eqref{ba:dvphi}-\eqref{ba:Yphi2}, via the standard continuity argument we then conclude
\begin{proposition}
\label{prop:continuityarg}
For a sufficiently small $|U_0|$ , in the region
\begin{equation*}
D_{\infty,r_0}:= [0,U_0)\times [v_0, \infty)\cap\{r\geq r_0\},
\end{equation*}
there exists a smooth solution $(r,\widehat{\Omega}^2, \phi)$ to the spherically symmetric Einstein-scalar field system \eqref{ESlam} arising from initial data prescribed as in Section \ref{sec:initialdata}. Furthermore, in $D_{\infty,r_0}$, the solution satisfies the estimates \eqref{ba:dvphi}--\eqref{ba:Yphi2} and all the estimates in Sections \ref{estimate in R delta} and \ref{estimate in N delta r0}.
\end{proposition}

\subsection{Lower bound estimates for the scalar field}
In this subsection, we present the lower bound estimates for $\partial_u\phi$ and $\partial_v\phi$ in the region $D_{\infty,r_0}$. 
\begin{proposition}
\label{prop:lowboundsposr0}
Let $\epsilon>0$ be suitably small and $|u_0|$ be suitably large. Then there exists a sufficiently large $v_1$ (depending on $r_0$, $\epsilon$, $D_1$, $\Delta_1'$) greater than $v_0$ such that for all $(u,v)$ in $\{r\geq r_1\}$ with $v\geq v_1$ and $r_S>r_1\geq r_0$, there hold
\begin{align}
\label{eq:lowbounddvphi0}
r\partial_v\phi(u,v)\geq&\: (D_1-\epsilon) e^{-qv},\\
\label{eq:lowboundYphi0}
r^2 Y\phi(u,v)\geq&\: (D_1-\epsilon)(\f{\Lambda P(r_1)}{3r_1^2}+\frac{1}{2}r_1^{-1}(1-\Lambda r_1^2)+\epsilon_0 r_S^{-1})^{-1}e^{-qv},
\end{align}
where $Y\phi=-\frac{1}{\partial_u r}\partial_u\phi$.

Furthermore, along $\{r=r_1\}$, for $v\geq v_1$, we also have
\begin{equation}
\label{eq:lowboundduphi1}
r\partial_u\phi|_{r=r_1}\geq \f{1}{2}\frac{\Lambda P(r_1)}{\Lambda P(r_1)+\frac{3}{2}r_1(1-\Lambda r_1^2)+3\epsilon_0 r_1^2 r_S^{-1}}(D_1-\epsilon) e^{-qv}.
\end{equation}
\end{proposition}
\begin{proof}
First, since $Y\phi(U,v_1)$ is continuous in $U$ along $v=v_0$, hence by \eqref{eq:idlowboundYphi}, for suitably large $|u_0|$, we have that
\begin{equation}
\label{eq:Yphiv1}
r^2Y\phi(u,v_1)\geq \f{D_1}{\alpha_S-q}(1-2\epsilon) e^{-qv}.
\end{equation}
Invoking this to \eqref{propeq:Yphi}, for $v\geq v_1$ it then follows that $Y\phi\geq 0$.
Integrating \eqref{propeq:phi1} and employing \eqref{eq:estdataphi}, \eqref{eq:estdatar}, \eqref{ba:Yphi}, \eqref{eq:dvrestprop}, \eqref{eq:rA1}, for all $v\geq v_1$ we obtain that:
\begin{equation*}
\begin{split}
r\partial_v\phi(u,v)=&\:r\partial_v\phi(-\infty,v)+ \int_{-\infty}^u [\partial_vr \partial_ur Y\phi](u',v)\,du'\\
=&\:r\partial_v\phi(-\infty,v)-\int_{r(u,v)}^{r_{\mathcal{A}}(v)} \partial_v r Y\phi\Big|_{v'=v}(r')\,dr' -\int^{r_{\mathcal{H}^+}(v)}_{r_{\mathcal{A}}(v)} \partial_v r Y\phi\Big|_{v'=v}(r')\,dr'\\
\geq &\: r\partial_v\phi(-\infty,v) -\int^{r_{\mathcal{H}^+}(v)}_{r_{\mathcal{A}}(v)} \partial_v r Y\phi\Big|_{v'=v}(r')\,dr'\\
\geq &\: r\partial_v\phi(-\infty,v)- (|r_{\mathcal{H}^+}(v)-r_S|+ |r_{\mathcal{A}}(v)-r_S|) \sup_{-\infty\leq u'\leq u_{\mathcal{A}}(v)}|\partial_v r||Y\phi|(u',v)\\
\geq &\: D_1 e^{-qv}- C(\Delta_1,D_1,D_2,D_3,v_0,p) e^{-5pv}.
\end{split}
\end{equation*}
Noting that $q<5p$, we can whence choose $v_1$ to be sufficiently large (depending on $D_1$, $\Delta_1$, $\epsilon$) such that for all $v\geq v_1$, the estimate \eqref{eq:lowbounddvphi0} holds 
\begin{equation*} 
r\partial_v\phi(u,v)\geq (D_1-\epsilon) e^{-qv}.
\end{equation*}

Furthermore, for an arbitrary small constant $\epsilon_0>0$, we define 
\begin{equation*}
\delta_1:= (D_1-\epsilon)f(r_1, \epsilon_0)^{-1} 
\end{equation*}
with
\begin{equation*}
    f(r, \epsilon_0):=\f{\Lambda P(r)}{3r^2}+\frac{1}{2}r^{-1}{(1-\Lambda r^2)}+\epsilon_0 r_S^{-1}.
\end{equation*}
Here $f(r_1,\epsilon_0)$ is well-defined and positive since $r_1<r_S$. Moreover, via a direct computation, one can verify that it holds
\begin{equation} \label{fr1 as}
    f(r_1,\epsilon_0)\geq f(r, \epsilon_0)\geq f(r_S,\epsilon_0)=\alpha_S+\epsilon_0 r_S^{-1}, \quad \textrm{for any} \ r_1\le r\le r_S.
\end{equation}
Then, we can express \eqref{eq:lowbounddvphi0} as
\begin{equation}
\label{eq:auxestdvphi}
r\partial_v\phi(u,v)\geq \delta_1  f(r_1, \epsilon_0) e^{-qv}.
\end{equation}
We now estimate $r^2Y\phi$. Consider its equation of propagation \eqref{propeq:Yphi}:
\begin{equation*}
\partial_v(r^2 Y\phi)=\left[-\frac{1}{4}r^{-1}\left(\frac{\Omega^2}{-\partial_u r}\right)(1-\Lambda r^2)+2r^{-1}\partial_vr\right] r^2 Y\phi+r\partial_v\phi.
\end{equation*}
Observe first that by \eqref{eq:durest1prop}, \eqref{eq:dvrestprop}, \eqref{eq:durest1prop2} and \eqref{eq:dvrestprop2}, for $(u,v)\in\{r\geq r_0\}\cap\{v\geq v_1\}$, we have
\begin{equation*}
\begin{split}
&\left|\frac{1}{4}\frac{\Omega^2}{r\partial_u r}(1-\Lambda r^2)+2\frac{\partial_vr}{r}+\f{\Lambda P(r)}{3r^2}+\frac{1}{2}r^{-1}(1-\Lambda r^2)\right|\\
\leq&\: C_{r_0}(\Delta_1,\Delta_1',D_1, D_2,D_3,v_0, p)e^{-2pv}\\
\leq&\: C_{r_0}(\Delta_1,\Delta_1',D_1, D_2,D_3,v_0, p) e^{-2pv_1}=:\epsilon_0 r_S^{-1},
\end{split}
\end{equation*}
where $\epsilon_0>0$ is arbitrarily small if we choose $v_1$ to be suitably large. Then, following from \eqref{propeq:Yphi} and \eqref{eq:Yphiv1}, for $v\geq v_1$ it holds that
\begin{equation*}
\partial_v(r^2 Y\phi)\geq -f(r_1, \epsilon_0)r^2Y\phi+r\partial_v\phi.
\end{equation*}
This is equivalent to
\begin{equation*}
\partial_v(e^{f(r_1, \epsilon_0)v} r^2 Y\phi)\geq e^{f(r_1, \epsilon_0) v} r\partial_v\phi.
\end{equation*}
Integrating the above inequality, we deduce that
\begin{equation*}
\begin{split}
e^{f(r_1, \epsilon_0)v} r^2 Y\phi(u,v)\geq&\: e^{f(r_1, \epsilon_0)v_1} r^2 Y\phi(u,v_1)\\
&+\int_{v_1}^v e^{f(r_1, \epsilon_0)v'} r \partial_v\phi(u',v)\,dv'.
\end{split}
\end{equation*}
Employing \eqref{eq:Yphiv1} and \eqref{eq:auxestdvphi}, it follows that
\begin{equation*}
\begin{split}
r^2 Y\phi(u,v)\geq&\:  e^{f(r_1, \epsilon_0)(v_1-v)} r^2 Y\phi(u,v_1) +f(r_1, \epsilon_0)e^{-f(r_1, \epsilon_0)v}\cdot  \int_{v_1}^v e^{f(r_1, \epsilon_0)v'} \delta_1 e^{-qv'}\,dv'\\
\geq &\:e^{f(r_1, \epsilon_0)(v_1-v)} [\f{D_1}{\alpha_s-q}(1-2\epsilon)- \f{\delta_1 f(r_1, \epsilon_0)}{f(r_1, \epsilon_0)-q}]e^{-qv_1}+ \f{\delta_1 f(r_1, \epsilon_0)}{f(r_1, \epsilon_0)-q} e^{-qv}  \\
\geq &\: \f{\delta_1 f(r_1, \epsilon_0)}{f(r_1, \epsilon_0)-q} e^{-qv}.
\end{split}
\end{equation*}
Here, we use the following fact: for suitably small $\epsilon>0$, noting that by \eqref{fr1 as}, it holds
\begin{equation*}
    \f{D_1}{\alpha_s-q}(1-2\epsilon)> \f{\delta_1 f(r_1, \epsilon_0)}{f(r_1, \epsilon_0)-q}=\f{D_1-\epsilon}{f(r_1, \epsilon_0)-q}.
\end{equation*}

We hence conclude that
\begin{equation*}
r^2 Y\phi(u,v)\geq (D_1-\epsilon)(\f{\Lambda P(r_1)}{3r_1^2}+\frac{1}{2}r_1^{-1}{(1-\Lambda r_1^2)}+\epsilon_0 r_S^{-1})^{-1}e^{-qv}.
\end{equation*}
Finally, for $v_1$ suitably large compared to $r_0^{-1}$, together with \eqref{eq:durest2prop}, we arrive at
\begin{equation*}
r\partial_u\phi|_{r=r_1}=(-\partial_ur) rY\phi|_{r=r_1}\geq \f{1}{2}\frac{\Lambda P(r_1)}{\Lambda P(r_1)+\frac{3}{2}r_1{(1-\Lambda r^2)}+3\epsilon_0 r_1^2 r_S^{-1}}(D_1-\epsilon) e^{-qv}.
\end{equation*}
This finishes the proof of this proposition.
\end{proof}
Applying the above result, we can further show the following proposition.
\begin{proposition}
\label{prop:lowboundsposr}
For a sufficiently small $\epsilon>0$, there exist sufficiently large $v_1\geq v_0$ and $|u_1|>|u_0|$ (depending on $r_0$, $\epsilon$, $D_1$, $\Delta_1$, $\Delta_1'$), such that for all $(u,v)\in\{v\geq v_1, |u|\geq |u_1|, r_0\leq r\leq r_1\}$, the following estimates hold:
\begin{align}
\label{eq:lowbounddvphi}
r\partial_v\phi(u,v)\geq&\:\f{1}{2}\frac{\Lambda P(r_1)}{\Lambda P(r_1)+\frac{3}{2}r_1{(1-\Lambda r_1^2)}+3\epsilon_0 r_1^2 r_S^{-1}}(D_1-\epsilon) e^{-qv},\\
\label{eq:lowboundduphi2}
r\partial_u\phi(u,v)\geq&\: \f{1}{2}\frac{C_{r_0,r_1}\Lambda P(r_1)}{\Lambda P(r_1)+\frac{3}{2}r_1{(1-\Lambda r_1^2)}+3\epsilon_0 r_1^2 r_S^{-1}}(D_1-\epsilon) e^{-qv}.
\end{align}
\end{proposition}
\begin{proof}
We define
\begin{equation*}
\underline{\Phi}(s):=\min \Big \{\inf_{\{r(u,v)= s\}\cap\{|u|\geq |u_1|,v\geq v_1\}} e^{qv} r\partial_v\phi , \inf_{\{r(u,v)= s\}\cap\{|u|\geq |u_1|,v\geq v_1\}}  e^{-qu} r\partial_u\phi \Big \}.
\end{equation*}
Here we require $r_0\leq s\leq r_1$, $|u_1|>|u_0|$ and $v_{r_1}(u)>v_1$ for all $|u|\geq |u_1|$. See the following figure.
 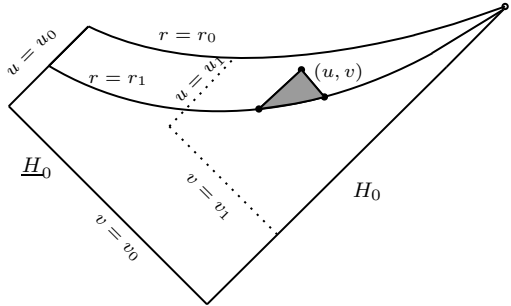
\begin{figure}[ht]
	\begin{center}
\begin{tikzpicture}[x=0.75pt,y=0.75pt,yscale=-1,xscale=1]

\draw [line width=0.75]    (100,100.6) -- (200,200.6) ;
\draw [line width=0.75]    (200,200.6) -- (320.8,80.8) ;
\draw    (100,100.6) -- (120.26,80.24) -- (140.06,60.45) ;
\draw [color={rgb, 255:red, 0; green, 0; blue, 0 }  ,draw opacity=1 ][fill={rgb, 255:red, 0; green, 0; blue, 0 }  ,fill opacity=1 ][line width=0.75]    (350.34,50.26) ;
\draw [shift={(350.34,50.26)}, rotate = 0] [color={rgb, 255:red, 0; green, 0; blue, 0 }  ,draw opacity=1 ][line width=0.75]      (0, 0) circle [x radius= 1.34, y radius= 1.34]   ;
\draw    (140.23,60.29) .. controls (180.34,80) and (236.91,81.43) .. (304.69,64.72) ;
\draw    (120.26,80.24) .. controls (174.4,112.72) and (246,110.72) .. (313.7,73.84) ;
\draw    (140.23,60.29) -- (120.26,80.24) ;
\draw    (304.4,64.72) .. controls (330.34,57.71) and (326.91,58.29) .. (350.34,50.26) ;
\draw    (320.8,80.8) -- (349.6,51.73) ;
\draw [fill={rgb, 255:red, 155; green, 155; blue, 155 }  ,fill opacity=1 ]   (313.41,73.84) -- (349.6,51.73) ;
\draw    (226.16,102) -- (247.62,82.21) ;
\draw [shift={(226.16,102)}, rotate = 317.32] [color={rgb, 255:red, 0; green, 0; blue, 0 }  ][fill={rgb, 255:red, 0; green, 0; blue, 0 }  ][line width=0.75]      (0, 0) circle [x radius= 1.34, y radius= 1.34]   ;
\draw    (247.62,82.21) -- (259.2,95.92) ;
\draw [shift={(259.2,95.92)}, rotate = 49.82] [color={rgb, 255:red, 0; green, 0; blue, 0 }  ][fill={rgb, 255:red, 0; green, 0; blue, 0 }  ][line width=0.75]      (0, 0) circle [x radius= 1.34, y radius= 1.34]   ;
\draw [shift={(247.62,82.21)}, rotate = 49.82] [color={rgb, 255:red, 0; green, 0; blue, 0 }  ][fill={rgb, 255:red, 0; green, 0; blue, 0 }  ][line width=0.75]      (0, 0) circle [x radius= 1.34, y radius= 1.34]   ;
\draw  [dash pattern={on 0.84pt off 2.51pt}]  (180.96,110.8) -- (235.2,165.12) ;
\draw  [dash pattern={on 0.84pt off 2.51pt}]  (180.96,110.8) -- (214,75.92) ;

\draw [fill={rgb, 255:red, 155; green, 155; blue, 155 }  ,fill opacity=1 ]   
(247.6, 82.2)--(226.2, 102) .. controls (249.47,98.77) and (249.13,98.43) .. (259.2,95.92)--(247.6, 82.2);

\draw (174,61.54) node [anchor=north west][inner sep=0.75pt]  [font=\tiny]  {$r=r_{0}$};
\draw (96.65,83.14) node [anchor=north west][inner sep=0.75pt]  [font=\tiny,rotate=-315]  {$u=u_{0}$};
\draw (105.14,125) node [anchor=north west][inner sep=0.75pt]  [font=\scriptsize]  {$\underline{H}_{0}$};
\draw (145.69,152.42) node [anchor=north west][inner sep=0.75pt]  [font=\tiny,rotate=-45]  {$v=v_{0}$};
\draw (272.05,137.57) node [anchor=north west][inner sep=0.75pt]  [font=\scriptsize]  {$H_{0}$};
\draw (252.23,77.57) node [anchor=north west][inner sep=0.75pt]  [font=\tiny]  {$( u,v)$};
\draw (138.8,83.14) node [anchor=north west][inner sep=0.75pt]  [font=\tiny]  {$r=r_{1}$};
\draw (181.45,95.94) node [anchor=north west][inner sep=0.75pt]  [font=\tiny,rotate=-315]  {$u=u_{1}$};
\draw (191.69,132.02) node [anchor=north west][inner sep=0.75pt]  [font=\tiny,rotate=-45]  {$v=v_{1}$};

\end{tikzpicture}
\end{center}
\vspace{-0.2cm}
\caption{Relevant region for the lower bound estimate.}
	\label{fig:gronwall2}
\end{figure}

\noindent Integrating \eqref{propeq:phi1} in $u$ and \eqref{propeq:phi2} in $v$, and noting that $\partial_v\phi>0$ and $\partial_u\phi>0$ in $\{r\geq r_0\}\cap\{v\geq v_1\}$ guaranteed by Proposition \ref{prop:lowboundsposr0}, we derive
\begin{equation*}
\begin{split}
\underline{\Phi}(r(u,v))\geq &\: \underline{\Phi}(r_1) +\min \Bigg \{ \int_{u_{r_{\delta}+\eta}(v)}^u r^{-1}(-\partial_vr) e^{qv}r\partial_u\phi(u',v)\,du' ,\\
&\int_{v_{r_{\delta}+\eta}(u)}^v r^{-1}(-\partial_ur)e^{-qu} |r\partial_v\phi|(u,v')\,dv'\Bigg\}\\
= &\:  \underline{\Phi}(r_1)+\min \Bigg \{ \int_{u_{r_{\delta}+\eta}(v)}^u r^{-1}\f{-\partial_vr}{-\partial_ur}\cdot (-\partial_ur)e^{q(v+u)}\cdot e^{-qu}r\partial_u\phi(u',v)\,du' ,\\
&\int_{v_{r_{\delta}+\eta}(u)}^v r^{-1}\f{-\partial_ur}{-\partial_vr}\cdot (-\partial_vr)e^{-q(v+u)}\cdot e^{qv} r\partial_v\phi(u,v')\,dv' \Bigg\} \\
&\geq \underline{\Phi}(r_1).
\end{split}
\end{equation*}
Together with Lemma \ref{lm:relatuvconstr} and Proposition \ref{prop:lowboundsposr0}, for $|u_0|$ sufficiently large, we then obtain \eqref{eq:lowbounddvphi} and \eqref{eq:lowboundduphi2}
\begin{align*}
r\partial_v\phi(u,v)\geq&\: \f{1}{2}\frac{\Lambda P(r_1)}{\Lambda P(r_1)+\frac{3}{2}r_1{(1-\Lambda r_1^2)}+3\epsilon_0 r_1^2 r_S^{-1}}(D_1-\epsilon) e^{-qv},\\
r\partial_u\phi(u,v)
\geq &\f{1}{2}\frac{C_{r_0,r_1}\Lambda P(r_1)}{\Lambda P(r_1)+\frac{3}{2}r_1{(1-\Lambda r_1^2)}+3\epsilon_0 r_1^2 r_S^{-1}}(D_1-\epsilon) e^{-qv}.
\end{align*}
\end{proof}
In particular, we set $C_{low}:=\inf\limits_{r_0\leq r_1\leq r_\delta}\f{1}{2}\frac{\min \{1,C_{r_0,r_1}\}\Lambda P(r_1)}{\Lambda P(r_1)+\frac{3}{2}r_1(1-\Lambda r_1^2)+3\epsilon_0 r_1^2 r_S^{-1}}$. The above proposition hence implies
\begin{equation} \label{r0 lowerbddvphi}
	r^2\partial_v\phi(u_{r_0}(v'),v')\geq C_{low}r_0(D_1-\epsilon) e^{-qv},
	\end{equation}
	\begin{equation} \label{r0 lowerbdduphi}
	r^2\partial_u\phi(u_{r_0}(v'),v)\geq C_{low}r_0(D_1-\epsilon) e^{-qv}.
	\end{equation}

\section{Estimates near spacelike singularity}
\label{sec:estnearsing}
In this section, we proceed to derive quantitative estimates near the spacelike singularities which will lead to the ultimate curvature blow-up estimates in Theorem \ref{main thm}.

For the rest of this section, we first fix a parameter $r_0>0$ and set it to be sufficiently small. Our estimates will be carried out in the region
$$\left\{|u|\geq |u_1|+2v_{r=0}(u_1), \, v\geq \max\{v_1, \t v_1\}\right\}\cap \left\{r(u,v)\leq r_0\right\}$$
with $u_1, v_1$ given in Proposition \ref{prop:lowboundsposr}. Furthermore, we claim that there exists a sufficiently large $\t v_1$, such that for $v\geq \t v_1$ the following estimate along $r=r_0$ holds
\be\label{additional v}
|\partial_v \log(r\O^2)(u_{r_0}(v),v)|\leq\f{1}{r_0}.
\ee
This is achieved via using the estimates in Proposition \ref{prop:metricest2} and Proposition  \ref{prop:durdvrr0}. Note that in Proposition \ref{prop:durdvrr0}, the estimate \eqref{eq:prelimestOmega} can be reformulated as
\begin{equation}\label{log r O middle}
|\partial_v \log(r\O^2)+\f{1}{r^2}(r\partial_v r+\f{\Lambda P(r)}{6})-\f{1}{2r}+\f{\Lambda r}{2}|(u_{r_0}(v),v)\leq C_{r_0}\cdot C(\Delta_1',D_1,D_2,D_3,v_0,p,q) e^{-2pv}.
\end{equation}
Employing triangle inequalities, this implies
\begin{equation*}
\begin{split}
&|\partial_v \log(r\O^2)(u_{r_0}(v),v)|\\
\leq& |\partial_v \log(r\O^2)+\frac{1}{r^2}(r\partial_v r+\frac{\Lambda P(r)}{6})-\frac{1}{2r}+\frac{\Lambda r}{2}|(u_{r_0}(v),v)+\f{1}{r_0^2}|r \partial_vr+\frac{\Lambda P(r)}{6}|(u_{r_0}(v),v)\\
&+(\frac{1}{2r_0}-\frac{\Lambda r_0}{2}).
\end{split}
\end{equation*}
Combining this with the estimate \eqref{eq:dvrestprop2} for $|r\partial_vr+\frac{\Lambda P(r)}{6}|$, and choosing $v\geq \t v_1$ with $\t v_1$ large, we thus arrive at \eqref{additional v}.

\subsection{Preliminary Estimates for $r\partial_u r$ and $r\partial_v r$}
Based on the obtained estimates in Section \ref{Section7}, via a monotonicity argument, we derive the following estimates for $r\partial_u r$ and $r\partial_v r$ in the region $\{r\leq r_0\}$.
\begin{proposition}\label{1st partial r}
There exists a sufficiently small $\epsilon>0$, such that for any $(u,v)$ satisfying $|u|\geq |u_1|+2v_{r=0}(u_1)$ and $r(u,v)\leq r_0$, the following estimates hold
\be\begin{split}\label{rvr1}
|(r\partial_v r+\frac{\Lambda}{6} P(r))(u,v)|\leq& C(\Delta_1',D_1,D_2,D_3,v_0,p) e^{-2pv}\\
&+[1+C(\Delta_1',D_1,D_2,D_3,v_0,p) e^{-2pv}]\cdot [r_0-r(u,v)]\\
\leq& \epsilon ,
\end{split}\ee
\be\begin{split}\label{rur1}
|(r\partial_u r+\frac{\Lambda}{6} P(r))(u,v)|\leq& C_{\delta,r_0}(C(\Delta_1',D_1,D_2,D_3,v_0,p)+\epsilon \Delta_2)e^{-2p|u|}\\
&+[1+C(\Delta_1,D_2,D_3,v_0) e^{-2pv}]\\
&\quad\times [1+C_{\delta,r_0} (C(\Delta_1,D_2,D_3,v_0)+\epsilon \Delta_2)e^{-2p|u|}] \\
&\quad\times [r_0-r(u,v)]\\
\leq& \epsilon.
\end{split}\ee
\end{proposition}
\begin{proof}
For $r\partial_v r(u, v)$, we first rewrite \eqref{propeq:r1} as
\bes
\partial_u(r\partial_v r+\frac{\Lambda}{6}P(r))=-\f14\O^2(1-\Lambda r^2)+(\frac{\Lambda}{2}r^2-\f12)\partial_u r.
\ees
Integrating this equation with respect to $u$, we obtain
\be\label{r partial v r1}
\begin{split}
&(r\partial_v r+\frac{\Lambda}{6}P(r))(u,v)\\
=&(r\partial_v r+\frac{\Lambda}{6} P(r))(u_{r_0}(v),v)-\int_{u_{r_0}(v)}^u\f{\O^2}{4\partial_u r}(1-\Lambda r^2)\partial_u r d{\color{black}u'}-\f12\int^u_{u_{r_0}(v)}(1-\Lambda r^2)\partial_u r d{\color{black}u'}\\
=&(r\partial_v r+\frac{\Lambda}{6} P(r))(u_{r_0}(v),v)-\int^u_{u_{r_0}(v)}\bigg(\f{\O^2}{4\partial_u r}+\f12\bigg)(1-\Lambda r^2) \partial_u r d{\color{black}u'}.
\end{split}
\ee

\noindent Along $r=r_0$, applying Proposition \ref{prop:metricest2}, we bound the first term as
\begin{align*}
\left|r \partial_vr+\frac{\Lambda}{6}P(r)\right|(u_{r_0}(v),v)\leq&\: C(\Delta_1',D_1,D_2,D_3,v_0,p) e^{-2pv}.
\end{align*}

\noindent We then estimate the term $\f{\O^2}{-\partial_u r}$. By reformulating \eqref{conseq:r1} as
$$\partial_u \bigg(\log\f{\O^2}{-\partial_u r}\bigg)=\f{r}{\partial_u r}(\partial_u \phi)^2\leq 0,$$
we observe the following monotonicity
\begin{equation} \label{monotonicity}
0\leq \f{\O^2(u,v)}{-\partial_u r(u,v)}\leq \f{\O^2(u_{r_0}(v),v)}{-\partial_u r(u_{r_0}(v),v)} .
\end{equation}
Combining \eqref{monotonicity} with the estimate \eqref{eq:durest1prop2} in Proposition \ref{prop:metricest2}, and noting the fact that $1-\Lambda r^2 \leq 1$, we derive
\bes\begin{split}
&|\int^u_{u_{r_0}(v)}\bigg(\f{\O^2}{4\partial_u r}+\f12\bigg)(1-\Lambda r^2)(u',v)\cdot \partial_u r(u',v) du'|\\
=&|\int^{r(u,v)}_{r_0}\bigg(\f{\O^2}{4\partial_u r}+\f12\bigg)(1-\Lambda r^2)(u',v) dr(u',v) |\\
\leq&\int_{r(u,v)}^{r_0}\bigg(|\f{\O^2}{4\partial_u r}|(u',v)+\f12\bigg) dr(u',v)\leq\int_{r(u,v)}^{r_0}\bigg(|\f{\O^2}{4\partial_u r}|(u_{r_0}(v),v)+\f12\bigg) dr(u',v)\\
\leq&|\int_{r(u,v)}^{r_0}[1+C(\Delta_1',D_1,D_2,D_3,v_0,p) e^{-2pv}]\cdot dr(u',v)|\\
\leq&[1+C(\Delta_1',D_1,D_2,D_3,v_0,p) e^{-2pv}]\cdot [r_0-r(u,v)].
\end{split}\ees
Putting these estimates back to (\ref{r partial v r1}), for $r(u,v)\leq r_0$, we thus obtain
\bes\begin{split}
|(r\partial_v r+\frac{\Lambda}{6} P(r))(u,v)|\leq& C(\Delta_1',D_1,D_2,D_3,v_0,p) e^{-2pv}\\
&+[1+C(\Delta_1',D_1,D_2,D_3,v_0,p) e^{-2pv}]\cdot [r_0-r(u,v)].
\end{split}\ees
Hence, for sufficiently large $v$ and sufficiently small $r_0$, we prove
$$|r\partial_v r(u,v)+\frac{\Lambda}{6} P(r)|\leq \e .$$
Analogously, employing Proposition \ref{prop:durdvrr0}, $\left|r\partial_u r+\frac{\Lambda}{6} P(r)\right|$ obeys a similar estimate.
\end{proof}
Recall that in Proposition \ref{Prop 4.1}, for any $\t q=(u_{\t q}, v_{\t q})\in \mathcal{S}$ we have showed that $\lim_{u\rightarrow u_{\t q}} r\partial_v r(u, v_{\t q})$ and $\lim_{v\rightarrow v_{\t q}} r\partial_u r(u_{\t q}, v)$ exist. Now, according to Proposition \ref{1st partial r}, we further denote
\be\label{f q}
\lim_{u\rightarrow u_{\t q}} r\partial_v r(u, v_{\t q})=-M+f_2(v_{\t q}) \mbox{ and }  \lim_{v\rightarrow v_{\t q}} r\partial_u r(u_{\color{black} \t q}, v)=-M+f_1(u_{\t q}).
\ee
Repeating the proof in Proposition \ref{improved ru rv}, via estimates (\ref{rvr1}) (\ref{rur1}) and the triangle inequalities, we deduce the following proposition.
\begin{proposition} \label{global refined dr} 
For sufficiently small $r_0>0$ and any $(u,v)\in J^-(\t q)\cap\{r(u,v)\leq r_0\}$,  there holds
\begin{equation}\label{An Zhang r}
\begin{split}
|r\partial_u r(u,v)+M-f_1(u_{\t q})|\leq& 2r(u,v)^{\f{1}{8}},\\
|r\partial_v r(u,v)+M-f_2(v_{\t q})|\leq& 2r(u,v)^{\f{1}{8}},
\end{split}
\end{equation}
where $f_1(u)$ and $f_2(v)$ are continuous functions satisfying
\be\label{1st f1 f2}
|f_1(u_{\t q})|\leq \e +r_0^{\f{1}{8}}\leq 2\e , \quad |f_2(v_{\t q})|\leq \e +r_0^{\f{1}{8}}\leq 2\e.
\ee
\end{proposition}
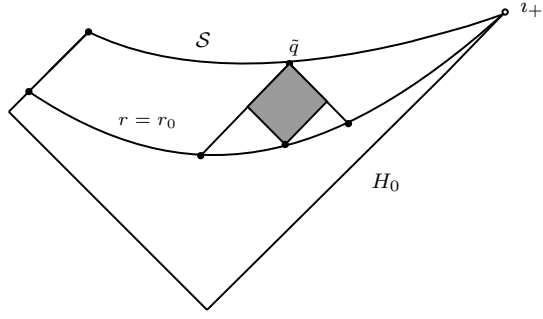
\begin{center}

\begin{tikzpicture}[x=0.75pt,y=0.75pt,yscale=-1,xscale=1]

\draw [line width=0.75]    (100,100.6) -- (200,200.6) ;
\draw [line width=0.75]    (200,200.6) -- (320.8,80.8) ;
\draw    (100,100.6) -- (120.26,80.24) -- (140.06,60.45) ;
\draw [color={rgb, 255:red, 0; green, 0; blue, 0 }  ,draw opacity=1 ][fill={rgb, 255:red, 0; green, 0; blue, 0 }  ,fill opacity=1 ][line width=0.75]    (350.34,50.26) ;
\draw [shift={(350.34,50.26)}, rotate = 0] [color={rgb, 255:red, 0; green, 0; blue, 0 }  ,draw opacity=1 ][line width=0.75]      (0, 0) circle [x radius= 1.34, y radius= 1.34]   ;
\draw    (140.23,60.29) .. controls (180.34,80) and (236.91,81.43) .. (304.69,64.72) ;
\draw [shift={(140.23,60.29)}, rotate = 26.17] [color={rgb, 255:red, 0; green, 0; blue, 0 }  ][fill={rgb, 255:red, 0; green, 0; blue, 0 }  ][line width=0.75]      (0, 0) circle [x radius= 1.34, y radius= 1.34]   ;
\draw    (110.13,90.42) .. controls (189.4,143.43) and (260.47,130.77) .. (349.16,51.45) ;
\draw [shift={(110.13,90.42)}, rotate = 33.77] [color={rgb, 255:red, 0; green, 0; blue, 0 }  ][fill={rgb, 255:red, 0; green, 0; blue, 0 }  ][line width=0.75]      (0, 0) circle [x radius= 1.34, y radius= 1.34]   ;
\draw    (140.23,60.29) -- (120.26,80.24) ;
\draw    (304.4,64.72) .. controls (330.34,57.71) and (325.74,59.48) .. (349.16,51.45) ;
\draw    (320.8,80.8) -- (349.16,51.45) ;
\draw    (241.55,76.31) -- (271.32,106.59) ;
\draw [shift={(271.32,106.59)}, rotate = 0] [color={rgb, 255:red, 0; green, 0; blue, 0 }  ][fill={rgb, 255:red, 0; green, 0; blue, 0 }  ][line width=0.75]      (0, 0) circle [x radius= 1.34, y radius= 1.34]   ;
\draw [shift={(241.55,76.31)}, rotate = 45.49] [color={rgb, 255:red, 0; green, 0; blue, 0 }  ][fill={rgb, 255:red, 0; green, 0; blue, 0 }  ][line width=0.75]      (0, 0) circle [x radius= 1.34, y radius= 1.34]   ;
\draw    (241.55,76.31) -- (196.8,122.77) ;
\draw [shift={(196.8,122.77)}, rotate = 133.93] [color={rgb, 255:red, 0; green, 0; blue, 0 }  ][fill={rgb, 255:red, 0; green, 0; blue, 0 }  ][line width=0.75]      (0, 0) circle [x radius= 1.34, y radius= 1.34]   ;
\draw  [fill={rgb, 255:red, 155; green, 155; blue, 155 }  ,fill opacity=1 ] (241.55,76.31) -- (260.52,95.38) -- (239.41,117.11) -- (220.44,98.04) -- cycle ;
\draw    (239.41,117.11) ;
\draw [shift={(239.41,117.11)}, rotate = 0] [color={rgb, 255:red, 0; green, 0; blue, 0 }  ][fill={rgb, 255:red, 0; green, 0; blue, 0 }  ][line width=0.75]      (0, 0) circle [x radius= 1.34, y radius= 1.34]   ;

\draw (153.67,101.88) node [anchor=north west][inner sep=0.75pt]  [font=\tiny]  {$r=r_{0}$};
\draw (281.38,130.24) node [anchor=north west][inner sep=0.75pt]  [font=\scriptsize]  {$H_{0}$};
\draw (356.5,41.63) node [anchor=north west][inner sep=0.75pt]  [font=\scriptsize]  {$i_{+}$};
\draw (192.9,58.91) node [anchor=north west][inner sep=0.75pt]  [font=\scriptsize]  {$\mathcal{S}$};
\draw (239.67,62.53) node [anchor=north west][inner sep=0.75pt]  [font=\tiny]  {$\tilde{q}$};

\end{tikzpicture}
\end{center}
\begin{remark}
In a fixed small region near the spacelike singularity the estimates in Proposition \ref{improved ru rv} and in Section \ref{pf of local bds} are (locally) uniformly dependent on the values of the local limits $C_1$, $C_2$. Whereas, in this section, the constants $C_1$, $C_2$ are proved to be close to $M$, which renders the estimate \eqref{An Zhang r} to be a global one.
\end{remark}
\begin{remark}
The $r^\f18$ remainder in \eqref{An Zhang r} will be sharpened to $r^\f23$ in Section \ref{AG sec10}.
\end{remark}

\subsection{Estimates for global coordinates $u$ and $v$}
Here we derive a useful estimate about the coordinates $u$ and $v$.
\begin{proposition}\label{v u global}
For any $0\leq r_1\leq r_0$ and $|u|\geq |u_1|+2v_{r=0}(u_1)$, along $r=r_1$, there holds
\begin{equation} \label{est v u global}
\begin{split}
-[u-u_1-v_{r=0}(u_1)](1-6M^{-1}\e)\leq v_{r_1}(u)\leq -[u-u_1-v_{r=0}(u_1)](1+6M^{-1}\e),
\end{split}
\end{equation}
where $v_{r_1}(u)$ is required to satisfy $r_1=r(u, v_{r_1}(u))$.
\end{proposition}
\begin{proof}
First, we differentiate the equation $r_1^2/2=r(u, v_{r_1}(u))^2/2$ along $r=r_1$ and obtain
\begin{equation*}
\begin{split}
0=&(r\partial_u r)(u, v_{r_1}(u))+v'_{r_1}(u)\cdot (r\partial_v r)(u, v_{r_1}(u))\\
=&[(r\partial_u r)(u, v_{r_1}(u))+M-f_1(u)]-M+f_1(u)\\
&+v'_{r_1}(u)\cdot [(r\partial_v r)(u, v_{r_1}(u))+M-f_2(v)]+v'_{r_1}(u)\cdot[-M+f_2(v)].
\end{split}
\end{equation*}
By \eqref{An Zhang r} and \eqref{1st f1 f2}, this implies
$$-1-5M^{-1}\e-5M^{-1}{r_1}^{\f{1}{8}}\leq v_{r_1}'(u)\leq -1+5M^{-1}\e+5M^{-1}{r_1}^{\f{1}{8}}.$$
Integrating this inequality, we get
\begin{equation}\label{vr1-v0}
\begin{split}
&(u-u_1)[-1+5M^{-1}\e+5M^{-1}{r_1}^{\f{1}{8}}]+v_{r_1}(u_1)-v_{r=0}(u_1)\\
\leq& v_{r_1}(u)-v_{r=0}(u_1)\\
&\leq (u-u_1)[-1-5M^{-1}\e-5M^{-1}{r_1}^{\f{1}{8}}]+v_{r_1}(u_1)-v_{r=0}(u_1).
\end{split}
\end{equation}
Along $u=u_1$, note that there holds
\begin{equation*}
\begin{split}
-r_1^2(u_1, v_{r_1}(u_1))=&r^2(u_1, v_{r=0}(u_1))-r^2_1(u_1, v_{r_1}(u_1))\\
=&\int_{v_{r_1}(u_1)}^{v_{r=0}(u_1)}\partial_v[r^2](u_1, v')dv'\\
=&2\int_{v_{r_1}(u_1)}^{v_{r=0}(u_1)}\{[r\partial_v r(u_1,v')+M-f_2(v')]-M+f_2(v')\} dv'.
\end{split}
\end{equation*}
Thus, by \eqref{An Zhang r} and \eqref{1st f1 f2}, the above formula yields 
$$|v_{r_1}(u_1)-v_{r=0}(u_1)|\leq M^{-1}r_1^2.$$ 
Substituting this estimate in \eqref{vr1-v0}, we deduce
\begin{equation*}
\begin{split}
&(u-u_1)[-1+5M^{-1}\e+5M^{-1}{r_1}^{\f{1}{8}}]-M^{-1}r_1^2+v_{r=0}(u_1)\\
\leq& v_{r_1}(u)\\
&\leq (u-u_1)[-1-5M^{-1}\e-5M^{-1}{r_1}^{\f{1}{8}}]+M^{-1}r_1^2+v_{r=0}(u_1).
\end{split}
\end{equation*}
By taking $|u|\geq |u_1|+2v_{r=0}(u_1)$ and requiring $v_{r=0}(u_1)$ to be sufficiently large, the desired estimate \eqref{est v u global} follows.
\end{proof}

\subsection{Estimates for $\partial_u \phi(u,v)$ and $\partial_v \phi(u,v)$.} \label{phi near s}
In this subsection, we derive upper and lower bounds for $\partial\phi$. To begin with, we recall the following inequality in \cite{AG}.
\begin{proposition}[Reverse Gr\"onwall inequality \cite{AG}]
\label{prop:revgron}
Let $t_0\in \R$ and $\psi,\beta: [t_0,\infty)\to \R$ be positive continuous functions. Set $A>0$ to be a positive constant and assume that $\p(t)$ satisfies:
\be \label{appendix 1}
\p(t)\geq A+\int_{t_0}^t \b(s)\p(s)ds \quad \mbox{ for any } \quad t\geq t_0.
\ee
Then it holds
$$\p(t)\geq A\cdot e^{\int_{t_0}^t \b(s)ds}.$$
\end{proposition} 
Now we state our estimates for $\partial \phi$.
\begin{proposition}\label{phi}
For $(u,v)\in \{|u|\geq |u_1|+2v_{r=0}(u_1),v\geq v_1\}\cap\{r(u,v)\leq r_0\}$, we have
\be\label{6.4}
\begin{split}
r^2|\partial_v \phi|(u,v)\leq \tilde{D}_1 e^{-p|u-u_1-v_{r=0}(u_1)|},\\
r^2|\partial_u \phi|(u,v)\leq \tilde{D}_2 e^{-p|u-u_1-v_{r=0}(u_1)|},\\
r^2\partial_v \phi(u,v)\geq D_1'e^{-q |u-u_1-v_{r=0}(u_1)|},\\
r^2\partial_u \phi(u,v)\geq D_2'e^{-q |u-u_1-v_{r=0}(u_1)|}.
\end{split}
\ee
where $\tilde{D}_1=\tilde{D}_2=8\D_1' r_S$ with the constant $\D_1'$ defined in Proposition \ref{prop:baimpphinoshift}, and $D_1'=D_2'=\frac{1}{2}C_{low}r_0(D_1-\epsilon)$ as defined in \eqref{r0 lowerbddvphi} and \eqref{r0 lowerbdduphi}.
\end{proposition}

\begin{proof}
We consider $(u,v)\in J^-(\t q)$ with $\t q \in \mathcal{S}$. According to (\ref{f q}),we denote the limits $C_1(\t q):=M-f_1(u_{\t q})$ and $C_2(\t q):=M-f_2(v_{\t q})$ which are positive constants. We then reformulate \eqref{propeq:phi1} and \eqref{propeq:phi2} as
\begin{equation}\label{weighted partial v phi}
\partial_u \bigg(C_1 r\partial_v \phi \bigg)=\f{C_1}{C_2}\cdot \f{1}{r}\cdot \f{-\partial_v r}{\partial_u r}\cdot \bigg(C_2 r\partial_u \phi\bigg)\cdot \partial_u r,
\end{equation}
\begin{equation}\label{weighted partial u phi}
\partial_v \bigg(C_2 r\partial_u \phi \bigg)=\f{C_2}{C_1}\cdot\f{1}{r}\cdot \f{-\partial_u r}{\partial_v r}\cdot \bigg(C_1 r\partial_v \phi\bigg)\cdot\partial_v r.
\end{equation}
Note that by \eqref{An Zhang r} and \eqref{1st f1 f2}, there exist $h_1(u,v), h_2(u,v)$ such that
\bes
\f{C_1}{C_2}\cdot \f{-\partial_v r}{\partial_u r}=-1-h_1, \quad \f{C_2}{C_1}\cdot \f{-\partial_u r}{\partial_v r}=-1-h_2,
\ees
with $h_1, h_2$ being of order $O({r^{\f{1}{8}}})$. Then we further rewrite (\ref{weighted partial v phi}) and (\ref{weighted partial u phi}) as
\begin{equation}\label{weighted partial v phi 1}
\partial_u \bigg(C_1 r\partial_v \phi \bigg)=-\f{1+h_1}{r}\cdot \bigg(C_2 r\partial_u \phi\bigg)\cdot \partial_u r,
\end{equation}
\begin{equation}\label{weighted partial u phi 1}
\partial_v \bigg(C_2 r\partial_u \phi \bigg)=-\f{1+h_2}{r}\cdot \bigg(C_1 r\partial_v \phi\bigg)\cdot\partial_v r.
\end{equation}
We are now ready to prove \eqref{6.4}. Denote $L_r$ to be the $r$-level sets within $J^-(\t q)$. We define
$$\tilde{\Psi}(r):=\max\{\sup_{P\in L_r} | C_2\cdot r\partial_u \phi|(P), \sup_{Q\in L_r} |C_1\cdot r\partial_v \phi|(Q)\}.$$
\begin{center}
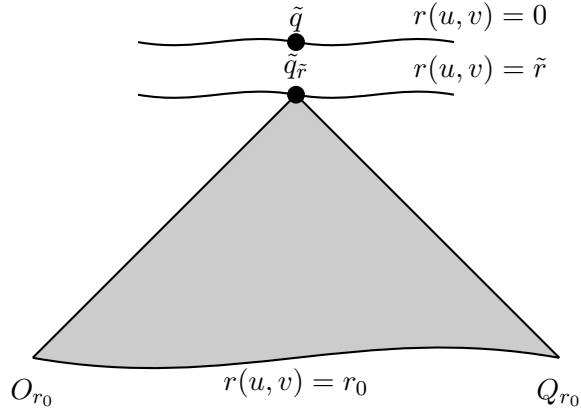
\begin{figure}[H]
\begin{minipage}[!t]{0.4\textwidth}
\begin{tikzpicture}[scale=0.7]
\filldraw[white, fill=gray!40] (0,0)--(-5,-5)to [out=-10, in=170] (5, -5)--(0,0);
\draw [white](4, 1)-- node[midway, sloped, above,black]{$r(u,v)=0$}(3, 1);
\draw [white](-2, 1)-- node[midway, sloped, above,black]{$\t q$}(2, 1);
\draw [white](-2, 0.1)-- node[midway, sloped, above,black]{$\t q_{\t r}$}(2, 0.1);

\draw [white](-6, -5.2)-- node[midway, sloped, below,black]{$O_{r_0}$}(-4, -5.2);

\draw [white](6, -5.2)-- node[midway, sloped, below,black]{$Q_{r_0}$}(4, -5.2);

\draw [white](4, 0)-- node[midway, sloped, above,black]{$r(u,v)=\t r$}(3, 0);

\draw[thick] (0,0)--(-5,-5);
\draw[thick] (0,0)--(5, -5);
\draw [thick] (-3,0) to [out=-10,in=170] (0,0);
\draw [thick] (0,0) to [out=-10,in=170] (3,0);
\draw [thick] (-3,1) to [out=-10,in=170] (0,1);
\draw [thick] (0,1) to [out=-10,in=170] (3,1);
\node[below] at (0, -5){$r(u,v)=r_0$};
\draw[thick] (-5, -5) to [out=-10, in=170] (5, -5);
\draw[fill] (0,0) circle [radius=0.15];
\draw[fill] (0,1) circle [radius=0.15];
\end{tikzpicture}
\end{minipage}
\hspace{0.05\textwidth}
\begin{minipage}[!t]{0.6\textwidth}
\end{minipage}
\caption{Region for estimates of $\partial_u \phi$ and $\partial_v \phi$.}
\end{figure}
\end{center}

Then for any $\t q_{\t r}\in J^-(\t q)\cap L_{\t r}$, via integrating (\ref{weighted partial v phi 1}) and \eqref{weighted partial u phi 1}, we have
\begin{equation}
\begin{split}
|C_1 r\partial_v \phi|(\t q_{\t r})\leq& \tilde{\Psi}(r_0)+ \int_{{u(Q_{r_0})}}^{u(\t q_{\t r})} -\f{1+h_1}{r}\cdot |C_2 r\partial_u \phi|\cdot \partial_u r \Bigr\vert_{v=v_{\t q_{\t r}}} \, du\\
=&\tilde{\Psi}(r_0)+\int_{r(Q_{r_0})}^{r(\t q_{\t r})}-\f{1+h_1}{r}\cdot |C_2 r\partial_u \phi| \Bigr\vert_{v=v_{\t q_{\t r}}} \, dr,\\
\end{split}
\end{equation}
and
\begin{equation}
|C_2 r\partial_u \phi|(\t q_{\t r})\leq \tilde{\Psi}(r_0)+\int_{r(O_{r_0})}^{r(\t q_{\t r})} -\f{1+h_2}{r}\cdot |C_1 r\partial_v\phi| \Bigr\vert_{u=u_{\t q_{\t r}}} \, dr.
\end{equation}
These two inequalities together yield   
$$\tilde{\Psi}(\t r)\leq \tilde{\Psi}(r_0)+\int_{r_0}^{\t r} -\f{1+\max\{h_1, h_2\}}{r}\cdot \tilde{\Psi}(r) \, dr.$$
Applying Gr\"onwall's inequality, we deduce that
\begin{align*}
\tilde{\Psi}(\tilde{r})\leq& \tilde{\Psi}(r_0)\times e^{\int_{\tilde{r}}^{r_0} \f{1+\max\{h_1, h_2\}}{r}} dr=\tilde{\Psi}(r_0)\times e^{-\ln \f{\tilde{r}}{r_0}+\int_{\tilde{r}}^{r_0} \f{\max\{h_1, h_2\}}{r} dr}\leq\f{2r_0\tilde{\Psi}(r_0)}{\tilde{r}}.
\end{align*}
This implies that for $(u,v)\in J^-(\t q)\cap\{r(u,v)\leq r_0\}$ 
\be\label{phi middle step}
C_2\cdot r^2|\partial_u \phi|(u,v)\leq 2r_0\tilde{\Psi}(r_0) \quad \text{and}\quad C_1\cdot r^2|\partial_v \phi|(u,v)\leq 2r_0\tilde{\Psi}(r_0).
\ee
We now apply the estimates obtained in Section \ref{Section7} to bound $r_0\tilde{\Psi}(r_0)$. By Proposition \ref{prop:baimpphinoshift}, along $r(u',v')=r_0$, there holds
\be
r_0^2|\partial_v\phi|(u_{r_0}(v'),v')\leq \frac{r_S}{2} \Delta_1' e^{-pv'}, \quad r_0^2|\partial_u \phi|(u_{r_0}(v'),v')\leq \frac{r_S}{2} \Delta_1' e^{-pv'}.
\ee
Then via Lemma \ref{lm:relatuvconstr} and Proposition \ref{v u global}, for $(u',v')$ satisfying $|u'|\geq |u_1|+2v_{r=0}(u_1)$, we derive 
$$|C_1\cdot r^2\partial_v\phi|(u_{r_0}(v'),v')\leq \frac{C_1 r_S}{2} \Delta_1' e^{-pv'}\leq C_1 r_S \Delta_1' e^{-p|u_{r_0}(v')-u_1-v_{r=0}(u_1)|} ,$$
$$|C_2\cdot r^2\partial_u\phi|(u_{r_0}(v'),v')\leq \frac{C_2 r_S}{2} \Delta_1' e^{-pv'}\leq C_2 r_S \Delta_1' e^{-p|u_{r_0}(v')-u_1-v_{r=0}(u_1)|}.$$
Notice that for $(u_{r_0}(v'), v')\in J^-\big((u,v)\big)$, we further have
$$e^{-p|u_{r_0}(v')-u_1-v_{r=0}(u_1)|}\leq e^{-p|u-u_1-v_{r=0}(u_1)|}.$$
Also note that $C_1(q)$ and $C_2(q)$ obey
$$(1-2\e)M\leq C_1(q)\leq (1+2\e)M,\quad (1-2\e)M\leq C_2(q)\leq (1+2\e)M.$$
Plugging these to (\ref{phi middle step}), we hence arrive at (\ref{6.4}), i.e.,
$$r^2|\partial_v \phi|(u,v)\leq\f{3(C_1+C_2)\cdot r_S\cdot\Delta_1'}{C_1}\cdot e^{-p|u-u_1-v_{r=0}(u_1)|}\leq\tilde{D}_1  e^{-p|u-u_1-v_{r=0}(u_1)|},$$
$$r^2|\partial_u \phi|(u,v)\leq\f{3(C_1+C_2)\cdot r_S\cdot\Delta_1'}{C_2}\cdot e^{-p|u-u_1-v_{r=0}(u_1)|}\leq\tilde{D}_2  e^{-p|u-u_1-v_{r=0}(u_1)|},$$
where $\tilde{D}_1=\tilde{D}_2=8\D_1' r_S$.

Next, we proceed to prove the lower bound estimates. Define
$$\Psi(r):=\min\{\inf_{P\in L_r} | C_2\cdot r\partial_u \phi|(P), \inf_{Q\in L_r} |C_1\cdot r\partial_v \phi|(Q)\}.$$
\noindent Integrating (\ref{weighted partial v phi 1}) and \eqref{weighted partial u phi 1}, we obtain
\begin{equation}
\begin{split}
C_1 r\partial_v \phi(\t q_{\t r})= C_1 r\partial_v \phi(Q_{r_0})+\int_{r(Q_{r_0})}^{r(\t q_{\t r})}-\f{1+h_1}{r}\cdot C_2 r\partial_u \phi \Bigr\vert_{v=v_{\t q_{\t r}}} \, dr,\\
C_2 r\partial_u \phi(\t q_{\t r})= C_2 r\partial_u \phi(O_{r_0})+\int_{r(O_{r_0})}^{\t r} -\f{1+h_2}{r}\cdot C_1 r\partial_v\phi \Bigr\vert_{u=u_{\t q_{\t r}}} \, dr.
\end{split}
\end{equation}
Combining these two inequalities and employing the reverse Gr\"onwall inequality (see Proposition \ref{prop:revgron}), we derive
\begin{align*}
\Psi(\tilde{r})\geq& \Psi(r_0)\times e^{\int_{\tilde{r}}^{r_0} \f{1+\min\{h_1,h_2\}}{r}} dr=\Psi(r_0)\times e^{-\ln \f{\tilde{r}}{r_0}+\int_{\tilde{r}}^{r_0} \f{1+\min\{h_1,h_2\}}{r} dr }\geq\f{r_0\Psi(r_0)}{2\tilde{r}}.
\end{align*}
Hence for $(u,v)\in J^-(\t q)\cap\{r(u,v)\leq r_0\}$, the following estimates hold
\be\label{Psi r tilde 2}
C_2\cdot r^2|\partial_u \phi|(u,v)\geq r_0\Psi(r_0)/2, \quad C_1 \cdot r^2|\partial_v \phi|(u,v)\geq r_0\Psi(r_0)/2.
\ee
Finally, to bound $r_0\Psi(r_0)$ from below, we employ Proposition \ref{prop:lowboundsposr} and recall that for $(u_{r_0}(v'), v')\in J^-\big((u,v)\big)$ it holds $e^{-qv'}\geq e^{-qv}$. Then along $r(u',v')=r_0$, we get
$$C_1\cdot r^2\partial_v\phi(u_{r_0}(v'),v')\geq C_1C_{low}r_0(D_1-\epsilon) e^{-qv},$$
$$C_2\cdot r^2\partial_v\phi(u_{r_0}(v'),v')\geq C_2 C_{low}r_0(D_1-\epsilon) e^{-qv}.$$
Substituting these estimates in (\ref{Psi r tilde 2}), by Proposition \ref{v u global}, we then conclude that
\begin{align*}
r^2\partial_v \phi(u,v)\geq&D_1'e^{-q |u-u_1-v_{r=0}(u_1)|},\\
r^2\partial_u \phi(u,v)\geq&D_2'e^{-q|u-u_1-v_{r=0}(u_1)|}.
\end{align*}
Here $D_1'=D_2'=\frac{1}{2}C_{low}r_0(D_1-\epsilon)$. This proves the current proposition.
\end{proof}

\subsection{Optimal estimates for $r\partial_v r(u,v)$ and $r\partial_u r(u,v)$} \label{AG sec10}
In line with our optimal estimates shown in Section \ref{subsec sharp dr}, we derive accurate asymptotic behaviours of $r\partial_u r(u,v)$ and $r\partial_v r(u,v)$.
\begin{proposition}\label{partial r}
For any point $(u,v)$, which satisfies $r(u,v)\leq r_0$, $|u|\geq |u_1|+2v_{r=0}(u_1)$ and $v$ being sufficiently large, the following estimates hold
\begin{equation}\label{addition dur}
|r\partial_u r(u,v)+M|\lesssim r(u,v)^{\f{2}{3}}+e^{-pv_{\t q}},
\end{equation}
\begin{equation}\label{addition dvr}
|r\partial_v r(u,v)+M|\lesssim r(u,v)^{\f{2}{3}}+e^{-p|u-u_1-v_{r=0}(u_1)|}.
\end{equation}
\end{proposition}
\begin{proof}
According to the argument in Section \ref{subsec sharp dr}, for any $(u,v)$ in the shadowed region below, we have the following estimates for $r\partial_v r(u,v)$ and $r\partial_u r(u,v)$:
\begin{equation}
|r\partial_u r(u,v)-r\partial_u r(u_{\t q}, v_{\t q})|\lesssim r(u,v)^{\f23},
\end{equation}
\begin{equation}
|r\partial_v r(u,v)-r\partial_v r(u_{\t q}, v_{\t q})|\lesssim r(u,v)^{\f23}.
\end{equation}
\begin{center}
\begin{figure}[ht]

\begin{tikzpicture}[x=0.75pt,y=0.75pt,yscale=-1,xscale=1]

\draw [line width=0.75]    (101,113.1) -- (211.71,233) ;
\draw [line width=0.75]    (211.71,233) -- (345.45,89.36) ;
\draw    (101,113.1) -- (123.43,88.69) -- (145.36,64.96) ;
\draw [color={rgb, 255:red, 0; green, 0; blue, 0 }  ,draw opacity=1 ][fill={rgb, 255:red, 0; green, 0; blue, 0 }  ,fill opacity=1 ][line width=0.75]    (378.16,52.74) ;
\draw [shift={(378.16,52.74)}, rotate = 0] [color={rgb, 255:red, 0; green, 0; blue, 0 }  ,draw opacity=1 ][line width=0.75]      (0, 0) circle [x radius= 1.34, y radius= 1.34]   ;
\draw    (145.54,64.76) .. controls (189.95,88.4) and (252.58,90.11) .. (327.61,70.08) ;
\draw [shift={(145.54,64.76)}, rotate = 28.02] [color={rgb, 255:red, 0; green, 0; blue, 0 }  ][fill={rgb, 255:red, 0; green, 0; blue, 0 }  ][line width=0.75]      (0, 0) circle [x radius= 1.34, y radius= 1.34]   ;
\draw    (112.22,100.9) .. controls (189.5,125.29) and (278.66,149.27) .. (376.85,54.18) ;
\draw    (145.54,64.76) -- (123.43,88.69) ;
\draw    (327.29,70.08) .. controls (356.02,61.68) and (350.91,63.8) .. (376.85,54.18) ;
\draw    (345.45,89.36) -- (376.85,54.18) ;
\draw    (218.43,85.56) -- (250.78,120.59) ;
\draw [shift={(250.78,120.59)}, rotate = 0] [color={rgb, 255:red, 0; green, 0; blue, 0 }  ][fill={rgb, 255:red, 0; green, 0; blue, 0 }  ][line width=0.75]      (0, 0) circle [x radius= 1.34, y radius= 1.34]   ;
\draw [shift={(218.43,85.56)}, rotate = 47.28] [color={rgb, 255:red, 0; green, 0; blue, 0 }  ][fill={rgb, 255:red, 0; green, 0; blue, 0 }  ][line width=0.75]      (0, 0) circle [x radius= 1.34, y radius= 1.34]   ;
\draw    (218.43,85.56) -- (186.97,119.64) ;
\draw [shift={(186.97,119.64)}, rotate = 0] [color={rgb, 255:red, 0; green, 0; blue, 0 }  ][fill={rgb, 255:red, 0; green, 0; blue, 0 }  ][line width=0.75]      (0, 0) circle [x radius= 1.34, y radius= 1.34]   ;
\draw  [fill={rgb, 255:red, 155; green, 155; blue, 155 }  ,fill opacity=1 ] (218.43,85.56) -- (232.97,101.12) -- (213.52,122.44) -- (198.98,106.88) -- cycle ;
\draw    (213.52,122.44) ;
\draw [shift={(213.52,122.44)}, rotate = 0] [color={rgb, 255:red, 0; green, 0; blue, 0 }  ][fill={rgb, 255:red, 0; green, 0; blue, 0 }  ][line width=0.75]      (0, 0) circle [x radius= 1.34, y radius= 1.34]   ;
\draw    (316.59,73.97) -- (331.61,90.23) ;
\draw [shift={(331.61,90.23)}, rotate = 0] [color={rgb, 255:red, 0; green, 0; blue, 0 }  ][fill={rgb, 255:red, 0; green, 0; blue, 0 }  ][line width=0.75]      (0, 0) circle [x radius= 1.34, y radius= 1.34]   ;
\draw    (316.59,73.97) -- (278.07,115.7) ;
\draw [shift={(278.07,115.7)}, rotate = 0] [color={rgb, 255:red, 0; green, 0; blue, 0 }  ][fill={rgb, 255:red, 0; green, 0; blue, 0 }  ][line width=0.75]      (0, 0) circle [x radius= 1.34, y radius= 1.34]   ;
\draw [shift={(316.59,73.97)}, rotate = 132.72] [color={rgb, 255:red, 0; green, 0; blue, 0 }  ][fill={rgb, 255:red, 0; green, 0; blue, 0 }  ][line width=0.75]      (0, 0) circle [x radius= 1.34, y radius= 1.34]   ;
\draw  [fill={rgb, 255:red, 155; green, 155; blue, 155 }  ,fill opacity=1 ] (316.59,73.97) -- (326.84,85.17) -- (311.84,101.28) -- (301.59,90.08) -- cycle ;
\draw    (310.73,101.28) -- (311.84,101.28) ;
\draw [shift={(311.84,101.28)}, rotate = 0] [color={rgb, 255:red, 0; green, 0; blue, 0 }  ][fill={rgb, 255:red, 0; green, 0; blue, 0 }  ][line width=0.75]      (0, 0) circle [x radius= 1.34, y radius= 1.34]   ;

\draw (135.88,98.49) node [anchor=north west][inner sep=0.75pt]  [font=\tiny]  {$r=r_{0}$};
\draw (99.12,90.63) node [anchor=north west][inner sep=0.75pt]  [font=\tiny,rotate=-315]  {$u=u_{1}$};
\draw (152.37,178.55) node [anchor=north west][inner sep=0.75pt]  [font=\tiny,rotate=-45]  {$v=v_{1}$};
\draw (299.33,159.07) node [anchor=north west][inner sep=0.75pt]  [font=\scriptsize]  {$H_{0}$};
\draw (385.73,43.91) node [anchor=north west][inner sep=0.75pt]  [font=\scriptsize]  {$i_{+}$};
\draw (129.63,45.64) node [anchor=north west][inner sep=0.75pt]  [font=\tiny]  {$( u_{1} ,v_{r=0}( u_{1}))$};
\draw (257.27,54.97) node [anchor=north west][inner sep=0.75pt]  [font=\scriptsize]  {$\mathcal{S}$};
\draw (311.76,58.31) node [anchor=north west][inner sep=0.75pt]  [font=\tiny]  {$\tilde{q} '$};
\draw (274.66,120.29) node [anchor=north west][inner sep=0.75pt]  [font=\tiny]  {$\tilde{q} '_{1}$};
\draw (329.58,72.79) node [anchor=north west][inner sep=0.75pt]  [font=\tiny]  {$\tilde{q} '_{2}$};
\draw (217.04,70.12) node [anchor=north west][inner sep=0.75pt]  [font=\tiny]  {$\tilde{q}$};
\draw (179.42,124.9) node [anchor=north west][inner sep=0.75pt]  [font=\tiny]  {$\tilde{q}_{1}$};
\draw (242.66,127.7) node [anchor=north west][inner sep=0.75pt]  [font=\tiny]  {$\tilde{q}_{2}$};

\end{tikzpicture}
\hspace{0.05\textwidth}
\caption{Asymptotic behaviours for $r\partial_u r$ and $r\partial_v r$.}
\label{asymp behav ru rv}
\end{figure}
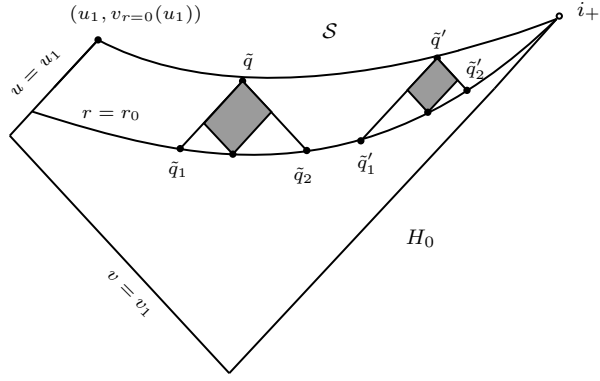
\end{center}

\noindent The task now is to give an optimal description of $r\partial_u r(u_{\t q}, v_{\t q})$ and $r\partial_v r(u_{\t q}, v_{\t q})$. We first rewrite equation \eqref{propeq:r1} as
\bes
\partial_u(r\partial_v r+\frac{\Lambda}{6} P(r))=\bigg(-\f14\O^2-\f12\partial_u r\bigg)(1-\Lambda r^2).
\ees
Integrating this equation, we obtain
\be\label{r partial v r}
(r\partial_v r+\frac{\Lambda}{6} P(r))(u,v)=(r\partial_v r+\frac{\Lambda}{6}P(r))(u_{r_0}(v),v)-\int^u_{u_{r_0}(v)}\bigg(\f{\O^2}{4\partial_u r}+\f12\bigg)(1-\Lambda r^2) \partial_u r du.
\ee
To control the integration term, we first estimate $\f{\O^2}{-\partial_u r}$ and we use the Raychauduri equation \eqref{conseq:r1} in the following form
$$\partial_u \bigg(\log\f{\O^2}{-\partial_u r}\bigg)=\f{r}{\partial_u r}(\partial_u \phi)^2.$$
Integrating this identity, it holds that
\be \label{rach rewrite}
\f{\O^2(u,v)}{-\partial_u r(u,v)}=\f{\O^2(u_{r_0}(v),v)}{-\partial_u r(u_{r_0}(v),v)}\cdot \exp\bigg(\int_{r_0}^{r(u,v)}\f{r^3 (\partial_u \phi)^2}{(r\partial_u r)^2}dr  \bigg).
\ee
For data along $r=r_0$, using Proposition \ref{prop:metricest2}, it holds
$$\left|- \Omega^{-2}\partial_u r-\frac{1}{2}\right|(u_{r_0}(v),v)\leq C(\Delta_1',D_1,D_2,D_3,v_0,p) e^{-2pv}.$$
For $v$ sufficiently large, we can write
$$\f{\O^2(u_{r_0}(v),v)}{4\partial_u r(u_{r_0}(v),v)}=-\f12+s(u,v)\cdot e^{-pv} \quad \text{with}\ 0\leq |s(u,v)|\ll 1.$$
For the exponential term on the RHS of \eqref{rach rewrite}, employing Proposition \ref{global refined dr}, Proposition \ref{v u global} and Proposition \ref{phi}, for $v$ sufficiently large, we have 
$$\exp\bigg(\int_{r_0}^{r(u,v)}\f{r^3 (\partial_u \phi)^2}{(r\partial_u r)^2}dr  \bigg)= \exp\bigg({2\t D_2^2\cdot l(u,v)\cdot e^{-2pv}}\cdot\ln\f{r(u,v)}{r_0}\bigg)=\bigg(\f{r(u,v)}{r_0}\bigg)^{{2\t D_2^2\cdot l(u,v)\cdot e^{-2pv}}}.$$
with $0\leq|l(u,v)|\lesssim 1$. Back to \eqref{rach rewrite}, we thus obtain
\begin{align*}
&\f{\O^2(u,v)}{4\partial_u r(u,v)}+\f12=\f{\O^2(u_{r_0}(v),v)}{4\partial_u r(u_{r_0}(v),v)}\cdot \exp\bigg(\int_{r_0}^{r(u,v)}\f{r^3 (\partial_u \phi)^2}{(r\partial_u r)^2}dr  \bigg)+\f12\\
=&\bigg(-\f12+s(u,v)\cdot e^{-pv}\bigg)\cdot \bigg[\bigg(\f{r(u,v)}{r_0}\bigg)^{{2\t D_2^2\cdot l(u,v)\cdot e^{-2pv}}}-1\bigg]+s(u,v)\cdot e^{-pv}.
\end{align*}
Therefore, for the integral on the RHS of (\ref{r partial v r}), we prove
\begin{align*}
&\Big|\int^u_{u_{r_0}(v)}\bigg(\f{\O^2}{4\partial_u r}+\f12\bigg)(1-\Lambda r^2) \partial_u r du\Big|=\Big|\int^{r(u,v)}_{r_0}\bigg(\f{\O^2}{4\partial_u r}+\f12\bigg)(1-\Lambda r^2) dr \Big|\\
\leq&\Big|\int^{r(u,v)}_{r_0} \bigg[\bigg(\f{r(u',v)}{r_0}\bigg)^{{2\t D_2^2\cdot l(u',v)\cdot e^{-2pv}}}-1\bigg] dr\,\Big|+\sup_{u_{r_0}(v)\leq u'\leq u}|s(u',v)|\cdot e^{-pv}\cdot [r_0-r(u,v)].
 \end{align*}
\noindent Denote $x=\f{r(u',v)}{r_0}$ and note that $|{2\t D_2^2 \cdot l(u,v)\cdot e^{-2pv}}|\leq {3\t D_2^2 \cdot e^{-2pv}}$. By decomposing $l(u,v)$ into negative part and non-negative part, we now arrive at
\begin{align*}
&\Big|\int^{r(u,v)}_{r_0} \bigg[\bigg(\f{r(u',v)}{r_0}\bigg)^{{2\t D^2_1\cdot l(u',v)\cdot e^{-2pv}}}-1\bigg] dr\Big|\\
\leq &\int^{r(u,v)}_{r_0} \bigg[\bigg(\f{r(u',v)}{r_0}\bigg)^{{-3\t D^2_1\cdot e^{-2pv}}}-1\bigg] dr+\int^{r(u,v)}_{r_0} \bigg[-\bigg(\f{r(u',v)}{r_0}\bigg)^{{3\t D_2^2\cdot e^{-2pv}}}+1\bigg] dr\\
=&r_0\int^{\f{r(u,v)}{r_0}}_{1} (x^{{-3\t D_2^2\cdot e^{-2pv}}}-1) dx-r_0\int^{\f{r(u,v)}{r_0}}_{1} (x^{{3\t D_2^2\cdot e^{-2pv}}}-1) dx\\
=&r_0\cdot \bigg(\f{x^{1-3\t D_2^2\cdot e^{-2pv}}}{1-3\t D_2^2\cdot e^{-2pv}} \vert^{x=\f{r(u,v)}{r_0}}_{x=1}  \bigg)-r(u,v)+r_0-r_0\cdot \bigg(\f{x^{1+3\t D_2^2\cdot e^{-2pv}}}{1+3\t D^2_1\cdot e^{-2pv}}\vert^{x=\f{r(u,v)}{r_0}}_{x=1}  \bigg)+r(u,v)-r_0\\
=&r^{\f12}\cdot r_0^{\f12}\cdot \bigg(\f{[\f{r(u,v)}{r_0}]^{\f12-3\t D_2^2\cdot e^{-2pv}}}{1{-3\t D_2^2\cdot e^{-2pv}}}  \bigg)-r^{\f12}\cdot r_0^{\f12}\cdot \bigg(\f{[\f{r(u,v)}{r_0}]^{\f12+3\t D_2^2\cdot e^{-2pv}}}{1+{3\t D_2^2\cdot e^{-2pv}}}  \bigg)+r_0\cdot \f{-6\t D_2^2\cdot e^{-2pv}}{(1{-3\t D_2^2\cdot e^{-2pv}})(1+{3\t D_2^2\cdot e^{-2pv}})}.
\end{align*}
This implies
\begin{equation*}
\Big|\int^{r(u,v)}_{r_0} \bigg[\bigg(\f{r(u',v)}{r_0}\bigg)^{{2\t D^2_2\cdot l(u',v)\cdot e^{-2pv}}}-1\bigg] dr\Big|\leq 2r_0^{\f12}\cdot r(u,v)^{\f12}+7r_0\t D_2^2 e^{-2pv}.
\end{equation*}
\noindent Substituting the above estimate to (\ref{r partial v r}) and using \eqref{eq:dvrestprop2}, we hence obtain 
\bes
\begin{split}
&\left|\bigg(r\partial_v r+\frac{\Lambda}{6} P(r)\bigg)(u,v)\right|\\
=&\left|\bigg(r\partial_v r+\frac{\Lambda}{6} P(r)\bigg)(u_{r_0}(v),v)-\int^u_{u_{r_0}(v)}\bigg(\f{\O^2}{4\partial_u r}+\f12\bigg)(1-\Lambda r^2) \partial_u r du\right|\\
\leq &C(\Delta_1',D_1,D_2,D_3,v_0,p) e^{-2pv}+2r_0^{\f12}\cdot r^{\f12}+7r_0 \t D_2^2\cdot e^{-2pv}+r_0 \cdot e^{-pv}.
\end{split}
\ees
In particular, at $\t q=(u_{\t q}, v_{\t q})\in \mathcal{S}$ with $r(u_{\t q}, v_{\t q})=0$, for sufficiently large $v_{\t q}$, there holds
\begin{equation} \label{uniform rdr}
|(r\partial_v r)(u_{\t q}, v_{\t q})+M|\lesssim e^{-pv_{\t q}}.
\end{equation}
This completes the proof of (\ref{addition dvr}). Estimate (\ref{addition dur}) can be derived in the same way.
\end{proof}
The above proposition and estimates \eqref{addition dur}-\eqref{addition dvr} immediately imply
\begin{corollary}
Along $\mathcal{S}$ towards the timelike infinity $i^+$, both $r\partial_u r$ and $r\partial_v r$ converge to their corresponding Schwarzschild-de Sitter value $-M$ at an inverse exponential rate $e^{-pv}$, i.e., as $v\to \infty$,
\begin{align*}
 r\partial_u r|_{\mathcal{S}}(v)\rightarrow -M,\\
 r\partial_u r|_{\mathcal{S}}(v)\rightarrow -M.
\end{align*}
\end{corollary}

\subsection{Estimate for $\O^2(u,v)$} With the aid of estimates for $\partial_u \phi$ and $\partial_v \phi$ in Section \ref{phi near s}, we now derive lower and upper bounds for $\O^2(u,v)$ in this subsection. We begin with the derivatives of $\log(r\O^2)$.
\begin{proposition}\label{Omega 02}
For $(u,v)\in \left\{|u|\geq |u_1|+2v_{r=0}(u_1), v\geq \max\{v_1, \t v_1\}\right\}\cap \left\{r(u,v)\leq r_0\right\}$, it hold
\be\label{rOmega2 v}
 \f{{D}'_1 {D}'_2 }{3M r^2}e^{-2qv}-\f{2}{r(u,v)}\leq -\partial_v\log(r\O^2)(u,v)\leq \f{3\t D_1 \t D_2  }{M r^2}e^{-2pv}+\f{2}{r(u,v)},
\ee
\be\label{rOmega2 u}
\f{{D}'_1 {D}'_2 }{3Mr^2}e^{-2qv}-\f{2}{r(u,v)}\leq -\partial_u\log(r\O^2)(u,v)\leq \f{3\t D_1 \t D_2  }{M r^2}e^{-2pv}+\f{2}{r(u,v)},
\ee
where $\t D_1, \t D_2, D_1', D_2'$ are defined in Proposition \ref{phi}.
\end{proposition}

\begin{proof}
We first integrate equation \eqref{propeq:Omegarescale1} and obtain
$$-\partial_v \log(r\O^2)(u,v)=-\partial_v \log(r\O^2)(u_{r_0}(v),v)+\int_{u_{r_0}(v)}^u\bigg(-\f{\O^2}{4r^2}(1+\Lambda r^2)+2\partial_u \phi\partial_v \phi\bigg)(u',v)du'.$$
Along $r=r_0$, for $v\geq \t v_1$, according to \eqref{additional v}, it holds $$|\partial_v\log(r\O^2)(u_{r_0}(v),v)|\leq\f{1}{r_0}.$$ 
Then, via Proposition \ref{global refined dr}, Proposition \ref{v u global} and Proposition \ref{phi}, we deduce
\begin{align*}
\int_{u_{r_0}(v)}^u 2|\partial_u \phi\cdot&\partial_v \phi|(u',v) du'\leq \int_{u_{r_0}(v)}^u \f{3\t D_1 \t D_2 e^{-2pv}}{r'^4(u',v)}du'\\
=&e^{-2pv}\int_{r_0}^{r(u,v)} \f{1}{r'^3}\cdot \f{3\t D_1 \t D_2}{r'\partial_u r}dr' \leq \f{3\t D_1 \t D_2 }{M r^2  }e^{-2pv}
\end{align*}
and the lower bound estimate
\begin{align*}
\int_{u_{r_0}(v)}^u 2\partial_u \phi\cdot\partial_v \phi&(u',v) du'\geq \int_{u_{r_0}(v)}^u \f{{D}'_1{D}'_2 e^{-2qv}}{r'^4(u',v)}du'\\
=&e^{-2qv}\int_{r_0}^{r(u,v)} \f{1}{r'^3}\cdot \f{{D}'_1 {D}'_2 }{r'\partial_u r}dr' \geq \f{{D}'_1 {D}'_2 }{3M r^2 }e^{-2qv}.
\end{align*}
\noindent Noting that $\partial_u (\f{\O^2}{-\partial_u r})\leq 0$, together with \eqref{eq:durest1prop2}, we get
$$|\int_{u_{r_0}(v)}^u -\f{\O^2}{4r^2}(1+\Lambda r^2) du'|=\int_{u_{r_0}(v)}^u \f{\O^2}{-\partial_u r}\cdot \f{1}{4r^2}(1+\Lambda r^2)\cdot (-\partial_u r)du'\leq -\int_{r_0}^r\f{3}{4r^2}dr\leq \f{1}{r}.$$
Combining all the above estimates, we conclude that
\bes
 \f{{D}'_1 {D}'_2 }{3M r^2}e^{-2qv}-\f{2}{r(u,v)}\leq -\partial_v\log(r\O^2)(u,v)\leq \f{3\t D_1 \t D_2   }{Mr^2}e^{-2pv}+\f{2}{r(u,v)}.
\ees
The estimate (\ref{rOmega2 u}) can be proved in the same manner.
\end{proof}

The estimate of $r\O^2(u,v)$ then follows from Proposition \ref{Omega 02}.
\begin{proposition}\label{Omega 03}
For $(u,v)\in \left\{|u|\geq |u_1|+2v_{r=0}(u_1), v\geq \max\{v_1, \t v_1\}\right\}\cap \left\{r(u,v)\leq r_0\right\}$, there holds
$$\f{2M-r_0}{4}\cdot\f{r_0^{-{\f{6 \t D_1 \t D_2}{M^2(1-\e)}\cdot e^{-2p|u-u_1-v_{r=0}(u_1)|}}}}{r(u,v)^{-{\f{6 \t D_1 \t D_2}{M^2(1-\e)}\cdot e^{-2p|u-u_1-v_{r=0}(u_1)|}}}} \leq r\O^2(u,v)\leq 4\cdot(2M-r_0)\cdot \f{r_0^{-{\f{D_1' D_2'}{6M^2(1+\e)}e^{-2q|u-u_1-v_{r=0}(u_1)|}}}}{r(u,v)^{-{\f{D_1' D_2'}{6M^2(1+\e)}e^{-2q|u-u_1-v_{r=0}(u_1)|}}}}.$$
\end{proposition}
\begin{proof}
By Proposition \ref{Omega 02}, we have
\bes\begin{split}
&\log(r\O^2)(u,v)=\log(r\O^2)(u,v_{r_0}(u))+\int^v_{v_{r_0}(u)}\partial_v \log(r\O^2)(u,v')dv'\\
\geq&\log(r\O^2)(u,v_{r_0}(u))+\int^v_{v_{r_0}(u)}-\f{3\t D_1 \t D_2 }{ Mr(u,v')^2}e^{-2pv}-\f{2}{r(u,v')}dv'\\
\geq&\log(r\O^2)(u,v_{r_0}(u))+\int^v_{v_{r_0}(u)}-\f{6\t D_1 \t D_2 }{Mr(u,v')^2} e^{-2p|{u-u_1-v_{r=0}(u_1)}|}dv'-\f{2r_0}{M(1-\e)}\\
=&\log(r\O^2)(u,v_{r_0}(u))+\int^v_{v_{r_0}(u)}-\f{6\t D_1 \t D_2 }{Mr(r\partial_v r)}e^{-2p|u-u_1-v_{r=0}(u_1)|}\cdot \partial_v r dv'-\f{2r_0}{M(1-\e)}\\
\geq&\log(r\O^2)(u,v_{r_0}(u))+\int^{r(u,v)}_{r_0}\f{1}{ r}\cdot\f{6\t D_1 \t D_2}{M^2(1-\e)}\cdot e^{-2p|u-u_1-v_{r=0}(u_1)|}dr-\f{2r_0}{M(1-\e)}\\
=&\log(r\O^2)(u,v_{r_0}(u))+[\ln(\f{r}{r_0})]\cdot\f{6\t D_1 \t D_2}{M^2(1-\e)}\cdot e^{-2p|u-u_1-v_{r=0}(u_1)|}-\f{2r_0}{M(1-\e)}.
\end{split}\ees
\noindent When $r_0$ is sufficiently small, it then holds that
$$r\O^2(u,v)\geq \f12 (r\O^2)(u,v_{r_0}(u))\cdot [\f{r(u,v)}{r_0}]^{\f{6\t D_1 \t D_2}{M^2(1-\e)}\cdot e^{-2p|u-u_1-v_{r=0}(u_1)|}}.$$
Noting that by \eqref{eq:rescaledOmegaestprop}, $r\O^2(u,v_{r_0}(u))\geq \f{2M-r_0}{2},$ we thus obtain
$$r\O^2(u,v)\geq \f{2M-r_0}{4}\cdot [\f{r(u,v)}{r_0}]^{\f{6\t D_1 \t D_2}{M^2(1-\e)}\cdot e^{-2p|u-u_1-v_{r=0}(u_1)|}}.$$
The upper bound of $r\O^2(u,v)$ can be obtained in the same fashion. 
\end{proof}

\section{Blow-up of Kretschmann scalar (Theorem \ref{main thm})} 
In this section, we finish the proof of Theorem \ref{main thm} by deriving the accurate blow-up rate for the Kretschmann scalar near the timelike infinity. 

\subsection{Mass inflation}
Recall Proposition \ref{Omega 03}. A direct corollary of it is the \underline{\textit{mass inflation}} phenomena, i.e., the blow-up of the Hawking mass $m$ along the spacelike singularity $\mathcal{S}$. \begin{proposition}\label{mass inflation}
For $(u,v)\in \left\{|u|\geq |u_1|+2v_{r=0}(u_1), v\geq \max\{v_1, \t v_1\}\right\}\cap \left\{r(u,v)\leq r_0\right\}$ with $r_0$ being sufficiently small, we have the following lower and upper bound estimates for the Hawking mass
\be
\begin{split}
\f{M}{8}\cdot{\color{black}\Big[}\f{r_0}{r(u,v)}{\color{black}\Big]}^{\f{D_1' D_2'}{6M^2(1+\e)}\cdot e^{-2q|u-u_1-v_{r=0}(u_1)|}}\leq m(u,v)\leq \f{r(u,v)}{2}+8M\cdot \Big[\f{r_0}{r(u,v)}\Big]^{\f{6\t D_1 \t D_2}{M^2(1-\e)}\cdot e^{-2p|u-u_1-v_{r=0}(u_1)|}}.
\end{split}
\ee
Here $D_1'=D_2'=\frac{1}{2}C_{low}r_0(D_1-\epsilon)$ as defined in \eqref{r0 lowerbddvphi} and \eqref{r0 lowerbdduphi}.
\end{proposition}
\begin{proof}
By the definition of Hawking mass in \eqref{Hawking mass}, applying Proposition \ref{global refined dr} and Proposition \ref{Omega 03}, there hold
\begin{equation}\label{mass inflation lower bound}
\begin{split}
m(u,v)\geq&\f{r(u,v)}{2}\bigg(1+\f{2}{4\cdot 2M}\cdot [\f{r_0}{r(u,v)}]^{\f{D_1' D_2'}{6M^2[1+\e]}\cdot e^{-2q|u-u_1-v_{r=0}(u_1)|}}\cdot \f{M^2}{r(u,v)}\bigg)\\
\geq&\f{M}{8}\cdot [\f{r_0}{r(u,v)}]^{\f{D_1' D_2'}{6M^2(1+\e)}\cdot e^{-2q|u-u_1-v_{r=0}(u_1)|}},
\end{split}
\end{equation}
and
\begin{align*}
m(u,v)\leq&\f{r(u,v)}{2}\bigg(1+\f{2\cdot 4\cdot 4}{2M}\cdot [\f{r_0}{r(u,v)}]^{\f{6\t D_1 \t D_2}{M^2(1-\e)}\cdot e^{-2p|u-u_1-v_{r=0}(u_1)|}}\cdot \f{M^2}{r(u,v)}\bigg)\\
\leq&\f{r(u,v)}{2}+8M\cdot [\f{r_0}{r(u,v)}]^{\f{6\t D_1 \t D_2}{M^2(1-\e)}\cdot e^{-2p|u-u_1-v_{r=0}(u_1)|}}.
\end{align*}
\end{proof}

\subsection{Estimates for the Kretschmann scalar}
In this section, we derive upper and lower bound estimates for the Kretschmann scalar in Theorem \ref{main thm}. The proof relies on Proposition \ref{mass inflation} and the estimates near $\mathcal{S}$ obtained in Section \ref{sec:estnearsing}. As a conclusion, we have
\begin{proposition}\label{thm 8.1}
For $(u,v)\in \left\{|u|\geq |u_1|+2v_{r=0}(u_1), v\geq \max\{v_1, \t v_1\}\right\}\cap \left\{r(u,v)\leq r_0\right\}$, we have that the Kretschmann scalar obeys
$$\f{c(M)}{r^{6+\f{D_1' D_2'}{3M^2(1+\e)}\cdot e^{-2q}|u-u_1-v_{r=0}(u_1)|}}\leq R^{\a\b\mu\nu}R_{\a\b\mu\nu}(u,v) \leq \f{C(M)}{r(u,v)^{6+\f{24 \t D_1 \t D_2}{M^2(1-\e)}\cdot e^{-2p|u-u_1-v_{r=0}(u_1)|}}},$$
where $c(M), C(M)$ are constants depending only on $M$, and $\t D_1, \t D_2, D_1', D_2'$ are defined in Proposition \ref{phi}.
\end{proposition}

\begin{proof}
First, for the lower bound of the Kretschmann scalar, by \eqref{mass inequality} and \eqref{mass inflation lower bound}, we deduce
\begin{align*}
R^{\a\b\mu\nu}R_{\a\b\mu\nu}(u,v)\geq& \f{\big(M\cdot r_0^{\f{D_1' D_2'}{6M^2(1+\e)}\cdot e^{-2q|u-u_1-v_{r=0}(u_1)|}}\big)^2}{{4}{r^{6+\f{D_1' D_2'}{3M^2(1+\e)}\cdot e^{-2q|u-u_1-v_{r=0}(u_1)|}}}}\geq \f{c(M)}{r^{6+\f{D_1' D_2'}{3M^2(1+\e)}\cdot e^{-2q|u-u_1-v_{r=0}(u_1)|}}}.
\end{align*}
We then proceed to derive sharp upper bound for $R^{\a\b\mu\nu}R_{\a\b\mu\nu}$. Recall the following expression of the Kretschmann scalar in \eqref{Kretschmann2}:
\begin{align} 
\nonumber &R^{\alpha\beta {\color{black} \mu \nu}}R_{\alpha\beta {\color{black} \mu \nu}}\\ \nonumber
=&\f{4}{r^2 \O^4}\bigg(16\cdot(\f{\partial^2 r}{\partial u \partial v})^2+16\cdot \f{\partial^2 r}{\partial u^2}\cdot\f{\partial^2 r}{\partial v^2} \bigg)\\ \nonumber
&+\f{4}{r^2 \O^4}\bigg(-32\cdot \f{\partial^2 r}{\partial u^2}\cdot\partial_v r  \cdot \partial_v \log\O -32\cdot\f{\partial^2 r}{\partial v^2}\cdot\partial_u r \cdot \partial_u \log\O \bigg)\\ \label{Kretschmann3}
&+\f{4}{r^4 \O^4}\bigg( 16\cdot (\partial_v r)^2\cdot (\partial_u r)^2+64\cdot \partial_v r\cdot r^2\cdot \partial_u r\cdot \partial_u \log\O\cdot \partial_v\log\O\bigg)\\ \nonumber
&+\f{32}{r^4 \O^2}\cdot\partial_v r\cdot \partial_u r+\f{4}{\O^8}\bigg(16\cdot (\f{\partial^2 \O}{\partial v \partial u})^2\cdot \O^2-32\cdot \f{\partial^2 \O}{\partial v \partial u}\cdot \O \cdot \O^2 \cdot \partial_v \log\O \cdot \partial_u \log\O \bigg)\\ \nonumber
&+\f{64}{\O^4}\cdot (\partial_v \log\O)^2\cdot (\partial_u \log\O)^2+\f{4}{r^4}.
\end{align}
Among the terms on the RHS, all but $\O^{-2}$ are controlled in the same manner as in Section \ref{Kretschmann}. Compared with Section \ref{Kretschmann}, to obtain a more accurate blow-up rate, we employ the following improved estimate in Proposition \ref{Omega 03} to control $\O^{-2}$: 
\begin{align*}
\O^{-2}(u,v)\leq& \f{4 r_0^{\f{6\t D_1 \t D_2}{M^2(1-\e)}\cdot e^{-2p|u-u_1-v_{r=0}(u_1)|}}}{2M-r_0}\cdot\f{r(u,v)}{r(u,v)^{\f{6\t D_1 \t D_2}{M^2(1-\e)}\cdot e^{-2p|u-u_1-v_{r=0}(u_1)|}}}\\
\lesssim& \f{r(u,v)}{r(u,v)^{\f{6 \t D_1 \t D_2}{M^2(1-\e)}\cdot e^{-2p|u-u_1-v_{r=0}(u_1)|}}}.
\end{align*}

\noindent All the other terms in \eqref{Kretschmann3} can be tackled as in Section \ref{Kretschmann}. And we deduce that
\begin{align*}
R^{\alpha\beta\rho\sigma}R_{\alpha\beta\rho\sigma}
\lesssim&\f{1}{r(u,v)^{\f{12 \t D_1 \t D_2}{M^2(1-\e)}\cdot e^{-2p|u-u_1-v_{r=0}(u_1)|}}}\cdot\f{1}{r(u,v)^6}+\f{1}{r(u,v)^{\f{12\t D_1 \t D_2}{M^2(1-\e)}\cdot e^{-2p|u-u_1-v_{r=0}(u_1)|}}}\cdot\f{1}{r(u,v)^4}\\
&+\f{1}{r(u,v)^{\f{24\t D_1 \t D_2}{M^2(1-\e)}\cdot e^{-2p|u-u_1-v_{r=0}(u_1)|}}}\cdot\f{1}{r(u,v)^6}+\f{1}{r(u,v)^4}\\
\lesssim&\f{1}{r(u,v)^{6+\f{24 \t D_1 \t D_2}{M^2(1-\e)}\cdot e^{-2p|u-u_1-v_{r=0}(u_1)|}}}.
\end{align*}
This concludes the proof of Proposition \ref{thm 8.1}.
\end{proof}

\subsection{Precise version of Theorem \ref{main thm}}
In double-null foliation, we consider a Lorentzian metric $g$ of the following form
\begin{equation*}
g=-\widehat{\Omega}^2(U,v) dUdv+r^2(U,v)(d\theta^2+\sin^2\theta d\varphi^2).
\end{equation*}
Let $(r, \widehat{\Omega}^2, \phi)(U,v)$ be the solution to the spherically symmetric cosmological Einstein--scalar field system \eqref{ESlam}. We prescribe initial data along $H_0$ and $\underline{H}_0$ as in Section \ref{sec:initialdata}. In conclusion, combining the results of Propositions \ref{phi}, \ref{partial r}, \ref{Omega 03}, \ref{mass inflation} and \ref{thm 8.1}, we summarize Theorem \ref{main thm} as follows.
\begin{theorem}
\label{thm:precisemain}
The spherically symmetric cosmological Einstein-scalar field system \eqref{ESlam} admits a unique smooth solution $(r, \widehat{\Omega}^2, \phi)(U,v)$ in $[0,U_0]_U\times [v_0,\infty)_v\cap \{r>0\}$ with initial data prescribed as in Section \ref{sec:initialdata}.

Moreover, near the spacelike singularity $\mathcal{S}$, in Eddington-Finkelstein coordinates $(u,v)$, which satisfies $u=-{(\alpha_S)^{-1}} \log \left(\frac{{(\alpha_S)}^{-1}}{U}\right)$ and hence $\Omega^2=\frac{dU}{du}\widehat{\Omega}^2$, there exist positive constants $c$, $C$, $\rho$, $\sigma$ and a sufficiently small $r_0$, such that for $v_1>v_0$ being sufficiently large, the following bounds hold in the region $(-\infty,u_0]\times [v_1,\infty)\cap \{0<r<r_0\}$:
\begin{align*}
&|r\partial_ur+M|(u,v)\leq\: Cr^{\frac{2}{3}}+ Ce^{-pv},\\
&|r\partial_vr+M|(u,v)\leq\: Cr^{\frac{2}{3}}+ Ce^{-pv},\\
&c  r^{\sigma e^{-2pv}} \leq r\Omega^2(u,v)\leq\: C  r^{\rho e^{-2qv}},\\
&c  r^{-\rho e^{-2qv}} \leq m(u,v) \leq\: C  r^{-\sigma e^{-2pv}},\\
&\rho e^{-qv} \leq  |r^2\partial_v\phi|(u,v)\leq\:  \sigma  e^{-pv},\\
&\rho e^{-qv} \leq  |r^2\partial_u\phi|(u,v)\leq\: \sigma  e^{-pv},\\
&\frac{c}{r^{6+\rho e^{-2qv} }}\leq \sum_{\alpha,\beta,\mu,\nu=0}^3R^{\alpha\beta \mu \nu}[g]R_{\alpha \beta \mu \nu}[g]\leq \frac{ C}{r^{6+\sigma e^{-2pv} }}.
\end{align*}
Here, the constants $c$, $C$, $r_0$ depend only on $M$; the constants $\rho$ and $\sigma$ depend on $M, D_1, D_2, D_3$. 

\end{theorem}

\bibliographystyle{plain}
\bibliography{bibliography}

\end{document}